\documentclass[prd,superscriptaddress,amsfonts,amssymb,amsmath,showpacs]{revtex4-2}
\usepackage{bm}
\usepackage{amsfonts}
\usepackage{latexsym}
\usepackage[latin1]{inputenc}
\usepackage{graphicx}
\usepackage{amsmath}
\usepackage{palatino}
\usepackage{mathpazo}
\usepackage[british]{babel}
\usepackage{hhline}
\usepackage{multirow}
\usepackage{textcomp}
\usepackage{float}
\usepackage{booktabs}
\usepackage{dcolumn}
\usepackage{hhline}
\usepackage{multirow}
\usepackage{ragged2e}
\usepackage{hyperref}
\hypersetup{colorlinks,citecolor=blue}
\hypersetup{colorlinks=true,linkcolor=red,filecolor=magenta,    urlcolor=cyan}
\usepackage{amsmath}
\usepackage{xcolor}
\usepackage{orcidlink}
\usepackage{epsfig}
\usepackage{caption}
\usepackage{subcaption}
\usepackage{commath}
\def\jnl@style{\it}
\def\aaref@jnl#1{{\jnl@style#1}}

\def\aaref@jnl#1{{\jnl@style#1}}

\def\aj{\aaref@jnl{AJ}}                   
\def\apj{\aaref@jnl{ApJ}}                 
\def\apjl{\aaref@jnl{ApJ}}                
\def\apjs{\aaref@jnl{ApJS}}               
\def\apss{\aaref@jnl{Ap\&SS}}             
\def\aap{\aaref@jnl{A\&A}}                
\def\aapr{\aaref@jnl{A\&A~Rev.}}          
\def\aaps{\aaref@jnl{A\&AS}}              
\def\mnras{\aaref@jnl{Mon.~Not.~Roy.~Astron.~Soc.}}             
\def\prd{\aaref@jnl{Phys.~Rev.~D}}        
\def\prc{\aaref@jnl{Phys.~Rev.~C}}  
\def\prl{\aaref@jnl{Phys.~Rev.~Lett.}}    
\def\qjras{\aaref@jnl{QJRAS}}             
\def\skytel{\aaref@jnl{S\&T}}             
\def\ssr{\aaref@jnl{Space~Sci.~Rev.}}     
\def\zap{\aaref@jnl{ZAp}}                 
\def\nat{\aaref@jnl{Nature}}              
\def\aplett{\aaref@jnl{Astrophys.~Lett.}} 
\def\apspr{\aaref@jnl{Astrophys.~Space~Phys.~Res.}} 
\def\physrep{\aaref@jnl{Phys.~Rep.}}      
\def\physscr{\aaref@jnl{Phys.~Scr}}       
\def\commat{\aaref@jnl{Comm.~Math.~Phys.}}              
\def\science{\aaref@jnl{Science}}               
\def\cqg{\aaref@jnl{Classical Quant.~Grav.}}            
\def\jpcs{\aaref@jnl{JPCS}}                                     
\def\ijmpd{\aaref@jnl{Int.~J.~Mod.~Phys.~D}}                    
\def\grg{\aaref@jnl{Gen.~Relat.~Gravit.}}               
\def\rpp{\aaref@jnl{Rep.~Prog.~Phys.}}          
\def\npa{\aaref@jnl{Nucl.~Phys.~A}}        
\def\lrr{\aaref@jnl{Living Rev.~Rel.}}                   
\def\jcap{\aaref@jnl{J.~Cosmology Astropart.~Phys.}}    
\def\rmp{\aaref@jnl{Rev.~Mod.~Phys.}}   
\def\epjc{\aaref@jnl{Eur.~Phys.~J.~C}} 
\def\plb{\aaref@jnl{~Phy.~Lett.~B}} 
\def\mpla{\aaref@jnl{Mod.~Phy.~Lett.~A}} 
\def\arxiv{\aaref@jnl{arxiv.org}}

\allowdisplaybreaks[1]
\renewcommand{\arraystretch}{1.1}
\begin{document}
\color{black}       

\title{\bf  Late Time Phenomena in $f(T,\mathcal{T})$ Gravity Framework: Role of $H_0$ Priors}

\author{L.K. Duchaniya \orcidlink{0000-0001-6457-2225}}
\email{duchaniya98@gmail.com}
\affiliation{Department of Mathematics, Birla Institute of Technology and Science, Pilani, Hyderabad Campus, Jawahar Nagar, Kapra Mandal, Medchal District, Telangana 500078, India.}

\author{B. Mishra \orcidlink{0000-0001-5527-3565}}
\email{bivu@hyderabad.bits-pilani.ac.in}
\affiliation{Department of Mathematics, Birla Institute of Technology and Science, Pilani, Hyderabad Campus, Jawahar Nagar, Kapra Mandal, Medchal District, Telangana 500078, India.}

\begin{abstract}   
{\bf{Abstract:}} This study explored the behavior of the $f(T, \mathcal{T})$ cosmological model with the use of various data set combinations. We also compared the results for this model between the Pantheon+ (without SH0ES) and the Pantheon+\&SH0ES (with SH0ES) data sets. Additionally, we incorporated data from BAO along with $H_0$ priors. We observed that integrating SH0ES data points leads to a higher estimation of $H_0$ than Pantheon+ (without SH0ES). We perform an extensive MCMC analysis for each combination of data sets, providing constraints on the model parameters. We also computed the $\chi^2_{min}$ value for each combination of data sets to evaluate the chosen model against the standard $\Lambda$CDM model. Our primary finding is that the various dataset combinations in the $f(T, \mathcal{T})$ model we examined relate to a range of Hubble constants, which could contribute to reducing the cosmic tension associated with this parameter. Additionally, we investigate the evolution of matter fluctuations by solving the density contrast evolution equation numerically. We calculate numerical solutions for the weighted growth rate $f\sigma_8$ using these findings. We plotted the cosmological background parameters to check the behavior of the $f(T, \mathcal{T})$ model in late-time. Based on the behavior of these background cosmological parameters, we conclude that our selected models reflect the late-time cosmic dynamics of the Universe. 
\end{abstract}

\maketitle
\section{Introduction} \label{SEC-I}
The late-time cosmic behavior of the Universe has emerged as the most compelling enigmas in contemporary cosmology. Following an extended period of almost uniform expansion, observational evidence revealed that an unidentified energy component is driving spacetime expansion at an accelerating rate \cite{Riess:1998cb, Perlmutter:1998np}. The leading candidate for this phenomenon is dark energy (DE) \cite{Bennett_2003a, Tegmark_2004a}, which has a major share of approximately 70$\%$ of the total mass-energy budget of the Universe. As a consequence, it largely influences the ultimate fate of the Universe. The discovery of cosmic acceleration poses profound questions about our understanding of gravity and drives new investigations into the fundamental principles of physics. The \(\Lambda\)CDM model that describes the Universe as a composition of cold dark matter (CDM) and DE represented by a cosmological constant \(\Lambda\) has been corroborated by several cosmological observations \cite{Alam_2017ab, Scolnic_2018, Riess:2019cxk, Benisty_2021, Cawthon_2022DES, Gupta_2023}. However, the discrepancies in Hubble constant \(H_0\) and the structure growth parameter \(S_8\) pose additional challenges. The recent observed values of \(H_0\) show a notable rift between the value derived from early Universe observations such as cosmic microwave background (CMB) via the Planck satellite \cite{plank2013} and gleaned from local measurements utilizing Cepheid variables and Supernovae Type Ia \cite{Riess_2022panplus}. In particular, Riess et al. \cite{Riess_2022panplus} within the SH0ES Team have reported a Hubble constant estimate of \(73.30 \pm 1.04 \, \text{km s}^{-1} \, \text{Mpc}^{-1}\), based on Supernovae Type Ia observations. Whereas the H0LiCOW Collaboration \cite{wang_H0LiCOWmnras} derived a value of \(73.3^{+1.7}_{-1.8} \, \text{km s}^{-1} \, \text{Mpc}^{-1}\) through a strong gravitational lensing of quasars. Further, Freedman et al. \cite{Freedman_2019TRGB} reported a lower estimate of \(69.8 \pm 1.9 \, \text{km s}^{-1} \, \text{Mpc}^{-1}\), obtained using the tip of the red giant branch (TRGB) as a distance indicator. The Planck Collaboration \cite{Aghanim:2018eyx} provides a Hubble constant value of \(67.4 \pm 0.5 \, \text{km s}^{-1} \, \text{Mpc}^{-1}\), while Aboot et al. \cite{Abbott_2018mnras} suggest \(67.2^{+1.2}_{-1.0} \, \text{km s}^{-1} \, \text{Mpc}^{-1}\). This divergence between the estimates from early and late Universe measurements, known as \(H_0\) tension, raises significant questions in modern cosmology. Initially identified with the first release of Planck data \cite{plank2013}, the \(H_0\) tension has garnered increasing scrutiny and attention in recent studies \cite{Di_Valentino_2017, Aghanim:2018eyx, Di_Valentino_2021}. This inconsistency has led to exploring possible modifications to the standard cosmological model and considering new physics beyond \(\Lambda\)CDM. The $\sigma_8$ tension refers to a discrepancy of approximately $3\sigma$ between the values of $\sigma_8$, which quantifies the amplitude of matter clustering on scales of $8 \, h^{-1} \, \text{Mpc}$. In particular, the $\sigma_8$ value inferred from Cosmic Microwave Background (CMB) measurements by \cite{Aghanim:2018eyx} is higher than that derived from large-scale structure observations, such as those from SDSS/BOSS \cite{Alam_2017ab, Ata_2017BAO}. However, recent analyses from the KiDS-1000 collaboration \cite{Asgari_2021, Ruiz_Zapatero_2021} and the SRG/eROSITA all-sky survey \cite{Ghirardini_2024} demonstrate full consistency with the Planck 2018 results \cite{Aghanim:2018eyx}, suggesting that the previously reported $\sigma_8$ tension may be less significant than initially thought. 

General relativity (GR) encapsulates the interplay between matter, energy and the curvature of spacetime. However, modifications in GR have been inevitable to address the late-time behavior of the Universe. A simple approach to understanding this phenomenon is the introduction of cosmological constant, leading to the \(\Lambda\)CDM model. Otherwise,  GR can be refined by integrating extra terms or scalar fields into the gravitational action. These modifications may include higher-order curvature invariants or scalar fields coupled to the curvature. This may provide a robust theoretical framework that accommodates a concordance model for cosmology through various straightforward adjustments \cite{Nojiri2006, NOJIRI2007238a}. The modifications can be incorporated as the Einstein-Hilbert action \cite{Sotiriou:2008rp, Capozziello:2011et} to provide various cosmological phenomena. One such gravitational modification is the torsional equivalent formulation of GR, known as the Teleparallel Equivalent of GR (TEGR) \cite{Maluf:1994j, Aldrovandi:2013wha}. TEGR permits the formulation of second-order equations in four-dimensional spacetime and employs the Weitzenb$\ddot{o}$ck connection \cite{Weitzenbock1923}. This framework is characterized by four linearly independent tetrad fields, which serve as the orthonormal bases for the tangent space at each point in spacetime, with the torsion tensor being derived from the first derivatives of these tetrads. To note,  \( f(R) \) gravity \cite{Sotiriou:2008rp} presents the basic modification of GR, whereas \( f(T) \) \cite{Bengochea:2008gz, Ferraro:2008ey, Linder:2010py, Chen:2010va, Cai_2016, Duchaniya:2022rqu, Briffa_2024, Duchaniya_2024ab} gravity is the modification of TEGR. Further,  many variations of the torsion-influenced gravitational theory such as $f(T, T_\mathcal{G})$ ($T_\mathcal{G}$ refers to the Gauss-Bonnet term) \cite{Kofinas:2014owa, Kofinas:2014daa}, $f(T, B)$ \cite{Escamilla-Rivera:2019ulu, Kadam_2023ab} ($B$ signifies the boundary term), $f(T, \phi)$ \cite{Gonzalez-Espinoza:2020jss, Gonzalez-Espinoza:2021mwr, Duchaniya_2023noet, Duchaniya_2023tphi} ($\phi$ represents the scalar field) and $f(T, \mathcal{T})$ \cite{Harko_2014a, Momeni_2014, Jackson2016a, Junior_2016, Pace_2017, Duchaniya_2024tt} ($\mathcal{T}$ represents trace of energy-momentum tensor) have been proposed.

The \( H_0 \) tension refers to the observed discrepancies in the Hubble constant measurements in various observational methodologies, which challenges the standard \( \Lambda \)CDM model \cite{Vagnozzi_2020newph, Freedman_2021apjh0tension, Abdalla:2022yfr, Moresco_2022_H0, Brout_2022pan, Briffa_2022, Vagnozzi_2023, Briffa_2023mnras, Capozziello_2024_ten, divalentino2025cosmoversewhitepaperaddressing}. The motivation of this study is to examine the influence of $H_0$ priors in alleviating $H_0$ tension and late time behavior of the Universe in $f(T, \mathcal{T})$ gravitational theory \cite{Harko_2014a}. The gravitational Lagrangian can be developed in this framework using the torsion scalar $T$ functions and the trace of the energy-momentum tensor $\mathcal{T}$. The paper is organized as follows: In Sec.-\ref{Mathformalism}, we present a brief overview of the mathematical formalism of $f(T, \mathcal{T})$ gravity. The formalism of the cosmological data sets and the mathematical formalism of the statistical criteria for model comparison (AIC and BIC) is defined in Sec.-\ref{cosmological observation}. In Sec.-\ref{model}, we analyze the cosmological observations for different data set combinations and the $H_0$ priors. In Sec.-\ref{linearperturbationection}, we study the growth of matter overdensities under the quasi-static approximation for sub-horizon scales. In Sec.-\ref{cosmologicalparameters}, we present the evolution of the background cosmological parameters to analyze the behavior of the models at late times. We have given a conclusion on the findings in Sec.-\ref{conclusion}. The results of cosmological observation for the $\Lambda$CDM model are summarized in the appendix. 
\section{Mathematical formalism} \label{Mathformalism}
In teleparallel gravity (TG), the tetrad fields \( e^{A}_{\mu} \) act as dynamical variables in place of the usual metric tensor \( g_{\mu \nu} \) in GR. The metric tensor in TG can be represented as, 
\begin{equation}\label{1}
g_{\mu \nu}=\eta_{AB} e_{\mu}^{A} e_{\nu}^{B} \,.    
\end{equation}  
The Greek indices denote space-time coordinates, while the capital Latin indices indicate tangent space-time coordinates.
\( \eta_{AB} \) denotes the Minkowski space-time and the tetrad fields satisfy the orthogonality condition \( e^{\mu}_{\,\,\,\, A}  e^{B}_{\,\,\,\,\mu} = \delta_A^B \). Employing the Weitzenb$\ddot{o}$ck connection as a framework for \( f(T,\mathcal{T}) \) gravity, we can define
\begin{equation}\label{2}
\hat\Gamma^{\lambda}_{\nu \mu}\equiv e^{\lambda}_{A} (\partial_{\mu} e^{A}_{\nu}+ \omega^{A}_{\,\,\, B \mu} e^{B}_\nu),
\end{equation}
where \( \omega^{A}_{\,\,\, B \mu} \) represents a spin connection that guarantees invariance under Lorentz transformations, which stems directly from the indices of the tangent space. The most general spin connection that results in zero curvature is the purely inertial spin connection 
\begin{equation}\label{spinconnection}
\omega^{A}_{\,\,\, B \mu}  = \Lambda^{\,A}_C \partial_{\mu} \Lambda^{\,C}_B\,,   
\end{equation}
where $\Lambda^{\,A}_C$ is a local Lorentz transformation matrix. The following represents solely the effects of inertia, as it relies only on the selection of the frame. Nevertheless, it is not uniquely defined because it depends solely on the choice of the observer, signified by a Lorentz matrix. Specifically, different spin connections indicate various inertial effects for the observer, which do not influence the field equations. On the other hand, the spin connections used in GR are not flat \cite{misner1973gravitation} because they depend on the tetrads. In TG, the equations governing motion incorporate both gravitational and local degrees of freedom, depicted by the pair of tetrads and spin connections. Now, the torsion tensor characterized by the anti-symmetric part of the Weitzenb$\ddot{o}$ck connection can be described as,
\begin{equation}\label{3}
T^{\lambda}_{\mu \nu}\equiv\hat\Gamma^{\lambda}_{\nu \mu}-\hat\Gamma^{\lambda}_{\mu \nu}=e^{\lambda}_{A} \partial_{\mu} e^{A}_{\nu}-e^{\lambda}_{A} \partial_{\nu} e^{A}_{\mu}.    
\end{equation}

The torsion tensor exhibits covariance under diffeomorphisms and Lorentz transformations. Provided that the torsion tensor has been correctly contracted, the torsion scalar can be represented as,
\begin{equation}\label{4}
T \equiv \frac{1}{4} T^{\rho \mu \nu} T_{\rho \mu \nu}+\frac{1}{2} T^{\rho \mu \nu} T_{\nu \mu \rho}-T_{\rho \mu}^{~~\rho} T^{\nu \mu}_{~~\nu}.
\end{equation}

The action in TG is based on the teleparallel Lagrangian \( T \) and the \( f(T) \) gravity extends this Lagrangian to an arbitrary function \( f(T) \). The action of \( f(T,\mathcal{T}) \) gravity \cite{Harko_2014a} is,
\begin{equation}\label{5}
S = \frac{1}{16 \pi G}\int d^{4}x\,e\,[T+f(T,\mathcal{T})]+ \int d^{4}x\,e\,\mathcal{L}_{m}\,,   \end{equation}
where \( \mathcal{L}_{m} \) denotes matter Lagrangian and \( G \) represents the gravitational constant. The determinant of the tetrad field is expressed as \( e = \text{det}[e^{A}_{\,\,\,\,\mu}] = \sqrt{-g} \). Varying action \eqref{5} with respect to the tetrad field, one can obtain the gravitational field equations for \( f(T,\mathcal{T}) \) gravity as,
\begin{eqnarray}\label{6}
&&[e^{-1}\partial_{\mu}(e e^{\rho}_{A}S_{\rho}^{~\mu \nu})-e^{\lambda}_{A}T^{\rho}_{~\mu \lambda}S_{\rho}^{~ \nu\mu}](1+f_{T}) +e^{\rho}_{A}S_{\rho}^{~\mu \nu}[\partial_{\mu}(T)f_{TT} +\partial_{\mu}(\mathcal{T})f_{T\mathcal{T}}]+\frac{1}{4}e^{\nu}_{A}[T+f(T)]\nonumber \\ &&-f_\mathcal{T}\left(\frac{e^{\rho}_{A}T_{~\rho}^{~~\nu}+p e^{\rho}_{A}}{2}\right)=4 \pi G e^{\rho}_{A}T_{~\rho}^{~~\nu}.
\end{eqnarray}
For brevity, we denote \( f_T = \frac{\partial f}{\partial T} \), \( f_{TT} = \frac{\partial^2 f}{\partial T^2} \), \( f_{\mathcal{T}} = \frac{\partial f}{\partial \mathcal{T}} \) and \( f_{T\mathcal{T}} = \frac{\partial^2 f}{\partial T \partial \mathcal{T}} \). The total energy-momentum tensor can be represented as \( T_{~\rho}^{~~\nu} \). The superpotential, \( S_{\rho}^{~~\mu \nu} \equiv \frac{1}{2}(K^{\mu \nu}_{~~~\rho} + \delta^{\mu}_{\rho} T^{\alpha \nu}_{~~~\alpha} - \delta^{\nu}_{\rho} T^{\alpha \mu}_{~~~\alpha}) \). The contortion tensor in the superpotential can be defined as, \( K^{\mu \nu}_{~~~\rho} \equiv \frac{1}{2}(T^{\nu \mu}_{~~~\rho} + T_{\rho}^{~~\mu \nu} - T^{\mu \nu}_{~~~\rho}) \).

We consider the homogeneous and isotropic flat Friedmann-Lema\^{i}tre-Robertson-Walker (FLRW) space-time as,
 \begin{equation}\label{7}
ds^{2}=dt^{2}-a^{2}(t)[dx^2+dy^2+dz^2]\,,
\end{equation}
where \( a(t) \) denotes the scale factor that represents the rate of expansion in the spatial dimensions. From Eq.~\eqref{4}, the torsion scalar can be derived as,
\begin{equation}\label{8}
T=-6H^{2}\,,
\end{equation}
and the associated tetrad field can be expressed as \( e^{A}_{\mu} \equiv \text{diag}(1, a(t), a(t), a(t)) \). We can now derive the field equations for \( f(T,\mathcal{T}) \) gravity Eq.~\eqref{6} as,
\begin{eqnarray}
3H^2&=&8\pi G \rho_m-\frac{1}{2}(f+12 H^{2}f_T)+f_{\mathcal{T}}(\rho_m+p_m), \label{firstfriedmann}\\
\dot{H}&=&-4\pi G(\rho_{m}+p_{m})-\dot{H}(f_{T}-12 H^{2}f_{TT})-H(\dot{\rho}_{m}-3\dot{p}_{m}) f_{T \mathcal{T}}-f_{\mathcal{T}}\left(\frac{\rho_m+p_m}{2}\right).\label{secondfriedmann} 
\end{eqnarray}
An over dot represents ordinary derivative with respect to cosmic time \( t \).  The trace of the energy-momentum tensor, \( \mathcal{T} = \rho_{m} - 3 p_{m} \), where \( p_{m} \) denotes the matter pressure and \( \rho_{m} \) be the corresponding energy density term. The Friedmann Eqs.(\ref{firstfriedmann}-\ref{secondfriedmann}) can be written as,
\begin{eqnarray}
3 H^{2}&=& 8 \pi G(\rho_{m}+\rho_{DE}), \label{11}\\
-2\dot{H}&=&8 \pi G(\rho_{m}+p_{m}+\rho_{DE}+p_{DE}). \label{12}
\end{eqnarray}

From Eqs. (\ref{firstfriedmann}-\ref{12}), the energy density ($\rho_{DE}$) and  pressure ($p_{DE}$) for the DE component can be retrieved as,
\begin{eqnarray}
\rho_{DE}&\equiv&-\frac{1}{16 \pi G}\left[f+12 H^{2}f_{T}-2 f_{\mathcal{T}}(\rho_m+p_m)\right], \label{13} \\
p_{DE}&\equiv& (\rho_m+p_m)\left[\frac{1+\frac{f_{\mathcal{T}}}{8 \pi G}}{1+f_{T}-12 H^{2}f_{TT}+H \dfrac{d \rho_{m}}{dH}(1-3 c^{2}_{s})f_{T \mathcal{T}}}-1\right]+ \frac{1}{16 \pi G}[f+12 H^{2}f_{T}-2 f_{\mathcal{T}}(\rho_m+p_m)].\label{14} 
\end{eqnarray}
The behavior of the total EoS parameter ($\omega_{tot}$) and the deceleration parameter ($q$) are significant to analyze the late time behavior of the Universe and can be expressed as, 
\begin{eqnarray} \label{omegatot} 
\omega_{tot}&=& \frac{p_{m}+p_{DE}}{\rho_{m}+\rho_{DE}}\equiv -1-\frac{2\dot{H}}{3 H^{2}},\\
q &=& \frac{1}{2}(1+3 \omega_{tot})\,.\label{dece}
\end{eqnarray}

 \section{cosmological Observation data sets} \label{cosmological observation}    
 We have presented brief descriptions of the cosmological data sets and the methodology. The data sets to be used are Hubble, Supernovae Type Ia and BAO. The $H_0$ priors used are R21 and TRGB. We shall use the emcee package Ref \cite{Foreman_Mackey_2013} to execute a Markov Chain Monte Carlo (MCMC) analysis to integrate these data sets. MCMC is a robust sampling technique that extracts samples from the posterior distribution of the cosmological model. This method is widely used in Bayesian statistics to estimate model parameters and quantify uncertainties. A large sample size would be required to ensure the accuracy of the model in this process. We compute the one-dimensional parameter distribution that provides the posterior distribution of each parameter and the two-dimensional parameter distributions, revealing the covariance relationship between pairs of parameters. Our analysis culminates in generating MCMC corner plots at 1$\sigma$ and 2$\sigma$ confidence levels.\\

\textbf{Cosmic Chronometers (CC):} We employed 31 data points for the Hubble parameter derived from the cosmic chronometer (CC) methodology. This technique allows for directly analyzing the Hubble function across a range of redshifts up to $z \leq 2$. The strength of CC data lies in its ability to evaluate the age difference between two passively evolving galaxies that formed simultaneously but have a small difference in redshift, enabling the calculation of \( \frac{\Delta z}{\Delta t} \). Refs \cite{Jimenez_2002, Zhang_2014hz, Jimenez_2003cmb, Moresco_2016hubb, Simon_2005prd, M_Moresco_2012JCAP, Daniel_Stern_2010jcap, Moresco_2015mnras} provide the basis for the CC data points. The associated estimate for $\chi^{2}_{CC}$ is given as,
\begin{equation}\label{chisqure_hz}
\chi^2_{CC}= \Delta H(z_i, \Theta)^{T} \,C_{CC}^{-1} \,\Delta H(z_i, \Theta) \,,   
\end{equation}
where $\Delta H(z_i, \Theta)= H(z_i, \Theta)-H_{\text{obs}}(z_i)$ and $C_{CC}^{-1}$ is the covariance matrix provided in \cite{Moresco_2020covariance}. The term $H(z_i, \Theta)$ represents the theoretical values of the Hubble parameter at a specific redshift $z_i$, while $H_{\text{obs}}(z_i)$ denotes the observed values of the Hubble parameter at $z_i$.

\textbf{Pantheon+ Supernovae Type Ia (SNIa) dataset :} We utilize the 1701 data points obtained from the Pantheon+ sample \cite{Brout_2022panplus, Riess_2022panplus, Scolnic_2022panplus}, which records the apparent distances of 1550 different Supernovae Ia (SNIa) occurrences within a redshift range of $0.01 < z < 2.3$. The distance modulus function is the discrepancy between the apparent magnitude $m$ and the absolute magnitude $M$. The Pantheon+ dataset incorporates SH0ES measurements of Cepheid distances to simultaneously ascertain $H_0$ and $M$ without including a prior in the parameters. From this point forward, this data set will be called PN$^{+}$. In our evaluation, we examined the PN$^{+}$ dataset together with the SH0ES data point and without it. At a redshift $z_i$, the distance modulus function $\mu(z_i, \Theta)$ can be formulated as
\begin{equation}\label{modulus_function}
\mu(z_i, \Theta) = m - M = 5 \log_{10} \left[ D_L(z_i, \Theta) \right] + 25\,,    
\end{equation}
the luminosity distance $D_L(z_i, \Theta)$ can be formulated as
\begin{equation}\label{luminosity_distance}
D_L(z_i, \Theta) = c(1 + z_i) \int_0^{z_i} \frac{dz'}{H(z', \Theta)}\,,  
\end{equation}

where $c$ denotes the speed of light. To determine the chi-square ($\chi^2_{\text{SN}}$) value with the $PN^{+}\& SH0ES$  compilation, which includes 1701 Supernovae data points, we apply the following formula\cite{Conley_2010Apjs}:
\begin{equation}\label{chisquare_pantheon}
 \chi^2_{\text{SN}} = \left(\Delta\mu(z_i, \Theta)\right)^{T} C^{-1} \left(\Delta\mu(z_i, \Theta)\right)\,.
\end{equation}

In this context, $C$ refers to the covariance matrix incorporating systematic and statistical uncertainties in the measurements. Additionally, $\Delta\mu(z_i, \Theta) = \mu(z_i, \Theta) - \mu(z_i)_{\text{obs}}$ indicates the disparity between the predicted and observed distance modulus at the redshift $z_i$.

\textbf{BAO data set:} We also analyze a combined baryon acoustic oscillation (BAO) data set of distinct data points. We utilize a BAO data set that features observations from the Six-degree Field Galaxy Survey at an effective redshift of \( z_{\text{eff}} = 0.106 \) \cite{Beutler_2011baosixdegree}, the BOSS DR11 quasar Lyman-alpha measurements at \( z_{\text{eff}} = 2.4 \) \cite{du_Mas_des_Bourboux_2017} and the SDSS Main Galaxy Sample at \( z_{\text{eff}} = 0.15 \) \cite{Ross_2015}. Additionally, we incorporate measurements of \( H(z) \) and angular diameter distances obtained from the SDSS-IV eBOSS DR14 quasar survey at effective redshifts \( z_{\text{eff}} = \{0.98, 1.23, 1.52, 1.94\} \) \cite{Zhao_2018sdss_IV}, along with the consensus BAO measurements of the Hubble parameter and the corresponding comoving angular diameter distances from SDSS-III BOSS DR12 at \( z_{\text{eff}} = \{0.38, 0.51, 0.61\} \) \cite{Alam_2017ab}. Our analysis considers the complete covariance matrix for the sets of BAO data. To evaluate the BAO data set for the cosmological model, it is necessary to establish the Hubble distance \( D_H(z) \), the comoving angular diameter distance \( D_M(z) \) and the volume-average distance \( D_V(z) \).
\begin{eqnarray}\label{BAO_distances}
D_H(z) = \frac{c}{H(z)},\quad D_M(z) = (1 + z)D_A(z), \quad D_V(z) = \left[(1 + z)^2 D_A^2(z) \frac{z}{H(z)}\right]^{1/3}\,,
\end{eqnarray}

where $D_A(z) = (1+z)^{-2} D_L(z)$ denotes the angular diameter distance. To include the BAO findings in MCMC analyses, we need to take into account the pertinent combinations of parameters:
\begin{eqnarray} \label{parametercombination}
\mathcal{F}(z_i) = \bigg\{ \frac{D_V(z_i)}{r_s(z_d)}, \frac{r_s(z_d)}{D_V(z_i)}, D_H(z_i), D_M(z_i) \bigg( \frac{r_{s, \text{fid}}(z_d)}{r_s(z_d)} \bigg), H(z_i) \bigg( \frac{r_s(z_d)}{r_{s, \text{fid}}(z_d)} \bigg), D_A(z_i) \bigg( \frac{r_{s, \text{fid}}(z_d)}{r_s(z_d)} \bigg) \bigg\},
\end{eqnarray}

where \( r_s(z_d) \) represents the sound horizon at the drag epoch, while \( r_{s, \text{fid}}(z_d) \) indicates the fiducial sound horizon. To accomplish this, we calculated the comoving sound horizon \( r_s(z) \) at the redshift \( z_d \approx 1059.94 \) \cite{Aghanim:2018eyx}, which corresponds to the conclusion of the baryon drag epoch.
\begin{eqnarray}\label{sound_horizon}
r_s(z) = \int_{z}^{\infty} \frac{c_s(\tilde{z})}{H(\tilde{z})} d\tilde{z} = \frac{1}{\sqrt{3}} \int_{0}^{1/(1+z)}\frac{da}{a^2 H(a) \sqrt{1 + \left[\frac{3\Omega_{b,0}}{4\Omega_{\gamma,0}}\right] a}},
\end{eqnarray}

where the subsequent values are utilized: \(\Omega_{b,0} = 0.02242\) \cite{Aghanim:2018eyx}, \(T_0 = 2.7255 \, \text{K}\) \cite{Fixsen_2009temcmb} and a reference value of \(r_{s, \text{fid}}(z_d) = 147.78 \, \text{Mpc}\). The relevant estimate for \(\chi^{2}_{\text{BAO}}\) is given by \cite{Conley_2010Apjs},
\begin{equation}\label{chisquare_pantheon}
\chi^2_{\text{BAO}}(\Theta) = \left(\Delta \mathcal{F}(z_i, \Theta)\right)^T C^{-1}_{\text{BAO}} \Delta \mathcal{F}(z_i, \Theta)\,,
\end{equation}

where \(C_{\text{BAO}}\) denotes the covariance matrix for the chosen BAO data and \(\Delta \mathcal{F}(z_i, \Theta) = \mathcal{F}(z_i, \Theta) - \mathcal{F}_{\text{obs}}(z_i)\) indicates the discrepancy between the theoretical and measured values of \(\mathcal{F}\) at redshift \(z_i\).
\\

We will investigate how an $H_0$ prior affects the selected functional form \( f(T, \mathcal{T}) \) along with the previously described data set. We will consider the recent local measurement from SH0ES which gives $H_0 = 73.04 \pm 1.04 \, \text{km} \, \text{s}^{-1} \text{Mpc}^{-1}$ (R21) \cite{Riess_2022panplus}. The measurement using the tip of the red giant branch (TRGB) as a standard candle with \( H_0 = 69.8 \pm 1.9 \, \text{km} \, \text{s}^{-1} \, \text{Mpc}^{-1} \) \cite{Freedman_2019TRGB}.

We analyze the effectiveness of the cosmological model and dataset by computing their respective minimum $\chi^{2}_{\text{min}}$ values, which are derived from the maximum likelihood $L_{\text{max}}$, using the formula $\chi^{2}_{\text{min}} = -2 \ln L_{\text{max}}$. Additionally, we compare the model against the standard $\Lambda$CDM by applying the Akaike Information Criteria (AIC), which takes into account both the quality of the fit (evaluated via $\chi^{2}_{\text{min}}$) and the complexity of the model (indicated by the number of parameters $k$). The AIC is formulated as
\begin{equation}\label{AIC}
\text{AIC} = \chi^{2}_{min} + 2k,   
\end{equation}

A smaller AIC value indicates that a model provides a better fit for the data while considering its complexity. The AIC imposes a penalty on models with a greater number of parameters, even if they demonstrate a better fit to the data. Therefore, a model with a lower AIC is preferred over one with a higher AIC, as long as the difference in AIC is significant enough.

In addition, we examine the Bayesian Information Criterion (BIC), which is similar to AIC but gives more weight to the complexity of the model than AIC does, and is expressed as
\begin{equation}\label{BIC}
\text{BIC} = \chi^{2}_{min} + k \ln N \,.    
\end{equation}

In this scenario, $N$ denotes the quantity of samples within the observational data set. The BIC serves the same purpose as the AIC: to find a balance between the accuracy of the model concerning the data and its complexity. However, the BIC applies a stricter penalty on models with more parameters than the AIC due to its dependence on the logarithm of the sample size. Consequently, the penalty for adding parameters becomes more significant as the sample size increases. Comparing the BIC values of two models can help determine which model aligns better with the data, with lower BIC values being favored, provided there is a notable difference.

To assess the performance of models using different combinations of data sets, we calculate the differences in AIC and BIC between the model and the reference model, $\Lambda$CDM. The constrained parameters for the $\Lambda$CDM model for each combination of data sets are outlined in Appendix~\ref{LCDMapp}. Lower values of $\Delta$AIC and $\Delta$BIC suggest that the chosen model with the data set corresponds more closely to the $\Lambda$CDM model, indicating better performance. The quantities $\Delta$AIC, and $\Delta$BIC can be defined as,
\begin{align}
\Delta\text{AIC} &= \Delta \chi^{2}_{min} + 2\,\Delta k\,,\\
\Delta\text{BIC} &= \Delta \chi^{2}_{min} + \Delta k\, \ln N\,.
\end{align}
\section{cosmological model} \label{model}
In this section, we present and evaluate the results following the methodology described in Sec~\ref{cosmological observation}, using cosmological observations datasets. We showcase contour plots of the constrained parameters and their respective $1\sigma$ and $2\sigma$ uncertainties, accompanied by tables summarizing the final results. In all tables and posterior plots, we include results for the Hubble constant $H_0$, the current matter density parameter $\Omega_{m0}$ and the parameters of the models. This will enable us to examine how the independent data sets and cosmological models influence the Hubble tension. We will briefly highlight the most significant findings, emphasizing the PN$^+$ (without SH0ES) and PN$^+$\&SH0ES (with SH0ES) distinctions. In the Friedmann equations [Eq.\eqref{firstfriedmann}-Eq.\eqref{secondfriedmann}], we need to incorporate some functional form of \( f(T, \mathcal{T}) \). In this setting, we have considered the following form of \( f(T, \mathcal{T}) \) \cite{Harko_2014a}
\begin{equation} \label{modelconsidered}
f(T,\mathcal{T}) = \alpha T^{n} \mathcal{T} + \Lambda \,,  \end{equation}

where \( \alpha \neq 0\), \( n \neq 0 \) and \( \Lambda \) are arbitrary constants. At present, one can express the Friedmann Eq.~\eqref{firstfriedmann} as,
\begin{equation}\label{alphapresenttime}
\alpha = \frac{2-2\Omega_{m0}+\frac{\Lambda}{3 H_{0}^2}}{(1+2n)\Omega_{m0} (-6 H_{0}^2)^{-n}}\,,    
\end{equation}

where \( H_0 \) and \( \Omega_{m0} \) respectively represent the Hubble parameter and matter density parameter at present time. From Eq.~\eqref{alphapresenttime}, it can be inferred that the model parameter \( \alpha \) depends on other parameters such as \{\( H_0 \), \( \Omega_{m0} \), \( n \) and \( \Lambda \)\}. By defining the dimensionless Hubble parameter \( E(z)= \frac{H(z)}{H_0} \), the Friedmann Eq.~\eqref{firstfriedmann} for this model can be reformulated as
\begin{eqnarray}\label{modelHz}
E^{2}(z)= (1+z)^3 \Omega_{m0}-\frac{\Lambda}{6 H_{0}^2}+\bigg(1-\Omega_{m0}+ \frac{\Lambda}{6 H_{0}^2}\bigg) (1+z)^3 E^{2n}(z)\,.  
\end{eqnarray}

To ensure that the term \( \frac{\Lambda}{6 H_{0}^2} \) remains dimensionless, we define the parameter as \( \Lambda =H_{0}^2 \). Now, Eq.~\eqref{modelHz} becomes,
\begin{eqnarray}\label{conmodelhz}
E^{2}(z)= (1+z)^3 \Omega_{m0}-\frac{1}{6}+\left(1-\Omega_{m0}+ \frac{1}{6}\right) (1+z)^3 E^{2n}(z)\,.     
\end{eqnarray}

Eq.~\eqref{conmodelhz} is an implicit formulation for \(E(z)\). Considering that analytical solutions for Eq.~\eqref{conmodelhz} are impractical, we have adhered to the numerical methods to compute the parameters. The methodology used as described in Sec.-\ref{cosmological observation} and the results obtained are described below.

\begin{figure}[H]
     \centering
     \begin{subfigure}[b]{0.49\textwidth}
         \centering
         \includegraphics[width=\linewidth]{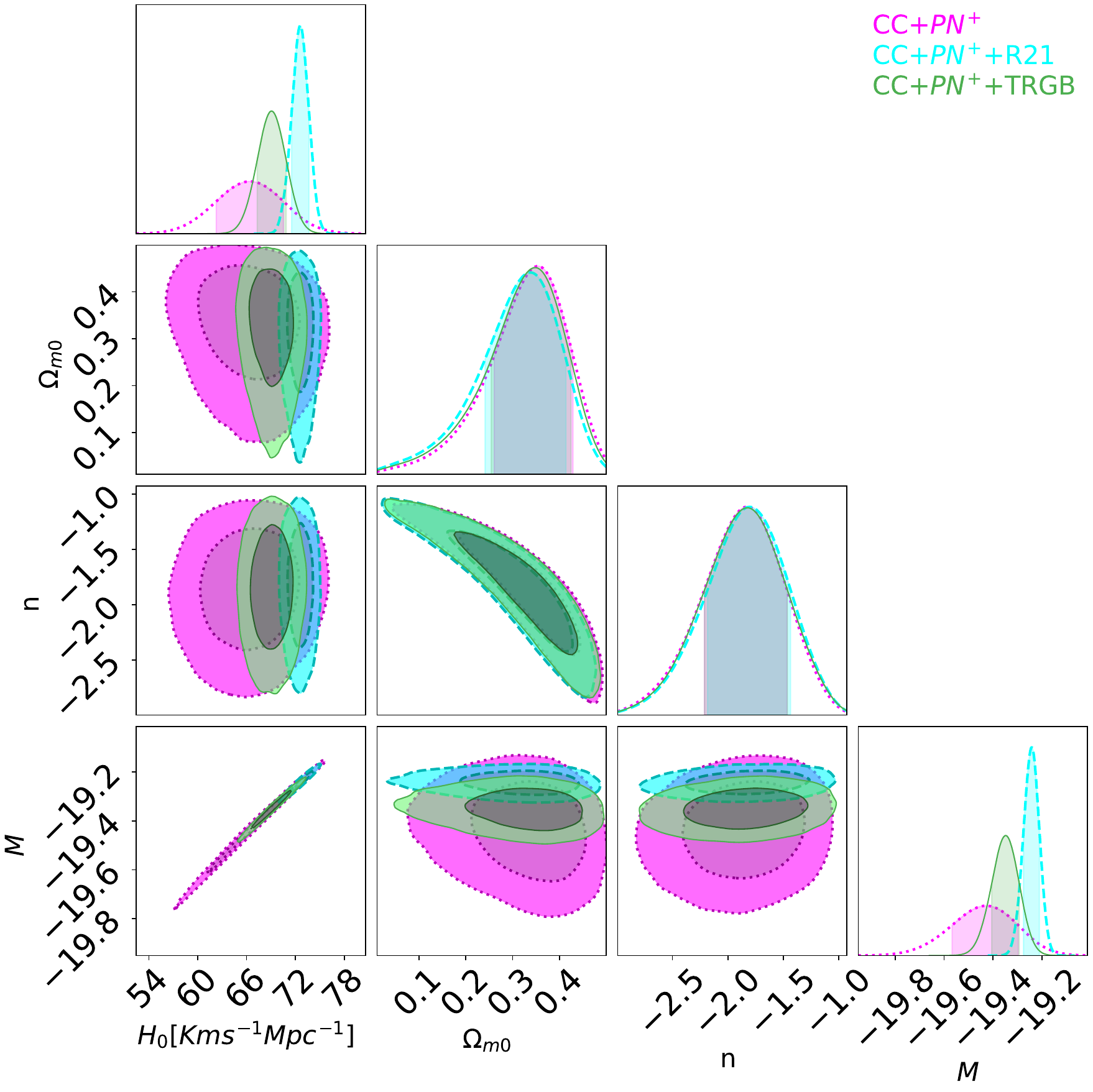}
          \caption{}
         \label{fig:CCMCMCplus}
     \end{subfigure}
     \hfill
     \begin{subfigure}[b]{0.49\textwidth}
         \centering
         \includegraphics[width=\linewidth]{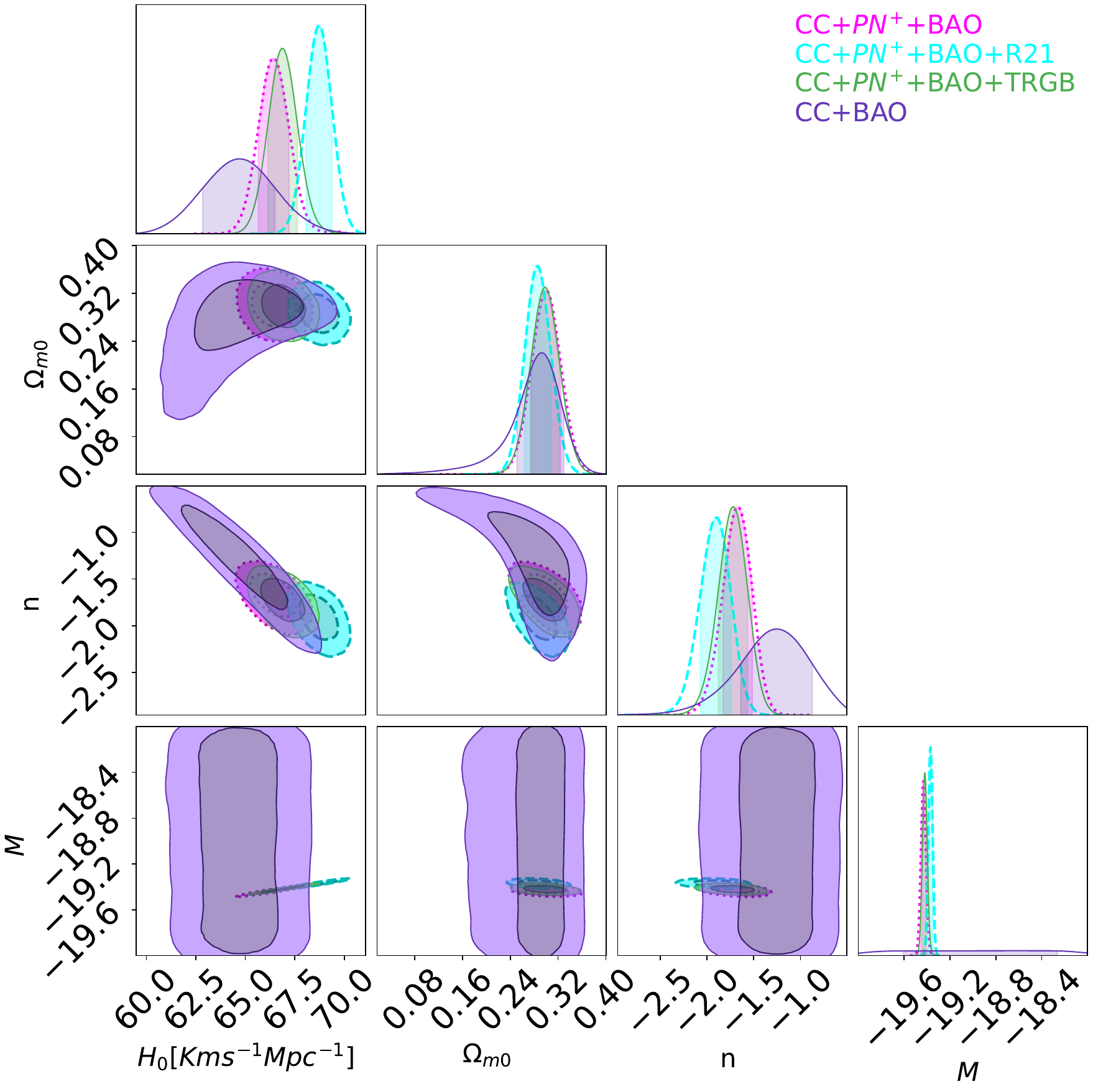} 
         \caption{}
         \label{fig:BAOMCMCplus}
     \end{subfigure}
\caption{The contour plot of $1\sigma$ and $2\sigma$ uncertainty regions and posterior distribution for the model parameters with the combination of data sets (a) CC, $PN^{+}$ (b) CC, $PN^{+}$ and BAO. The $H_0$ priors are: TRGB (Green) and R21 (Cyan).} 
\label{plusmodelMCMC}
\end{figure}

In Fig.-\ref{plusmodelMCMC}, we have presented the $1\sigma$ and $2\sigma$ confidence levels along with the posterior distributions for the parameters $H_0$, $\Omega_{m0}$, $n$ and $M$ using CC, $PN^{+}$ (without SH0ES) and BAO data sets in addition to the R21 and TRGB priors. It displays the marginalized posterior distributions for different combinations of parameters. The inner contours indicate the 68\% confidence level while the outer contours represent the 95\% confidence level. This visual representation facilitates a thorough evaluation of parameter uncertainties and correlations. The most noticeable aspect of the findings is the influence these priors exert on $H_0$ values, as it is clear that these priors generally lead to an increase in the Hubble constant value when contrasted with the scenario of no priors. Including priors decreases the estimated value of $\Omega_{m0}$, albeit to a lesser degree than the impact observed on $H_0$. This reduction is expected, as the priors effectively adjust the value of $H_0$. Table-\ref{tab:model_outputsmodelplus} summarizes the results derived from Fig.-\ref{plusmodelMCMC}. Notably, the highest value of $H_0$ is obtained from the data set combination CC+$PN^{+}$+R21, while the lowest is observed for CC+BAO. The diminished $H_0$ in the CC+BAO combination can be attributed to incorporating the BAO data set, which originates from early Universe measurements. Furthermore, the nuisance parameter $M$ remains unconstrained in the CC+BAO analysis due to the absence of the PN$^+$ data set. The determined value of \( H_0 \) for the CC+$PN^{+}$+R21 aligns with the elevated \( H_0 \) value reported by the SH0ES team (R22), which states \( H_{0} = 73.30\pm{1.04} \, \text{km s}^{-1} \text{ Mpc}^{-1} \) \cite{Riess_2022panplus}.

\begin{figure}[H]
     \centering
     \begin{subfigure}[b]{0.49\textwidth}
         \centering
         \includegraphics[width=\linewidth]{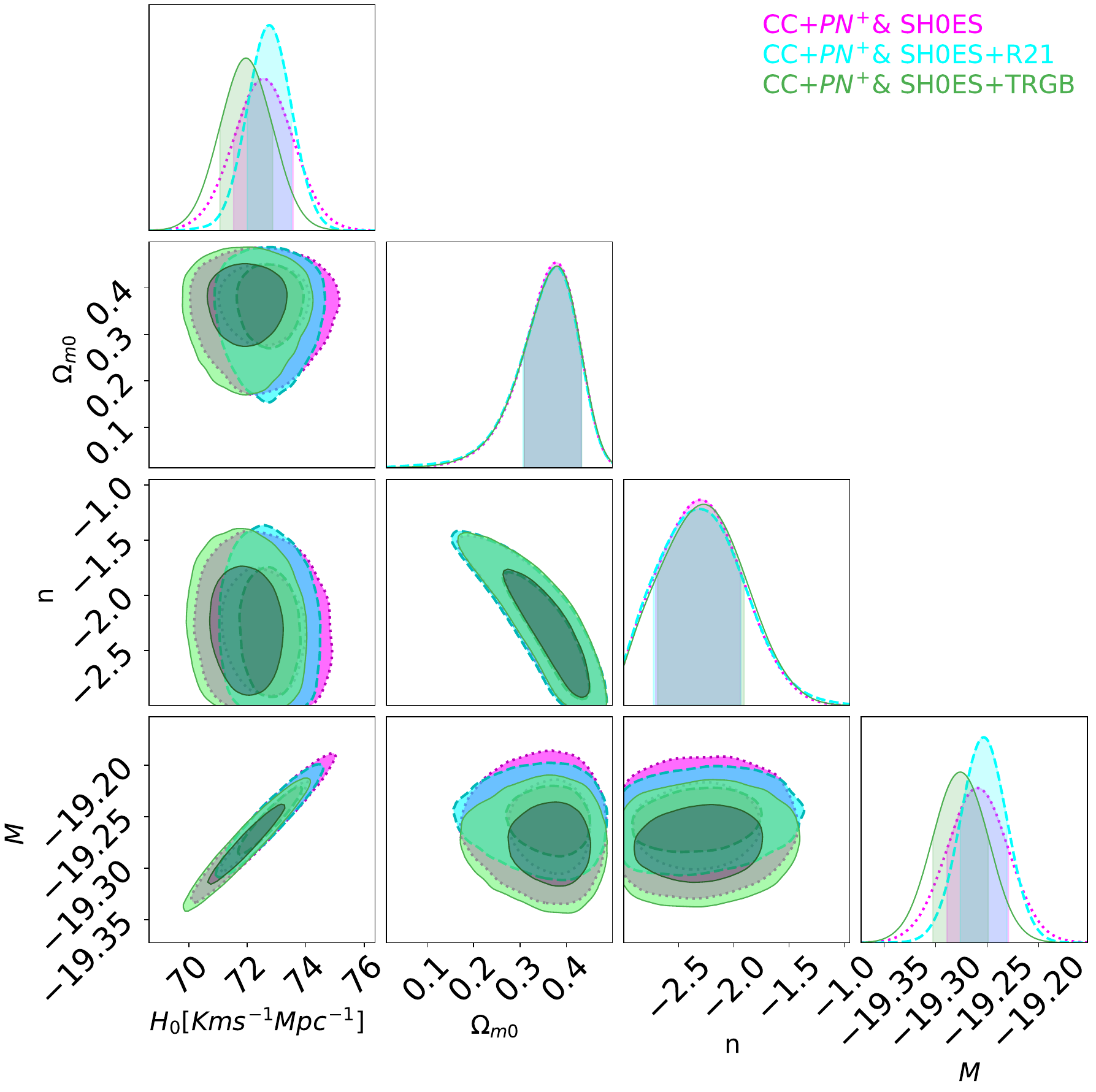}
          \caption{}
         \label{fig:CCMCMC}
     \end{subfigure}
     \hfill
     \begin{subfigure}[b]{0.49\textwidth}
         \centering
         \includegraphics[width=\linewidth]{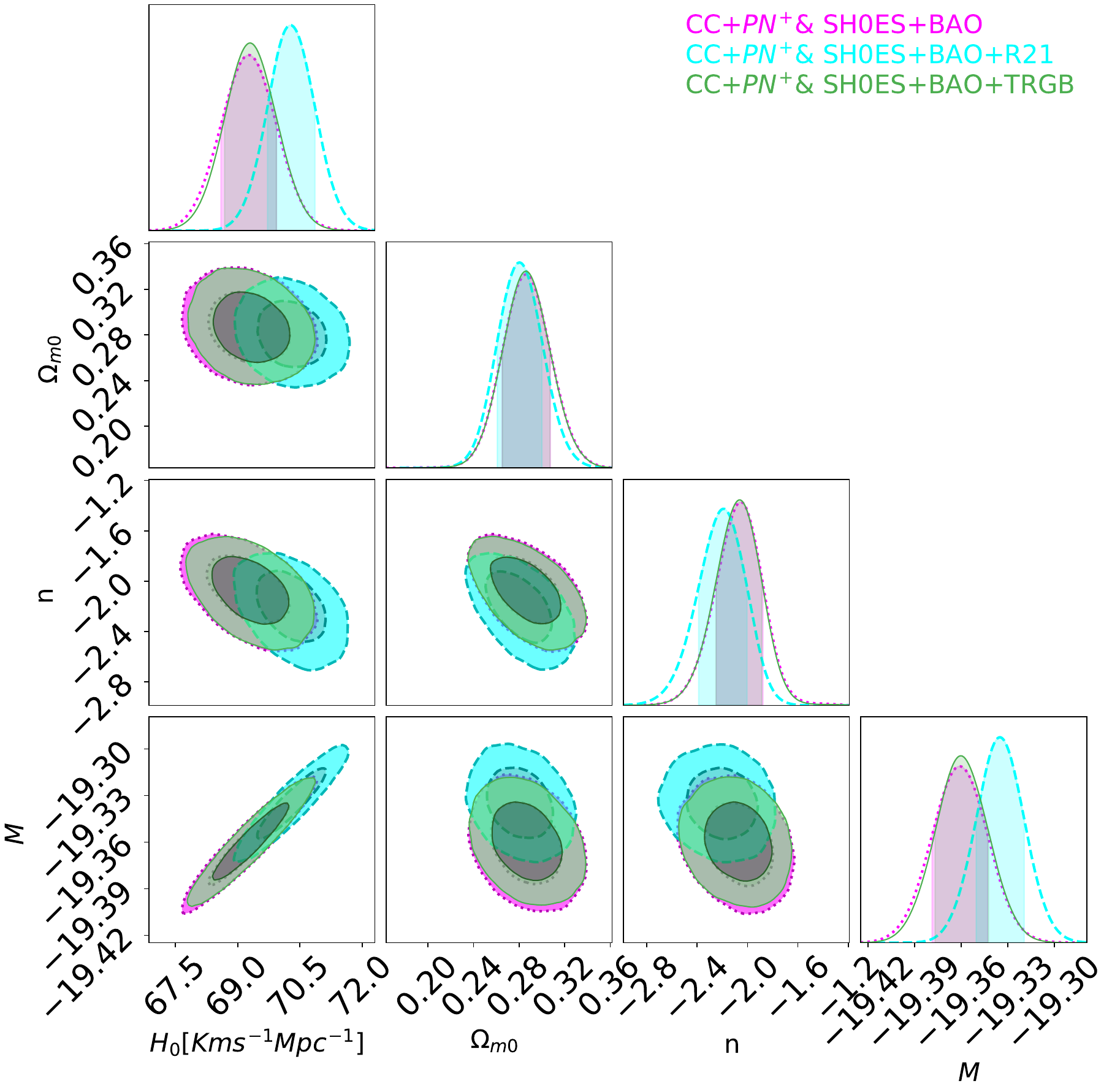} 
         \caption{}
         \label{fig:BAOMCMC}
     \end{subfigure}
\caption{The contour plot of $1\sigma$ and $2\sigma$ uncertainty regions and posterior distribution for the model parameters with the combination of data sets (a) CC, $PN^{+}\& SH0ES$ (b) CC, $PN^{+}\& SH0ES$ and BAO. The $H_0$ priors are: TRGB (Green) and R21 (Cyan).} 
\label{modelMCMC}
\end{figure}

In Fig.-\ref{modelMCMC}, we display the $1\sigma$ and $2\sigma$ confidence intervals alongside the posterior distributions for the parameters $H_0$, $\Omega_{m0}$, $n$ and $M$ utilizing the CC, $PN^{+}$\&SH0ES (with SH0ES), and BAO datasets, in addition to the R21 and TRGB priors. In Table-\ref{tab:model_outputsmodel}, it is observed that the highest \( H_{0} \) value is noted for the combination of data sets CC+$PN^{+}\& SH0ES$+R21, which is \( H_{0} = 72.74^{+0.77}_{-0.74} \, \text{km s}^{-1} \text{ Mpc}^{-1} \). The addition of SH0ES data points alongside the PN$^+$ data set raises the \( H_{0} \) value in comparison to using only the PN$^{+}$ data set (without SH0ES). Integrating the BAO data set results in a lower \( H_0 \) value than the combination of the PN$^{+}$\& SH0ES data sets. The value of \(H_0\) for inclusion with the BAO data set is consistent with the Planck Collaboration \cite{Aghanim:2018eyx}, which reports a Hubble constant of \(67.4 \pm 0.5 \, \text{km s}^{-1} \, \text{Mpc}^{-1}\). In contrast, Aboot et al. \cite{Abbott_2018mnras} propose a value of \(67.2^{+1.2}_{-1.0} \, \text{km s}^{-1} \, \text{Mpc}^{-1}\). In this research, we have analyzed the differences between the PN$^+$ (without SH0ES) and PN$^{+}$\& SH0ES (with SH0ES) datasets, also incorporating the BAO dataset and $H_0$ priors. The findings can be observed in Tables-\ref{tab:model_outputsmodel} and Table-\ref{tab:model_outputsmodelplus}. From the results, we concluded that including the SH0ES data points with the PN$^+$ dataset raises the $H_0$ value compared to PN$^+$, which in turn caused adjustments in $\Omega_{m0}$ and $n$ due to the change in the $H_0$ value. Our results for the Hubble constant \(H_0\) value from the data set combination PN\(^{+}\) (without SH0ES) and PN\(^{+}\)\& SH0ES (with SH0ES), along with the BAO data set combination, align with the findings presented by Brout et al.\cite{Brout_2022pan} for the PN\(^{+}\) (without SH0ES) and PN\(^{+}\)\& SH0ES (with SH0ES).
           
We have also computed the AIC and BIC values, providing a statistical foundation for selecting the appropriate model. Results related to the \(\Lambda\)CDM model can be found in the Appendix, particularly in Table-\ref{tab:model_outputsLCDM}. Lower values of \(\Delta\)AIC and \(\Delta\)BIC suggest that the model using the chosen data sets closely resembles the \(\Lambda\)CDM model. In Tables~\ref{tab:model_outputAICBICplus} and \ref{tab:model_outputAICBIC}, we present the statistical results including $\chi^{2}_{\text{min}}$, $\Delta$AIC, and $\Delta$BIC for PN$^+$ (without SH0ES) and PN$^+$\& SH0ES (with SH0ES), respectively. In this analysis, the values of \(\Delta \text{AIC}\) and \(\Delta \text{BIC}\) for the data set combinations for  CC+$PN^{+}$ and CC+$PN^{+}\& SH0ES$ that include \( H_{0} \) priors are notably lower than those for the BAO data set combinations that also incorporate \( H_{0} \) priors. The study indicates that our model for dataset combination of CC+$PN^{+}$ and CC+$PN^{+}\& SH0ES$ performs more effectively than when BAO data is included in combination with the standard $\Lambda$CDM model. In fact, for the scenario where the statistical distance deviates most from $\Lambda$CDM, specifically for the inclusion of the BAO data set, the statistical criteria yield a negative result, suggesting a slight preference for the model. In contrast, the other scenarios slightly inclined towards the standard cosmological model. The lower value of the  \(\Delta \text{AIC}\) and \(\Delta \text{BIC}\) values for the dataset combination with \( H_{0} \) priors signify that the particular data combination aligns more closely with the standard \(\Lambda\)CDM model.

\begin{table}[H]
\renewcommand{\arraystretch}{1.5}
    \centering
    \caption{The best-fit values of the parameters explored by MCMC analysis. The first column enumerates a combination of data sets with the \(H_0\) priors. The second column presents the constrained \(H_0\) values. The third column contains the constrained \(\Omega_{m0}\) values. The fourth and fifth columns represent the optimal \(n\) and \(M\) values respectively. \textbf{(Without SH0ES data points)}}
    \label{tab:model_outputsmodelplus}
      \begin{tabular}{cccccc}
        \hline
		Data set & $H_0[\text{km s}^{-1} \, \text{Mpc}^{-1}]$ & $\Omega_{m0}$  & n & M \\     
		\hline
		CC+$PN^{+}$ & $66.4\pm 4.1$ & $0.354^{+0.074}_{-0.094}$ & $-1.82^{+0.36}_{-0.39}$ & $-19.43^{+0.13}_{-0.14}$\\ 
		CC+$PN^{+}$+R21 & $72.6^{+1.1}_{-1.0}$ & $0.338^{+0.076}_{-0.096}$ & $-1.80^{+0.37}_{-0.38}$ & $-19.242^{+0.032}_{-0.033}$ \\ 
		CC+$PN^{+}$+TRGB & $69.1\pm 1.8$ & $0.347^{+0.077}_{-0.093}$ & $-1.82^{+0.36}_{-0.39}$ & $-19.348^{+0.055}_{-0.057}$ \\ 
		\cline{1-5}
       CC+$PN^{+}$+BAO & $66.43\pm 0.78$ & $0.300\pm 0.025$ & $-1.67^{+0.15}_{-0.16}$ & $-19.431^{+0.022}_{-0.023}$ \\ 
		CC+$PN^{+}$+BAO+R21 & $68.73^{+0.64}_{-0.67}$ & $0.285^{+0.023}_{-0.021}$ & $-1.90^{+0.16}_{-0.17}$ & $-19.370^{+0.019}_{-0.018}$ \\ 
		CC+$PN^{+}$+BAO+TRGB &$66.88^{+0.75}_{-0.74}$ & $0.298^{+0.025}_{-0.024}$ & $-1.72^{+0.15}_{-0.16}$ & $-19.419^{+0.022}_{-0.021}$ \\
        CC+BAO &$64.7\pm 1.8$ & $0.292^{+0.038}_{-0.041}$ & $-1.25\pm 0.38$ & - \\
  \hline
    \end{tabular}
\end{table}
\begin{table}[H]
\renewcommand{\arraystretch}{1.5} 
    \centering
    \caption{The statistical comparison between the chosen model and the standard \(\Lambda\)CDM model. Details regarding the \(\Lambda\)CDM model are given in {\color{blue}Appendix}. The first column enumerates the data sets including the \(H_0\) priors. The second column displays the values of \(\chi^{2}_{\text{min}}\). The third and fourth column respectively provides the value of AIC and BIC. The fifth and sixth column respectively illustrates the values of \(\Delta \text{AIC}\) and \(\Delta \text{BIC}\). \textbf{(Without SH0ES data points)}}
    \label{tab:model_outputAICBICplus}
      \begin{tabular}{cccccc}
        \hline
		data set& $\chi^{2}_{min}$  &AIC &BIC&$\Delta$AIC &$\Delta$BIC \\ 
		\hline
		CC+$PN^{+}$&1787.76 &1795.76 &  1800.71& -2.54&-1.31 \\ 
		CC+$PN^{+}$+R21& 1780.52 &1798.52 &  1803.48&-3.12 &-1.87\\ 
		CC+$PN^{+}$+TRGB&1788.41 &1796.41 &  1801.37&-2.79 &-1.55\\ 
		\cline{1-6}
       CC+$PN^{+}$+BAO&1798.26 &  1806.26 &  1811.08& -17.49&-16.28 \\ 
		CC+$PN^{+}$+BAO+R21& 1524.96 & 1832.96 &  1837.78&-0.2 &1\\ 
		CC+$PN^{+}$+BAO+TRGB&1801.01 &  1809.01 &  1813.83&-14.99 &-13.78\\
        	CC+BAO&15.98 &  23.98 &    28.80&-2.16 &-0.96\\
  \hline
    \end{tabular}
\end{table}
\begin{table}[H]
\renewcommand{\arraystretch}{1.5} 
    \centering
    \caption{The best-fit values of the parameters explored by MCMC analysis. The first column enumerates a combination of data sets with the \(H_0\) priors. The second column presents the constrained \(H_0\) values. The third column contains the constrained \(\Omega_{m0}\) values. The fourth and fifth columns represent the optimal \(n\) and \(M\) values respectively.\textbf{(With SH0ES data points)}}
    \label{tab:model_outputsmodel}
      \begin{tabular}{cccccc}
        \hline
		Data set  & $H_0[\text{km s}^{-1} \, \text{Mpc}^{-1}]$ & $\Omega_{m0}$  & n & M \\     
		\hline
		CC+$PN^{+}\& SH0ES$ & $72.5\pm 1.0$ & $0.379^{+0.053}_{-0.069}$ & $-2.30^{+0.36}_{-0.41}$ & $-19.260\pm 0.030$ \\ 
		CC+$PN^{+}\& SH0ES$+R21 & $72.74^{+0.77}_{-0.74}$ & $0.378^{+0.055}_{-0.071}$ & $-2.31^{+0.38}_{-0.42}$ & $-19.254^{+0.023}_{-0.022}$ \\ 
		CC+$PN^{+}\& SH0ES$+TRGB & $71.96^{+0.91}_{-0.90}$ & $0.381^{+0.052}_{-0.071}$ & $-2.27^{+0.36}_{-0.42}$ & $-19.276\pm 0.027$  \\ 
		\cline{1-5}
       CC+$PN^{+}\& SH0ES$+BAO & $69.26^{+0.68}_{-0.66}$ & $0.286^{+0.022}_{-0.021}$ & $-2.05^{+0.17}_{-0.20}$ & $-19.361^{+0.018}_{-0.017}$ \\ 
		CC+$PN^{+}\& SH0ES$+BAO+R21 & $70.28^{+0.59}_{-0.56}$ & $0.280^{+0.020}_{-0.019}$ & $-2.19^{+0.18}_{-0.20}$ & $-19.335\pm 0.015$    \\ 
		CC+$PN^{+}\& SH0ES$+BAO+TRGB &$69.30^{+0.63}_{-0.62}$ & $0.286\pm 0.021$ & $-2.06^{+0.18}_{-0.19}$ & $-19.360\pm 0.017$ \\
  \hline
    \end{tabular}
\end{table}
\begin{table}[H]
\renewcommand{\arraystretch}{1}
    \centering
    \caption{The statistical comparison between the chosen model and the standard \(\Lambda\)CDM model. Details regarding the \(\Lambda\)CDM model are given in {\color{blue}Appendix}. The first column enumerates the data sets including the \(H_0\) priors. The second column displays the values of \(\chi^{2}_{\text{min}}\). The third and fourth column respectively provides the value of AIC and BIC. The fifth and sixth column respectively illustrates the values of \(\Delta \text{AIC}\) and \(\Delta \text{BIC}\). \textbf{(With SH0ES data points)}}
    \label{tab:model_outputAICBIC}
      \begin{tabular}{cccccc}
        \hline
		data set& $\chi^{2}_{min}$  &AIC &BIC&$\Delta$AIC &$\Delta$BIC \\ 
		\hline
		CC+$PN^{+}\& SH0ES$&1538.29 &1546.29 &  1551.25& 1.07&2.32 \\ 
		CC+$PN^{+}\& SH0ES$+R21& 1538.41 &1546.41 &  1551.37&1.16 &2.41\\ 
		CC+$PN^{+}\& SH0ES$+TRGB&1539.99 &1547.99 &  1552.92&0.81 &2.03\\ 
		\cline{1-6}
       CC+$PN^{+}\& SH0ES$+BAO&1571.27 &  1579.27 &  1584.25& 6.1&7.35\\ 
		CC+$PN^{+}\& SH0ES$+BAO+R21& 1580.95 & 1588.95 &  1593.92&12.99 &14.25\\ 
		CC+$PN^{+}\& SH0ES$+BAO+TRGB&1571.35 &  1579.35 &  1584.32&5.85&7.09\\
  \hline
    \end{tabular}
\end{table}
\section{Linear Matter Perturbations and Large-Scale Structure Evolution}\label{linearperturbationection}
Cosmological models that do not account for interactions within the dark sector, the fundamental equation that governs the growth of matter perturbations in the linear regime at sub-horizon scales during the matter era is \cite{Lue_2004, Linder_2005growth, Uzan_2007, Gannouji_2009, Tsujikawa_2008, Basilakos_2013, Anagnostopoulos_2019},
\begin{equation}\label{deltarhom}
 \ddot{\delta}_m+2 \,H\, \dot{\delta}_m=4\pi G_{eff}\,\rho_m\,\delta\,, 
\end{equation}

where the matter overdensity can be defined as $\delta_m \equiv \frac{\delta \rho_m}{\rho_m}$. In the context of Eq.~\eqref{deltarhom}, the effective Newton's constant is introduced as $G_{eff}(a) = G \,\,P(a)$, where $G$ is the gravitational constant defined in the action of the theory. This formulation accounts for modifications to gravity. The underlying gravitational theory determines the specific form of $P(a)$. In case of GR, we find that $G_{eff}(a) = G$, which corresponds to $P(a) = 1$. Consequently, Eq.~\eqref{deltarhom} simplifies to yield the standard evolution equation for matter density perturbations \cite{Peebles1993}.

So, it is clear that we can use this general perturbation approach in the context of $f(T, \mathcal{T})$ cosmology, provided the form of $G_{eff}(a)$ is known, or equivalently $P(a)$, in $f(T, \mathcal{T})$ gravity. It can be demonstrated with relative ease that for $f(T, \mathcal{T})$ gravity \cite{Harko_2014a, Junior_2016}.
\begin{equation}\label{Geffective}
P(a)=\frac{G_{eff}}{G}= \frac{1}{1+f_T}\,\bigg(1+\frac{f_{\mathcal{T}}}{2}\bigg)\,.    
\end{equation}
Eq.~\eqref{deltarhom} can also be rewritten as
\begin{equation}\label{deltaloga}
   \delta''_{m}+\bigg(2+ \frac{H'}{H}\bigg)\, \delta'_{N}-\frac{3}{2G}\,G_{eff}\,\Omega_{m}\,\delta_{m} = 0.
\end{equation}

In Eq.~\eqref{deltaloga} prime $(')$ denotes the derivative for $log(a)$. By incorporating the growth rate of matter fluctuations defined as $f_{\delta}\equiv \frac{\delta'_{m}}{\delta_{m}}$, we can express Eq. \eqref{deltaloga} in a different form as,
\begin{equation}
 f_{\delta}'+f_{\delta}^2+\left(2+\frac{H'}{H}\right)f_{\delta}-\frac{3}{2G}\,G_{eff}\,\Omega_{m}=0\,. 
\label{GrowthRate}   
\end{equation}
Where $\Omega_m=\frac{\Omega_{m0}\,a^{-3}}{E^2(a)}$. The measurable $f\sigma_8(z)$ contrasts theoretical predictions with observations data and is specified as
\begin{equation}
f\sigma_{8}(z)\equiv f_{\delta}(a)\cdot \sigma(a)=\frac{\sigma_{8}}{\delta_m{(1)}}\, a\, \delta_m'(a)\,,     
\end{equation}

where $\sigma(a)=\sigma_{8} \delta_m(a)/\delta_m(1)$ represents the amplitude of fluctuations in the matter density within spheres measuring $8~h^{-1}~Mpc$ ($k\sim k_{\sigma_{8}}=0.125$~$h~Mpc^{-1}$), with $\sigma_{8}=\sigma(1)$ being significantly impacted by late-time cosmic expansion and dark energy models \cite{Sola:2017znb, Gonzalez-Espinoza:2018gyl}. Discrepancies in the $\Lambda$CDM framework, especially concerning $H_{0}$ and $\sigma_8$, arise because large-scale structure (LSS) observations indicate $f\sigma_{8}$ values are roughly $8\%$ lower than anticipated. This implies that the $\Lambda$CDM model might overestimate $\sigma_8$ given the same current growth rate $f_{\delta}(1)$ \cite{Gomez-Valent:2018nib}. Nevertheless, since $f_{\delta}(a)$ depends on the specific model \cite{DeFelice:2010aj}, a reduced growth rate might also enhance compatibility with the LSS observations.

\subsection{Numerical results}
\begin{figure}[H]
     \centering
     \begin{subfigure}[b]{0.49\textwidth}
         \centering
         \includegraphics[width=\linewidth]{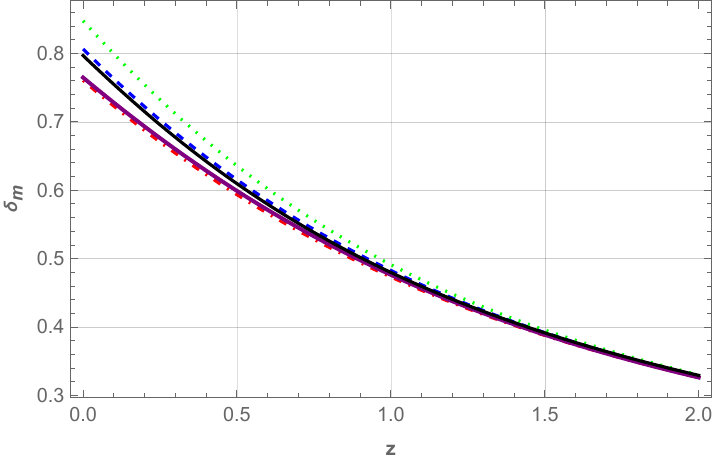}
         \caption{The graph displays the evolution of the matter density fluctuation, referred to as $\delta_m$, to the redshift $z$ across various dataset values, along with the corresponding evolution for the $\Lambda$CDM model.}
         \label{growthrate:CCOm}
     \end{subfigure}
     \hfill
     \begin{subfigure}[b]{0.49\textwidth}
         \centering
         \includegraphics[width=\linewidth]{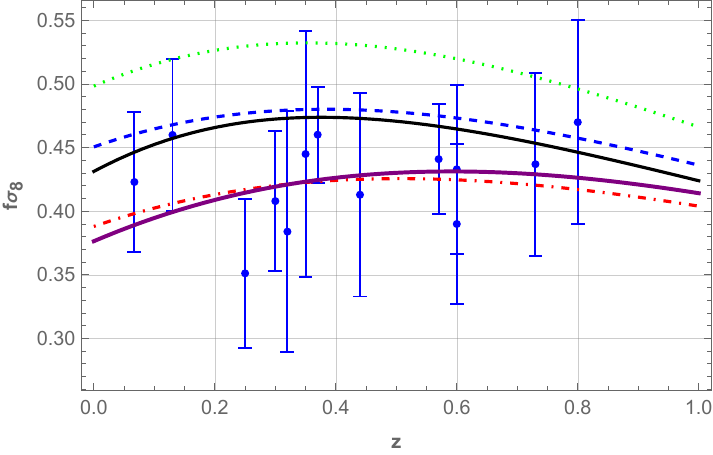}
         \caption{The graph displays the evolution of the weighted growth rate $f\sigma_8$ and the related progression for the $\Lambda$CDM model. In this figure, we have used the complete redshift-space distortion (RSD) $f\sigma_8$ dataset listed in Table II of Ref.~\cite{Kazantzidis:2018rnb}.}
         \label{sigma8:BAOom}
     \end{subfigure}
\caption{Evolution of the matter density perturbation $\delta_m$ and weighted growth rate $f\sigma_8$ with respect to redshift $z$ using the relation $a=\frac{1}{1+z}$. In this graph, the blue-dashed line represents dataset CC+PN$^+$, the red-dot dashed line corresponds to dataset CC+PN$^+$+BAO, the green-dotted line represents dataset CC+PN$^+$\&SH0ES, the purple-thick line corresponds to dataset CC+PN$^+$\&SH0ES+BAO and black-thinning line corresponds to $\Lambda$CDM model. } 
\label{growthratesigma8}
\end{figure}

The theoretical curves for matter density perturbations are shown in Fig.~\ref{growthrate:CCOm}, based on various selections of model parameters. For our analysis, we observe that $\sigma(a) \sim \delta_m(a)$ while $\delta_m(a) \approx a= \frac{1}{1+z}$ \cite{Kazantzidis:2018rnb,Gonzalez-Espinoza:2018gyl}. For the chosen model \eqref{modelconsidered}, from Fig.~\ref{growthrate:CCOm}, we determine the subsequent values: $\sigma_8 = 0.81$ (blue-dashed line), $\sigma_8 = 0.76$ (red-dot dashed line), $\sigma_8 = 0.85$ (green-dotted line), $\sigma_8 = 0.76$ (purple-thick line). For the $\Lambda$CDM model, we determine that $\sigma_8 = 0.8$ (black-thinning). These values serve as a good approximation that aligns with the findings shown in Refs.\cite{Aghanim:2018eyx, Kazantzidis:2018rnb, DiValentino_2018a, Poulin_2023}.  

In Fig.~\ref{sigma8:BAOom}, we illustrate the theoretical curves for the weighted linear growth rate \( f \sigma_8(z) \) corresponding to various selections of model parameters. We notice that for the red-dot dashed line and the purple-thick line, the values of \( f \sigma_8(z) \) are lower than the related standard from the \(\Lambda\)CDM model (black-thinning), suggesting that our model might reduce the \(\sigma_8\)-tension. For the selected model \eqref{modelconsidered}, we extract the following values from Fig.~\ref{sigma8:BAOom}: $f\sigma_8(0) \approx 0.45$ (blue-dashed line), $f\sigma_8(0) \approx 0.39$ (red-dot dashed line), $f\sigma_8(0) \approx 0.50$ (green-dotted line), and $f\sigma_8(0) \approx 0.38$ (purple-thick line). In the case of the $\Lambda$CDM model, we find that $f\sigma_8(0) \approx 0.43$ (black-thinning line). To measure this impact, we establish the precise relative difference \cite{Gomez-Valent:2018nib} as,
\begin{equation}
\Delta f\sigma_8 (z) \equiv 100 \times \frac{f\sigma_8(z)_{\text{model}} - f\sigma_8(z)_{\Lambda\text{CDM}}}{f\sigma_8(z)_{\Lambda\text{CDM}}},
\end{equation}

In Fig.~\ref{sigma8:BAOom}, the red-dot dashed line represents $f\sigma_8(0) \approx 0.39$, whereas for the $\Lambda$CDM model, we calculate $f\sigma_8(0) \approx 0.43$, resulting in a relative difference of $\Delta f\sigma_8(0) \approx 9\%$. The purple-thick line indicates that the relative difference with the $\Lambda$CDM model is $\Delta f\sigma_8(0) \approx 11\%$. Consequently, the projection indicated by the red-dot dashed line is approximately 9\% below the $\Lambda$CDM model prediction. The purple-thick line shows a deviation of roughly 11\% below the $\Lambda$CDM model. This result represents an improvement over the findings associated with the XCDM parametrization, as reported in Ref. \cite{Gomez-Valent:2018nib}.

\section{Dynamics of Cosmological Parameters} \label{cosmologicalparameters} 
\begin{figure}[ht]
     \centering
         \includegraphics[width=40mm]{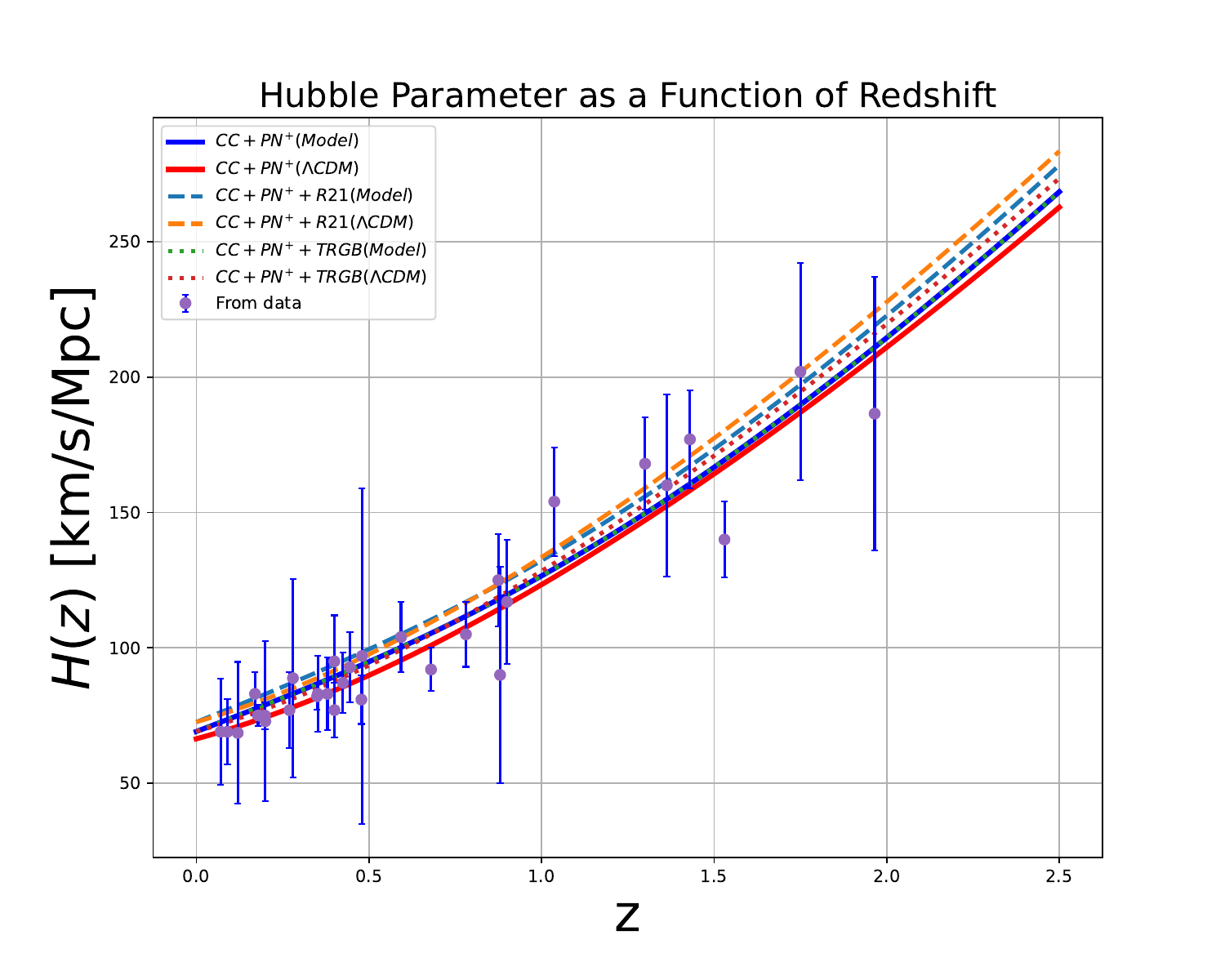}
          \includegraphics[width=40mm]{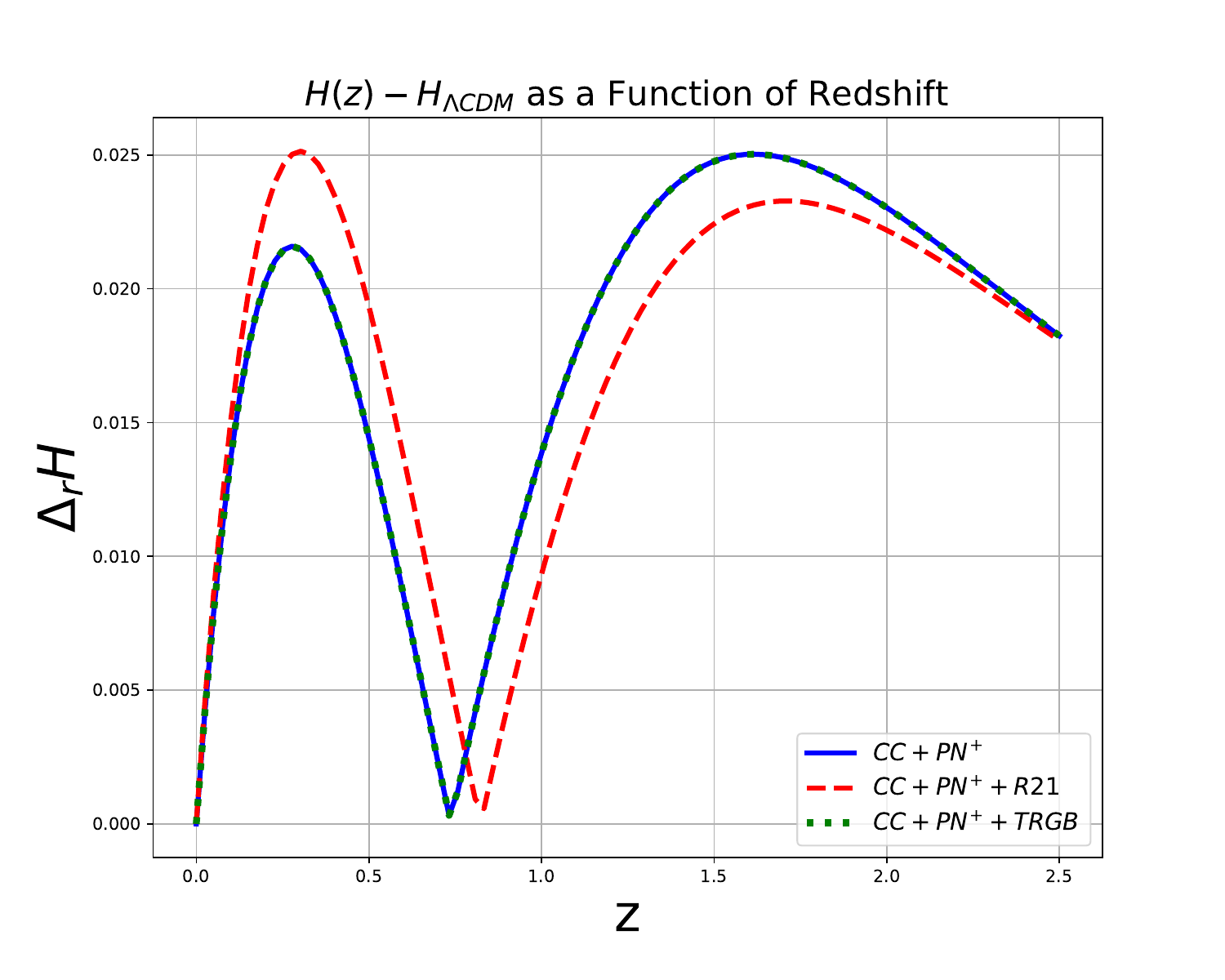}
         \includegraphics[width=40mm]{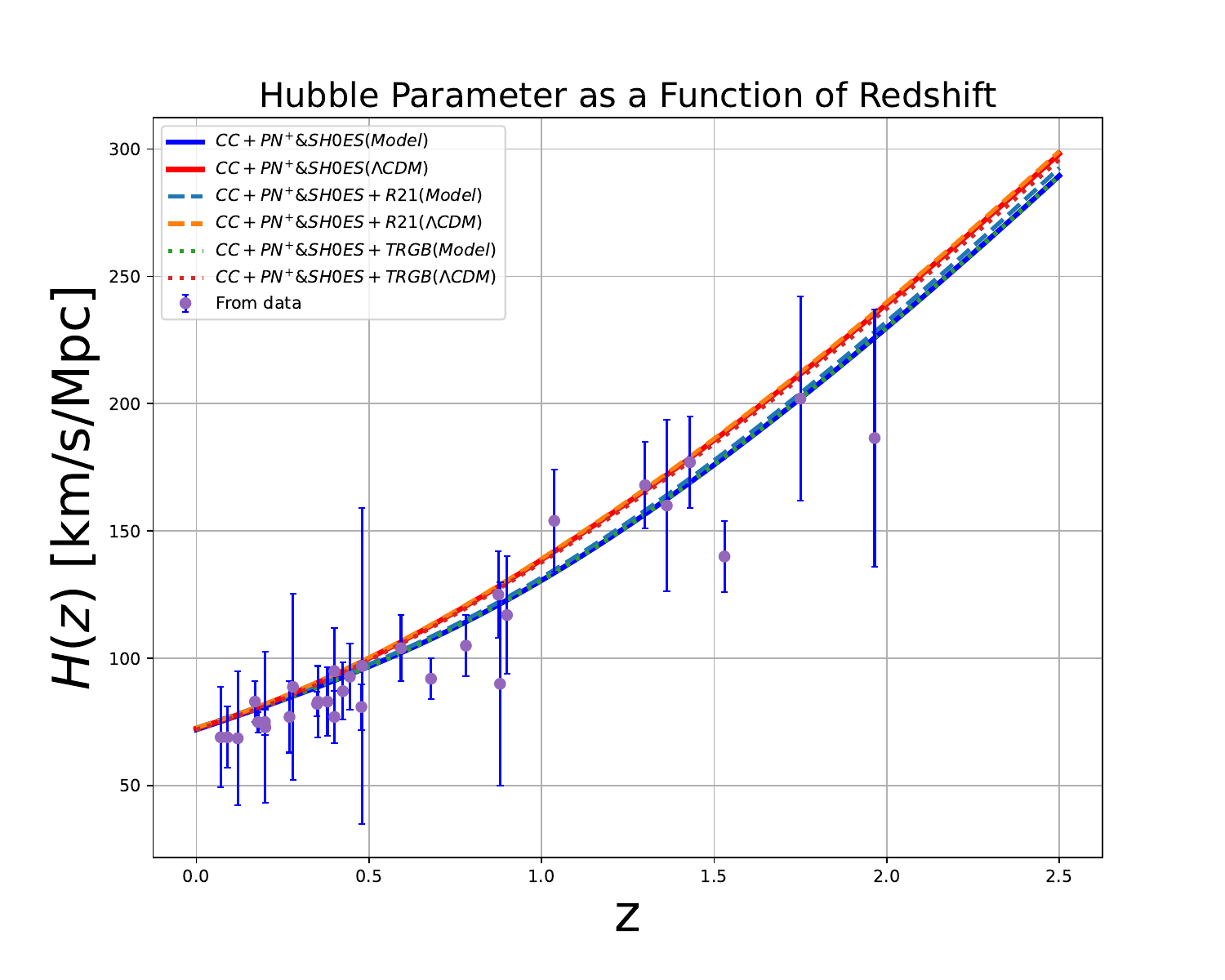}
          \includegraphics[width=40mm]{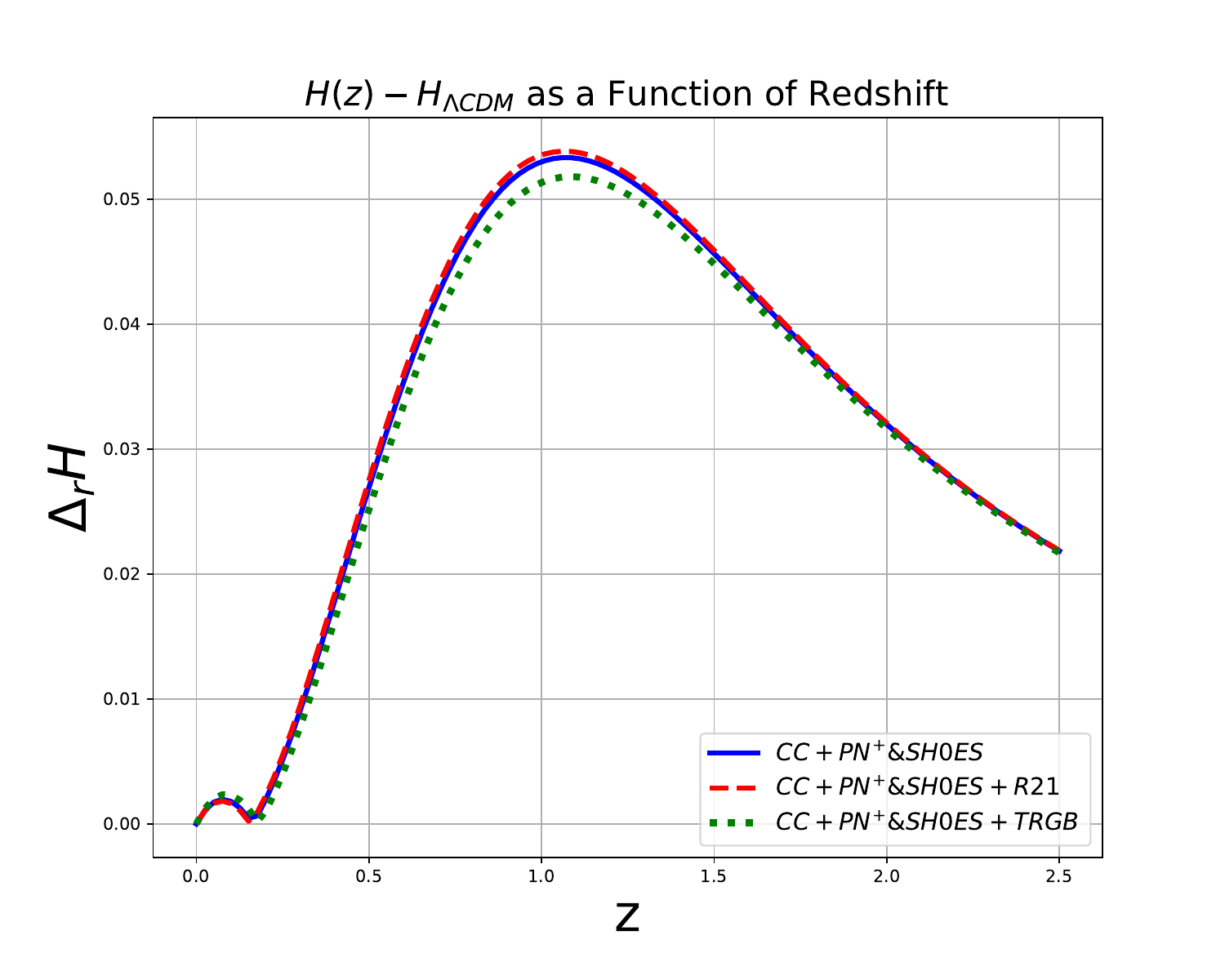}
\caption{Evolutionary behavior of Hubble parameter and comparative analysis of the evolution of the Hubble parameter between the selected model and the $\Lambda$CDM model in redshift for the data set combination: CC, PN$^{+}$ (without SH0ES) and  PN$^{+}$\&SH0ES (with SH0ES). The $H_0$ priors are: R21 and TRGB.} 
\label{plusFighubblediffer}
\end{figure}

In this section, we will examine the background cosmological parameters to investigate the behavior at the late time for the $f(T, \mathcal{T})$ model. Additionally, we will juxtapose these results with those from the standard $\Lambda$CDM model. Fig.~\ref{plusFighubblediffer}, displays the behavior of the Hubble parameter and comparative analysis of the evolution of the Hubble parameter between the selected model and the $\Lambda$CDM model for the PN$^+$ and the PN$^+$\& SH0ES data set combinations. It has been noted that the curves follow a similar pattern to that of the $\Lambda$CDM and remain well within the error margins. The analysis of these figures indicates that the model exhibits behavior consistent with that of the $\Lambda$CDM paradigm across the specified combination of data sets. We present the relative difference to demonstrate the differences between the selected and standard $\Lambda$CDM models.
\begin{equation}\label{relativedifferenceCC}
\Delta_{r} H(z) = \frac{\left| H_{\text{model}} - H_{\Lambda \text{CDM}} \right|}{H_{\Lambda \text{CDM}}} \,.
\end{equation}

\begin{figure}[ht]
     \centering
         \includegraphics[width=40mm]{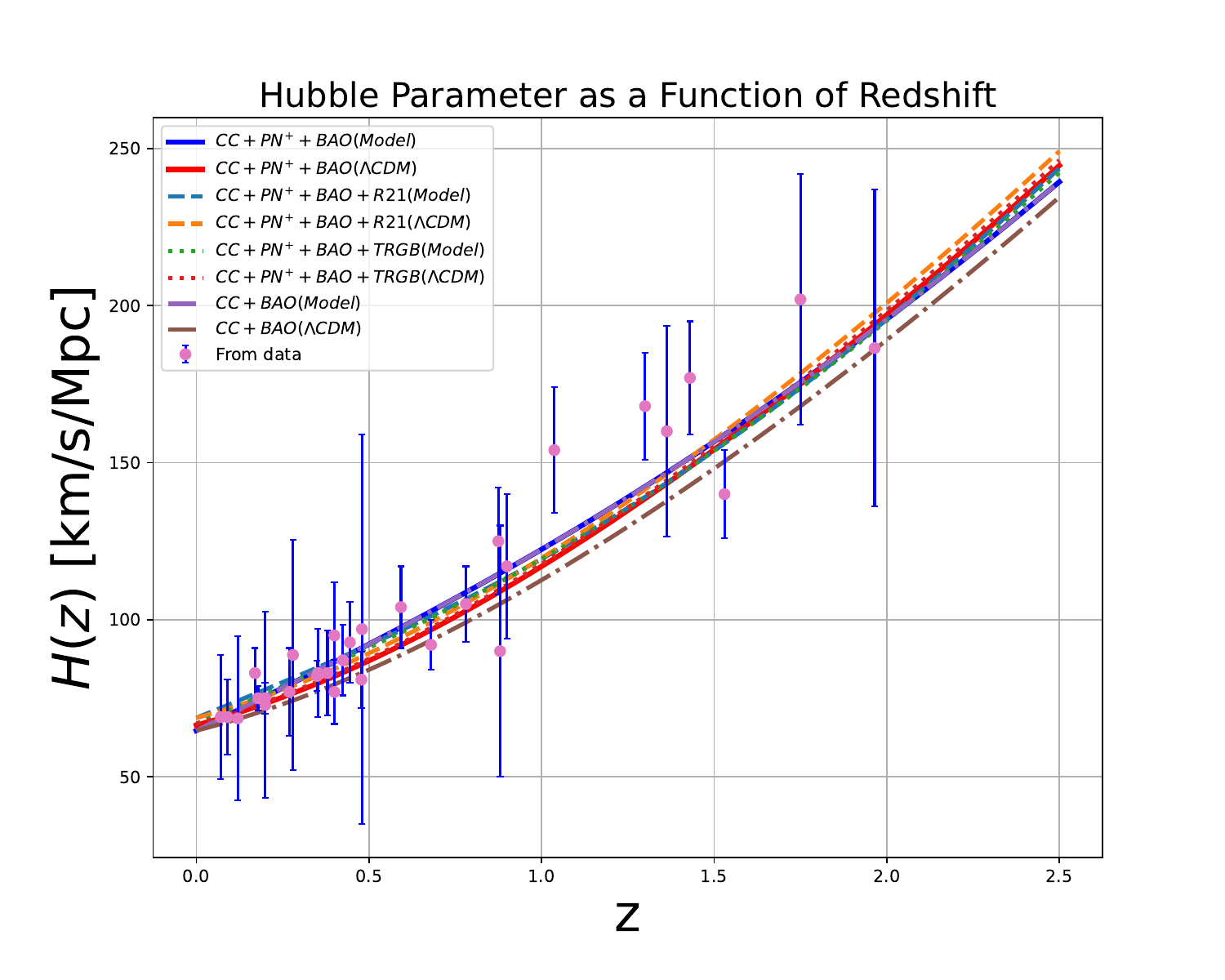}
         \includegraphics[width=40mm]{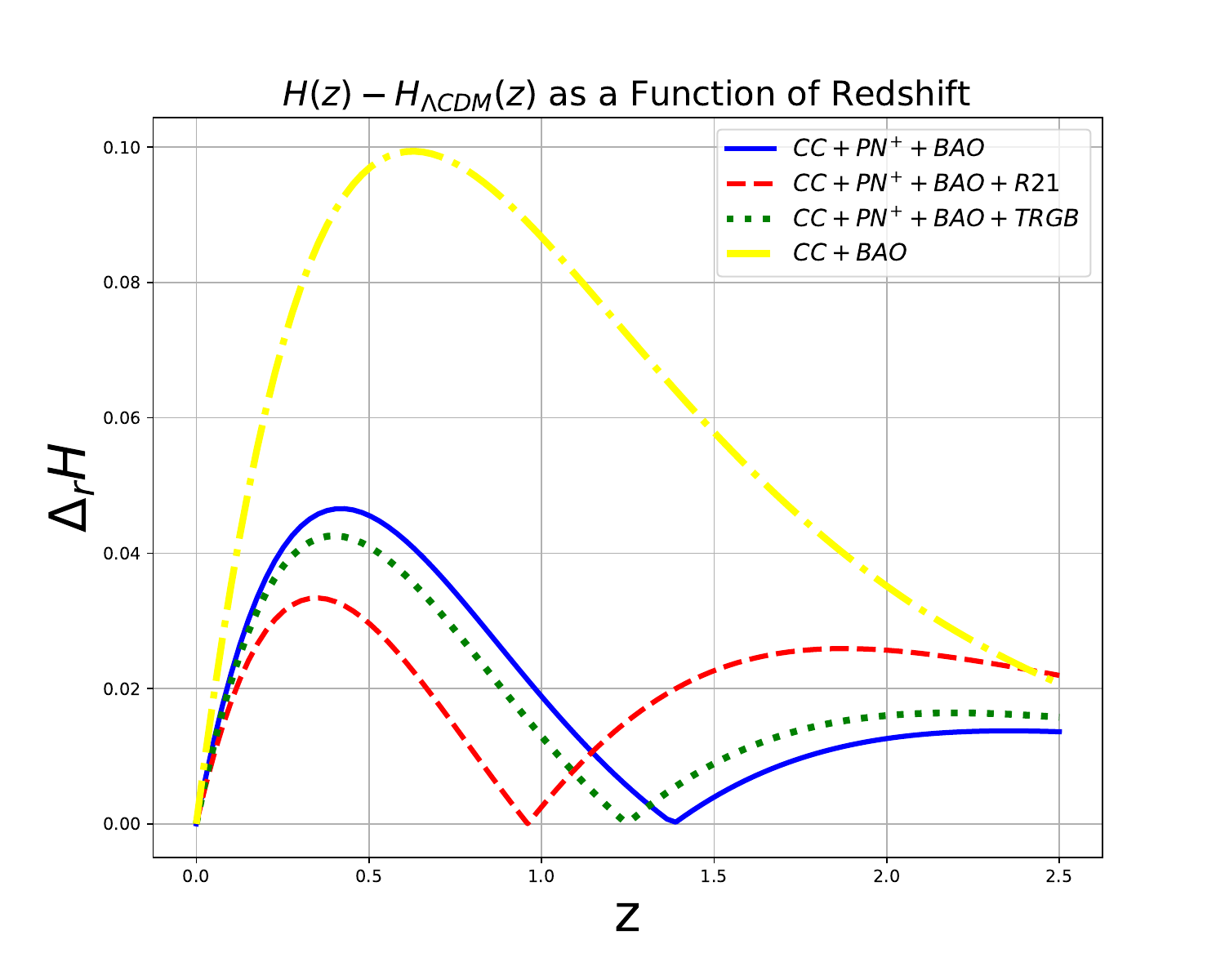}
       \includegraphics[width=40mm]{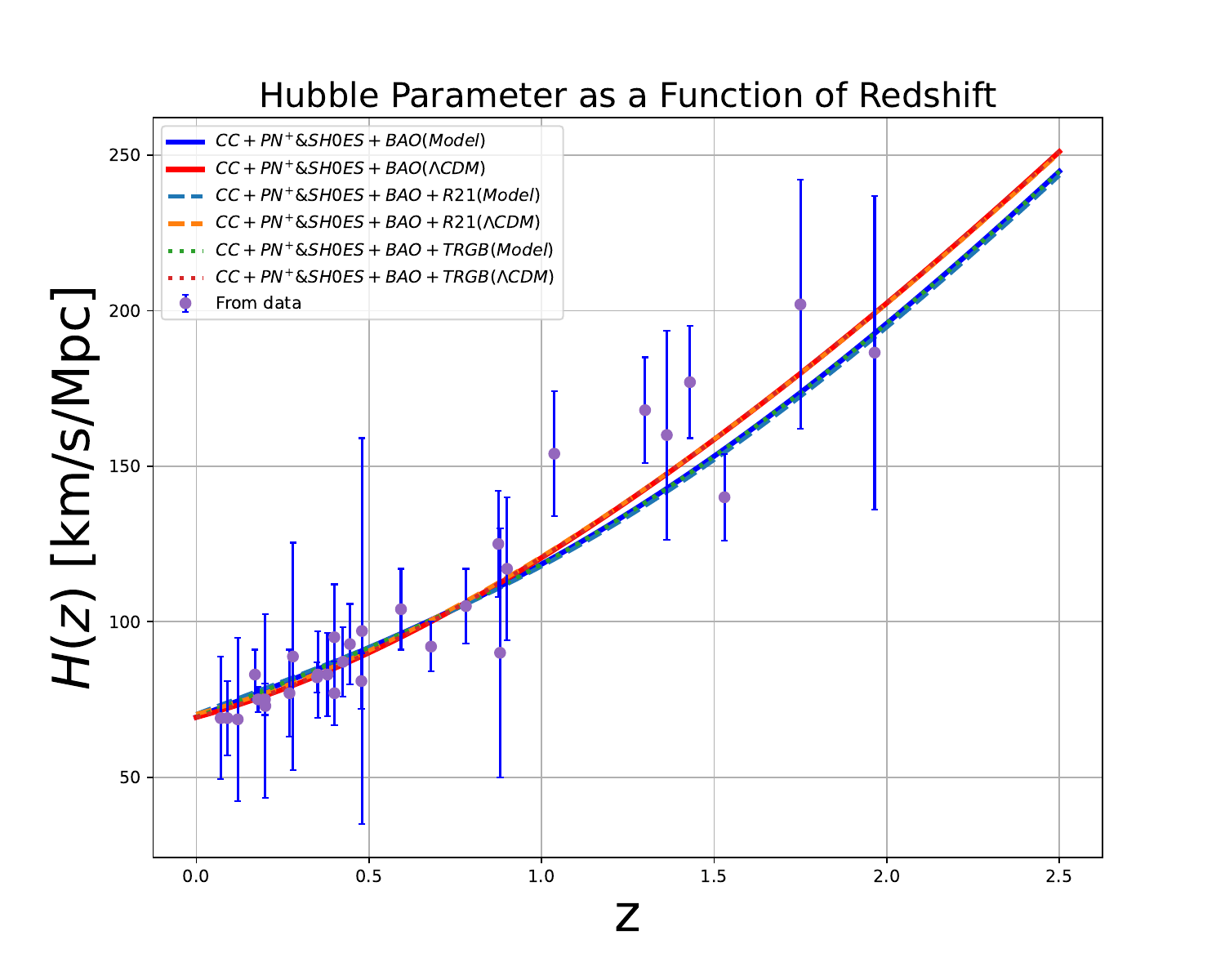}
        \includegraphics[width=40mm]{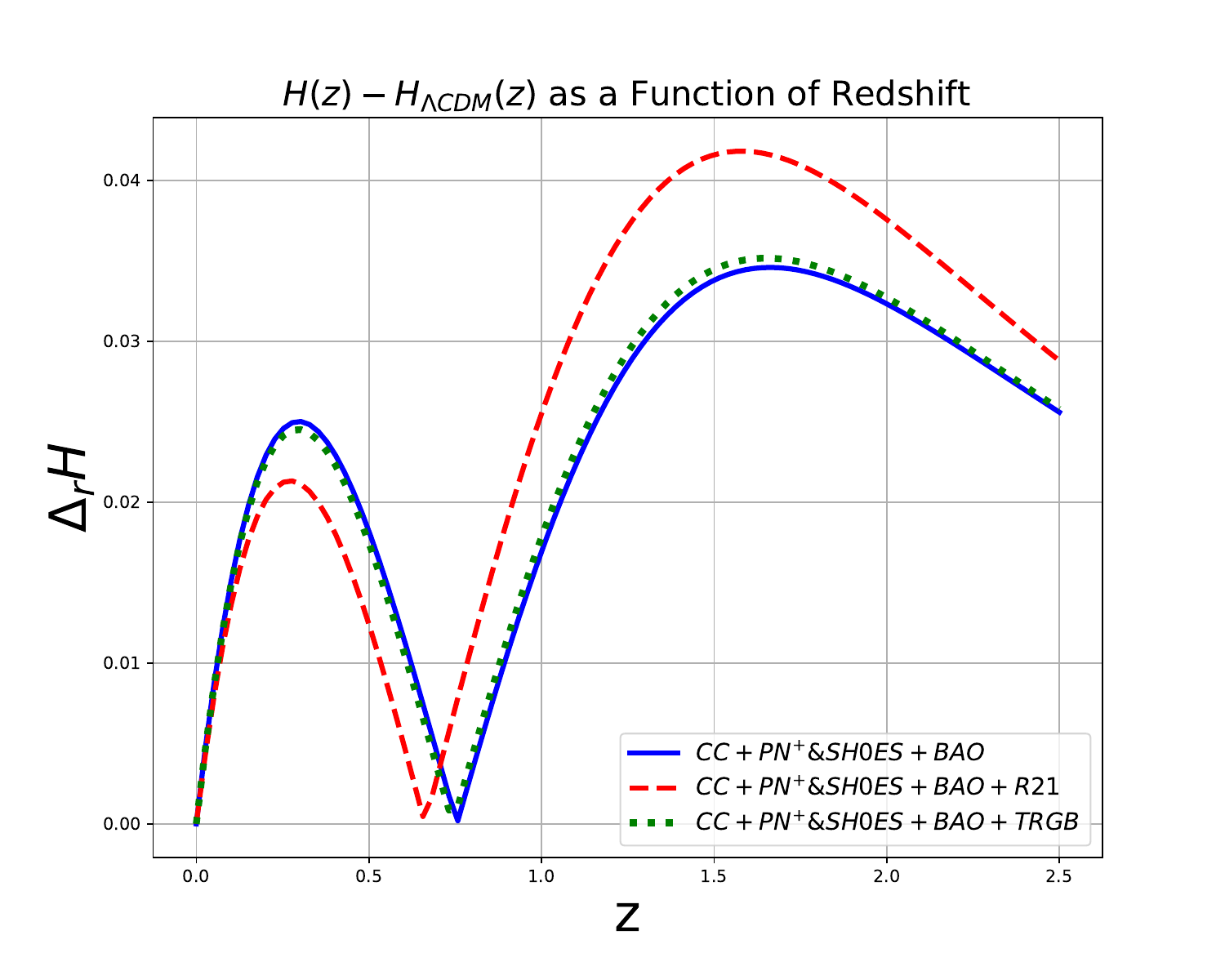}
\caption{Evolutionary behavior of Hubble parameter and comparative analysis of the evolution of the Hubble parameter between the selected model and the $\Lambda$CDM model in redshift for the data sets combination: CC, PN$^{+}$ (without SH0ES), PN$^{+}$\&SH0ES (with SH0ES) and BAO. The $H_0$ priors are: R21 and TRGB.} 
\label{plusFigbaohdiff}
\end{figure}

Additionally, Fig.~\ref{plusFigbaohdiff} showcase the evolution of the Hubble parameter and comparative analysis of the evolution of the Hubble parameter between the selected model and the $\Lambda$CDM model for various data sets, including the CC, PN$^{+}$, PN$^{+}$\& SH0ES and BAO, with the incorporation of $H_{0}$ priors.  

\begin{figure}[ht]
     \centering
         \includegraphics[width=40mm]{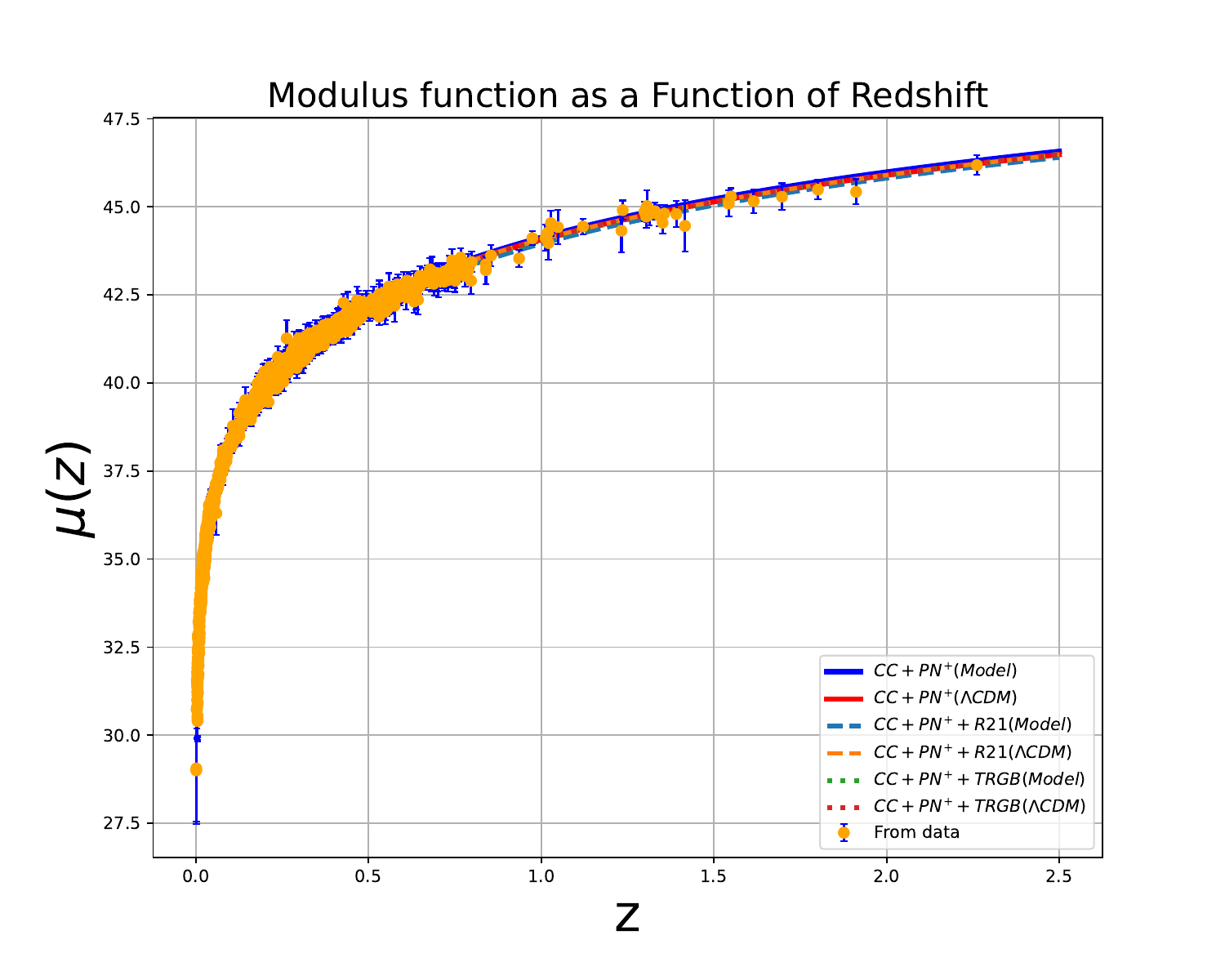}
         \includegraphics[width=40mm]{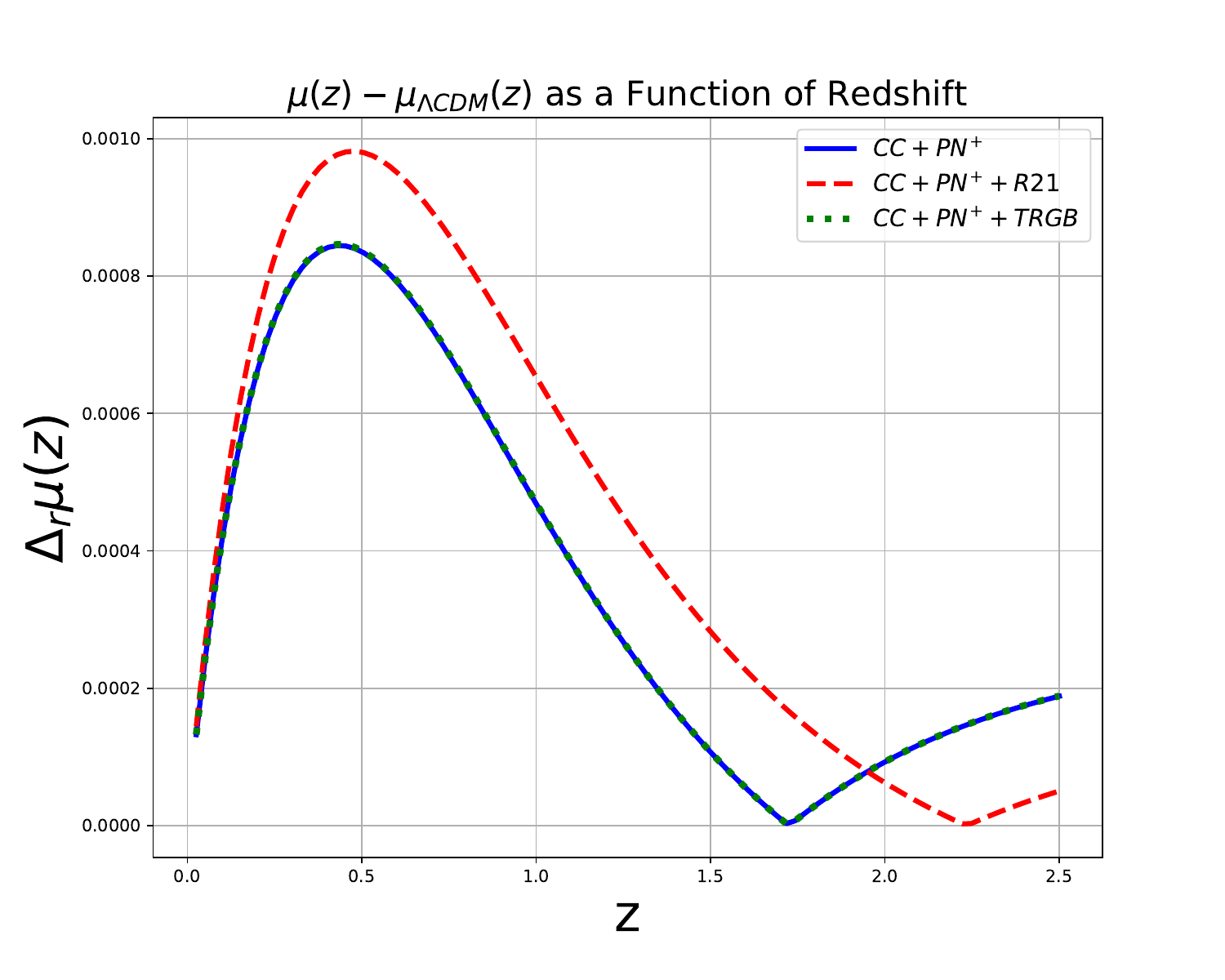}
        \includegraphics[width=40mm]{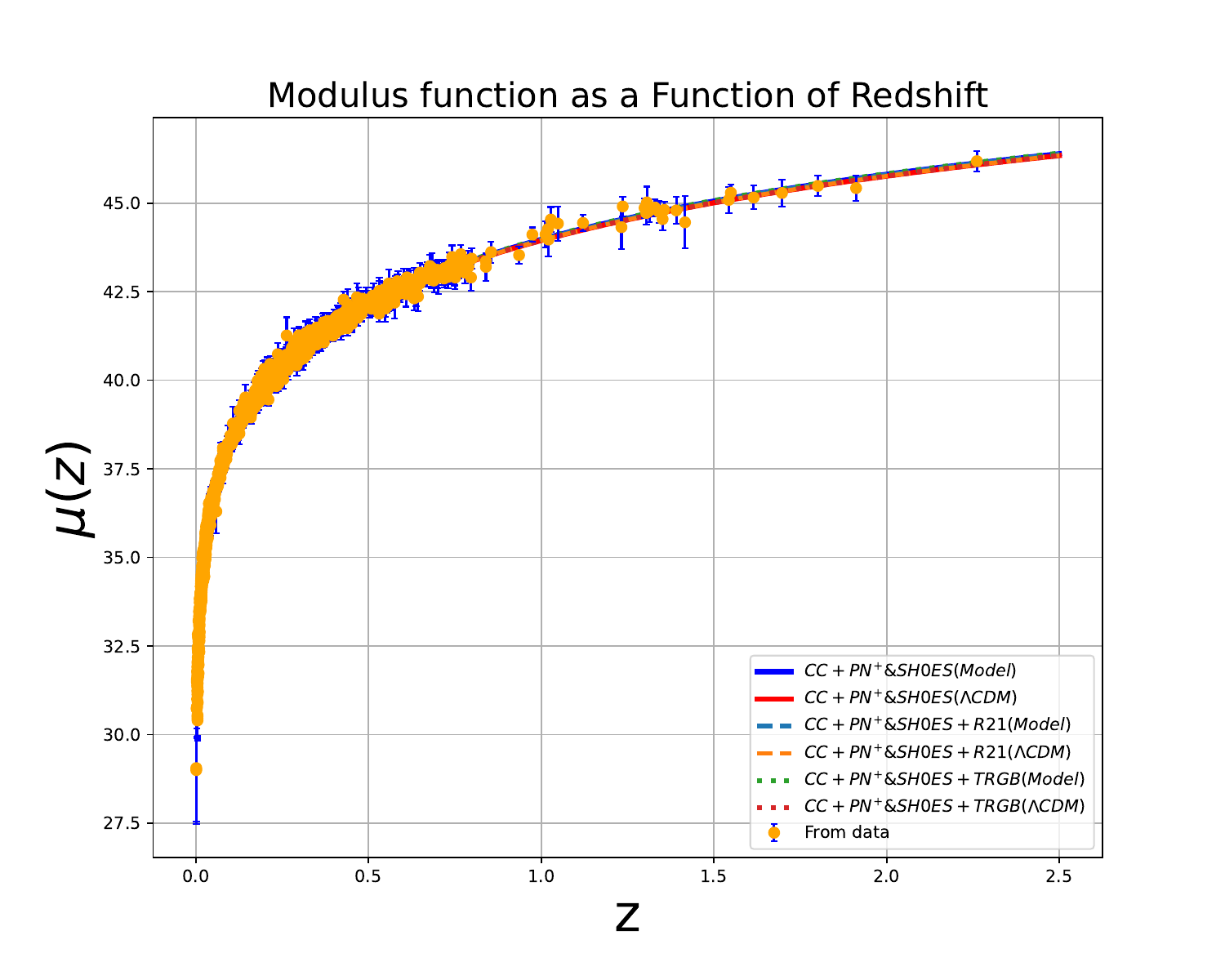}
        \includegraphics[width=40mm]{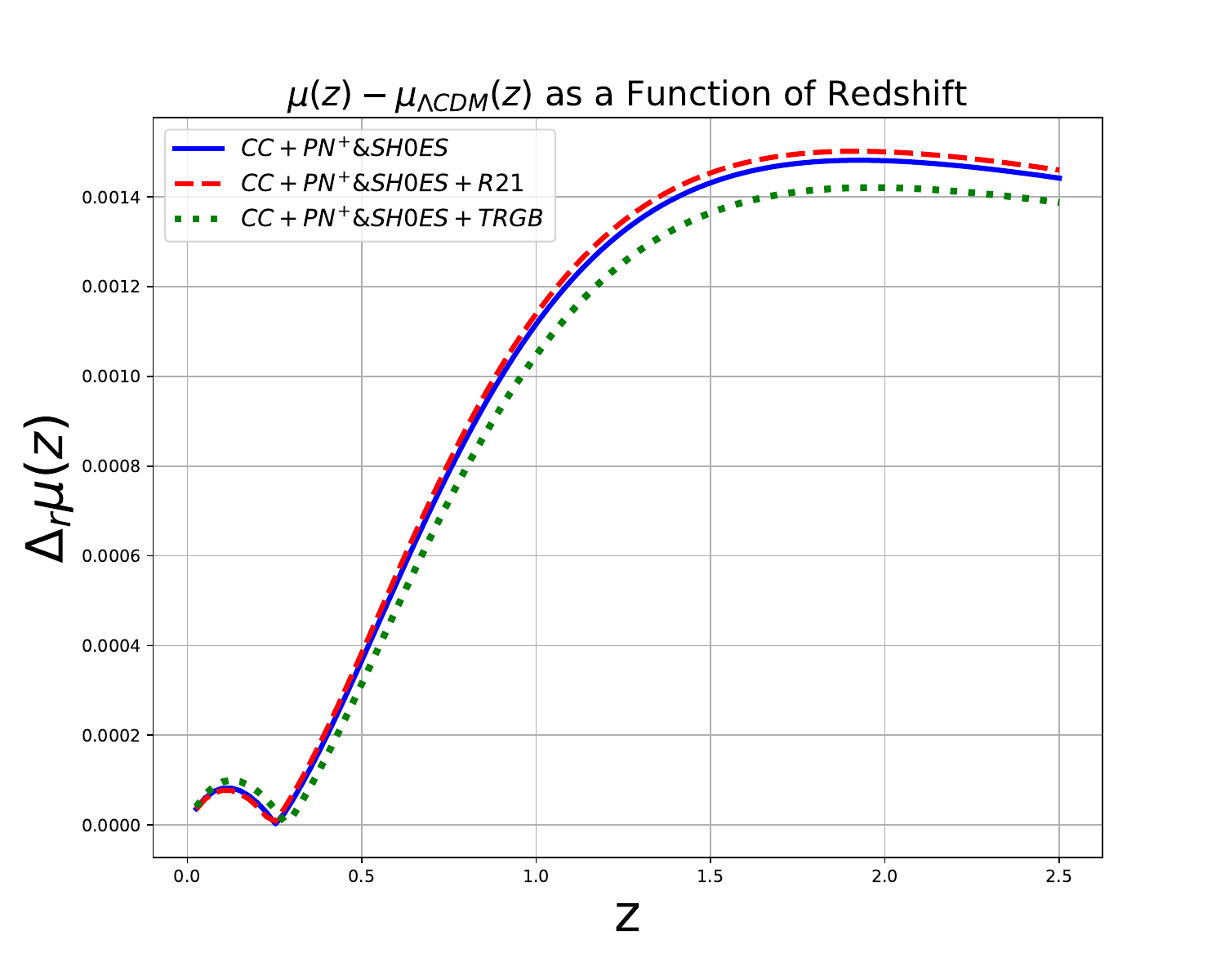}
\caption{Evolutionary behavior of distance modulus and comparative analysis of the evolution of the distance modulus function between the selected model and the $\Lambda$CDM model in redshift for the datasets combination: CC, PN$^{+}$ (without SH0ES) and  PN$^{+}$\&SH0ES (with SH0ES). The $H_0$ priors are: R21 and TRGB. } 
\label{plusFigmudulasdifference}
\end{figure}

In Figs.~\ref{plusFigmudulasdifference}, we illustrate the evolution of the distance modulus and comparative analysis of the evolution of the distance modulus function between the selected model and the $\Lambda$CDM model. For the selected cosmological model in comparison with the $\Lambda$CDM framework, utilizing a data set comprising 1701 data points from the SNIa observations. The analysis reveals a noteworthy concordance between the selected and $\Lambda$CDM models. The calculated relative difference can be defined as  
\begin{equation}\label{relativedifferencemodulus}
\Delta_{r}\mu(z) = \frac{\left| \mu_{\text{model}} - \mu_{\Lambda \text{CDM}} \right|}{\mu_{\Lambda \text{CDM}}} \,.
\end{equation}

\begin{figure}[ht]
     \centering
         \includegraphics[width=40mm]{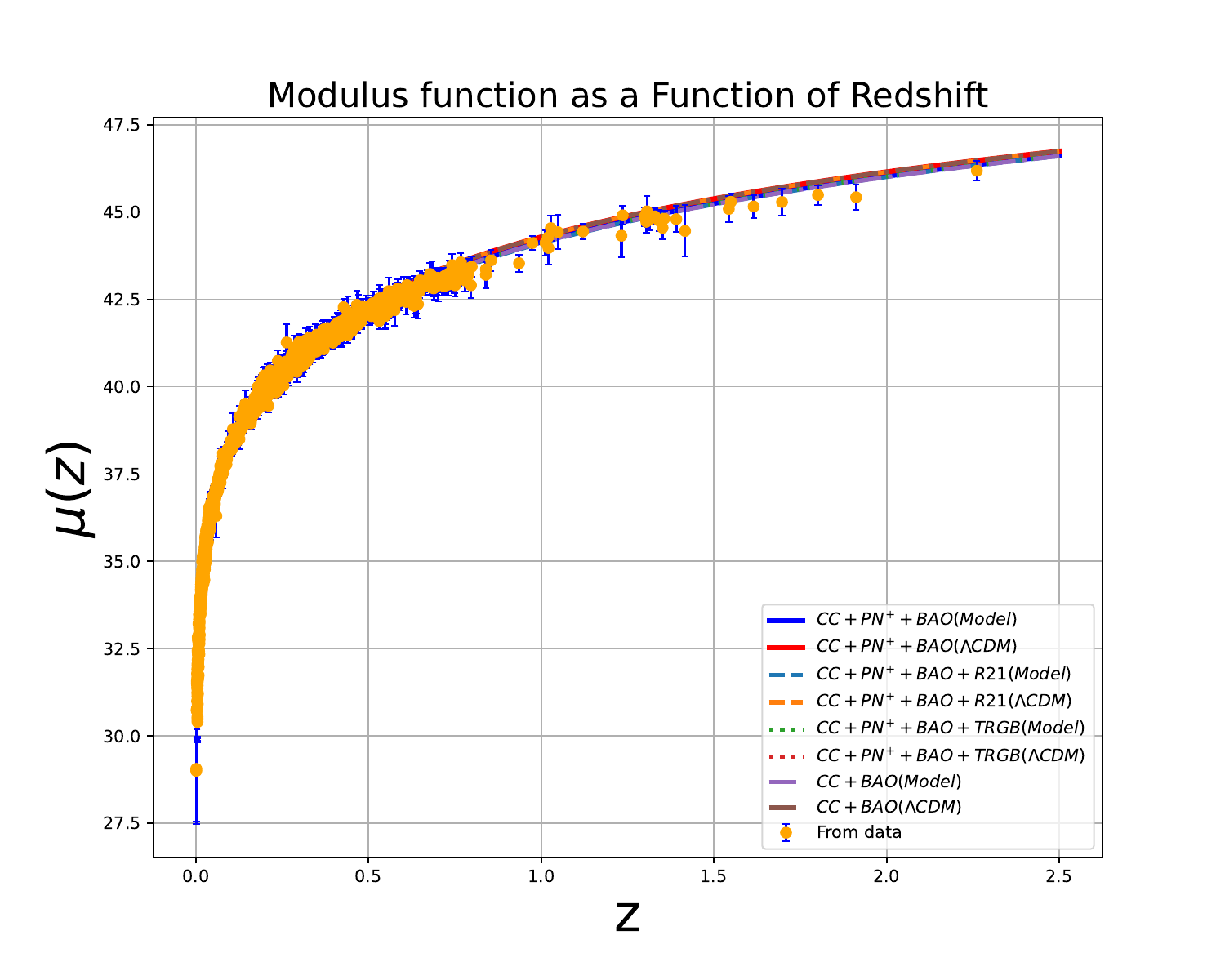}
          \includegraphics[width=40mm]{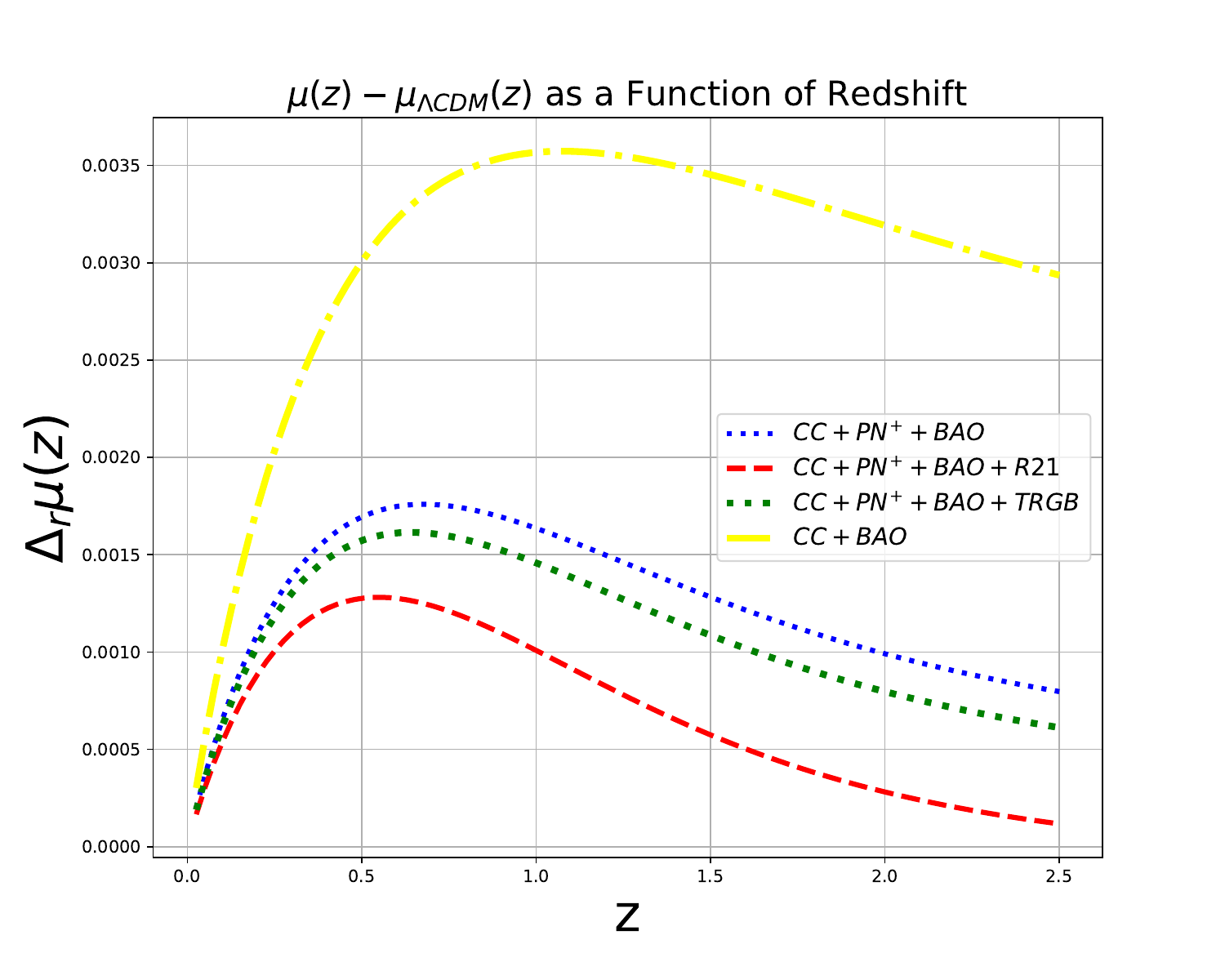}
       \includegraphics[width=40mm]{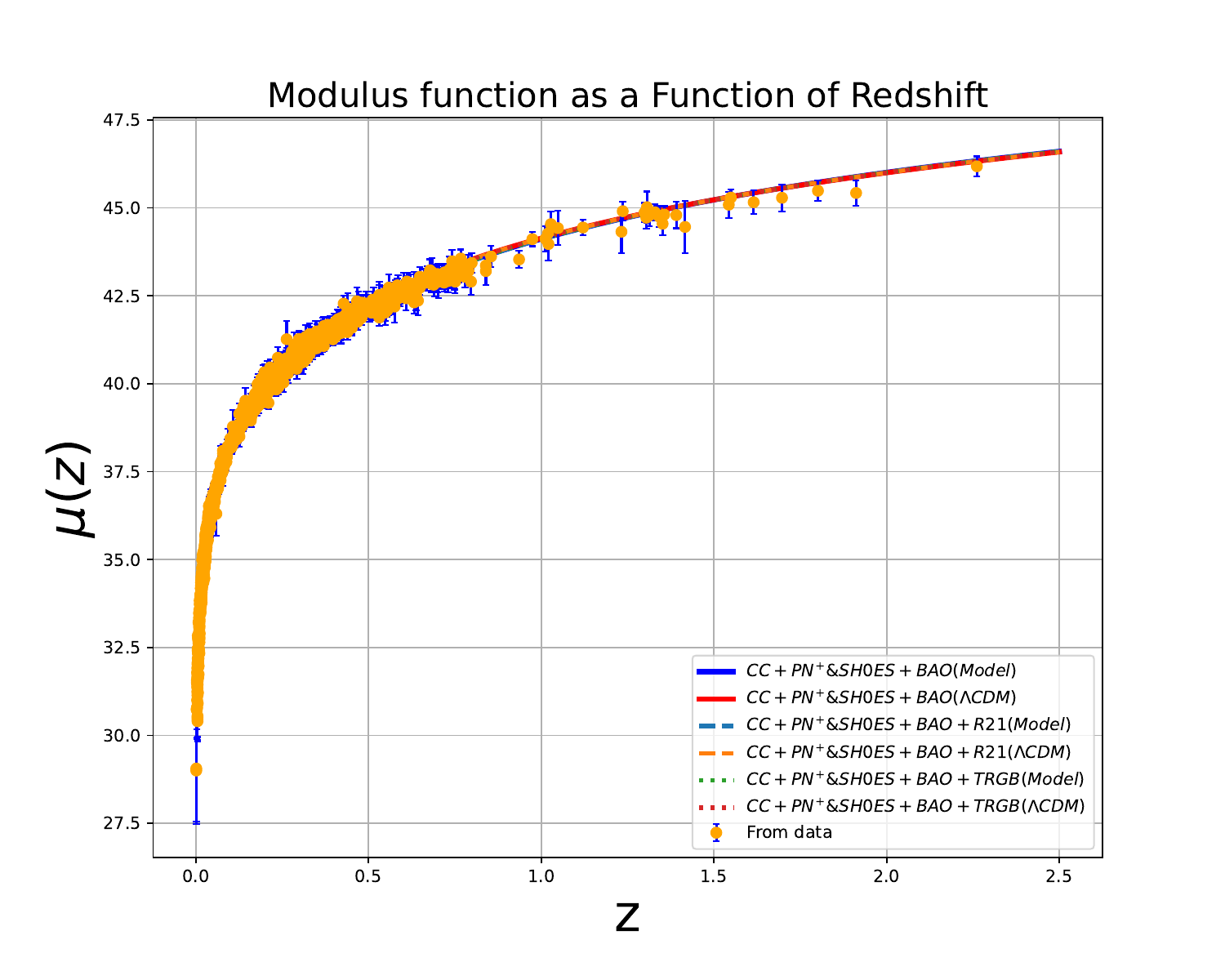}
        \includegraphics[width=40mm]{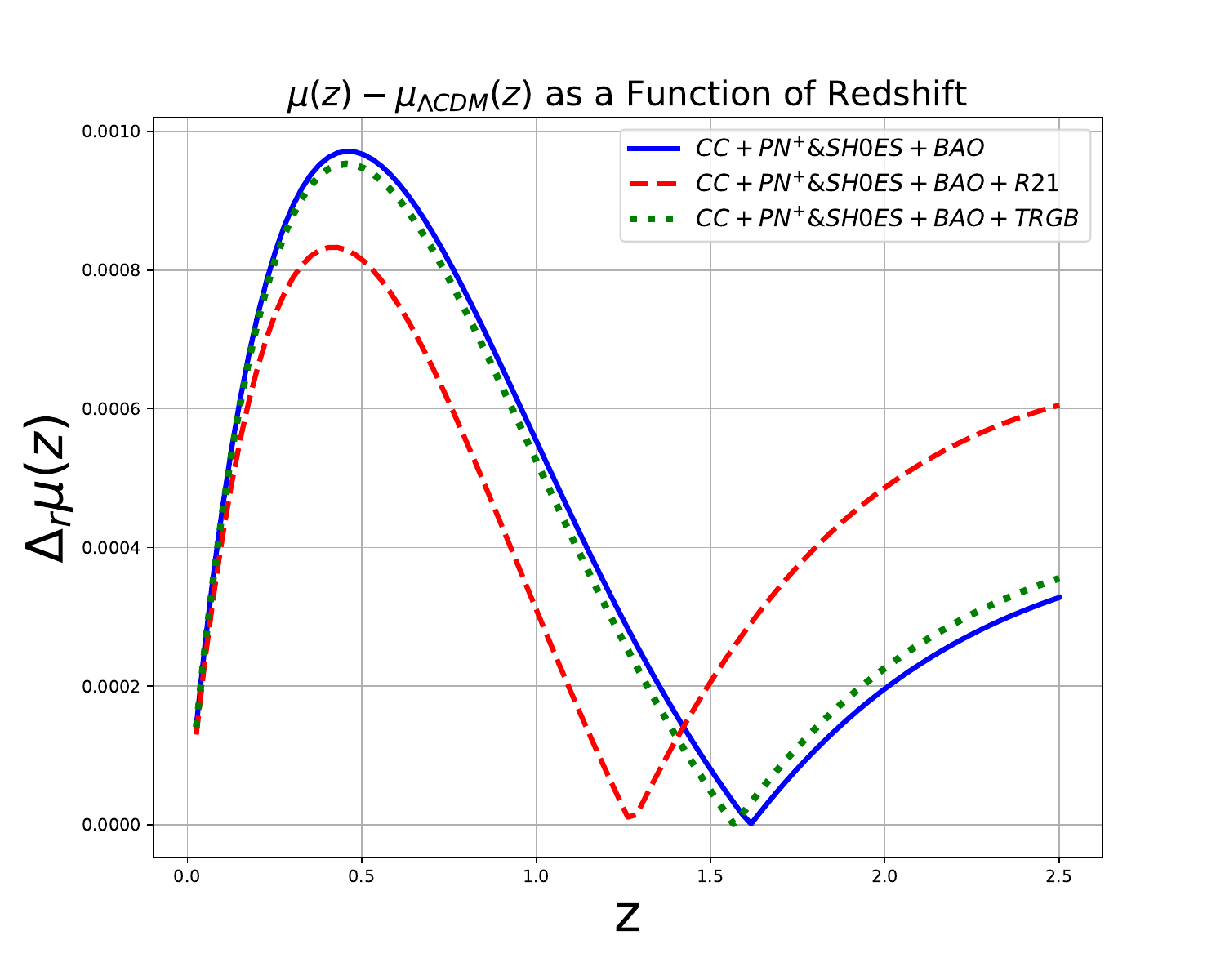}
\caption{Evolutionary behavior of distance modulus and comparative analysis of the evolution of the distance modulus function between the selected model and the $\Lambda$CDM model in redshift for the data sets combination: CC, PN$^{+}$ (without SH0ES), PN$^{+}$\&SH0ES (with SH0ES) and BAO. The $H_0$ priors are: R21 and TRGB. } 
\label{plusFigBAOmudulasdifference}
\end{figure}

Fig.~\ref{plusFigBAOmudulasdifference} illustrates the progression of the distance modulus function alongside the relative difference in distance modulus for the CC, PN$^{+}$, PN$^{+}$\& SH0ES and BAO data set, taking into account the $H_{0}$ priors. This observed trend is consistent with the patterns seen in Figs.~\ref{plusFigmudulasdifference}, highlighting similar behaviors across the different datasets.

The deceleration parameter, the total EoS and the matter-energy density as a function of redshift can be expressed as,
\begin{eqnarray}
q = -1 + \frac{(1+z) H^{'}(z)}{H(z)} \,, \\ 
\omega_{tot} = -1 + \frac{2 (1+z) H^{'}(z)}{3 H(z)} \,, \\ 
\Omega_{m} = \frac{\Omega{m0} (1+z)^3 H^2_{0}}{H^{2}(z)} \,.  
\end{eqnarray} 

\begin{figure}[ht]
     \centering
         \includegraphics[width=40mm]{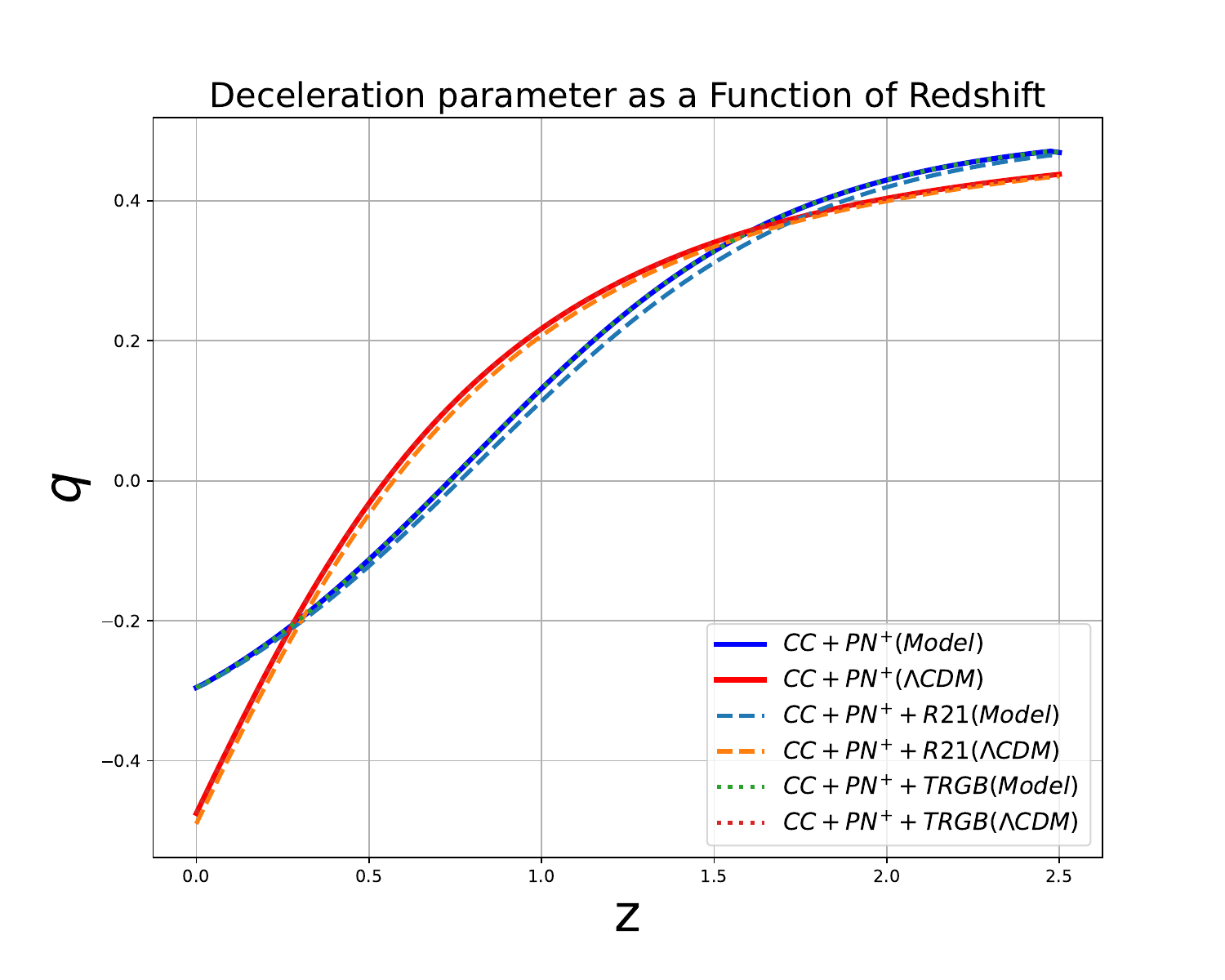}
         \includegraphics[width=40mm]{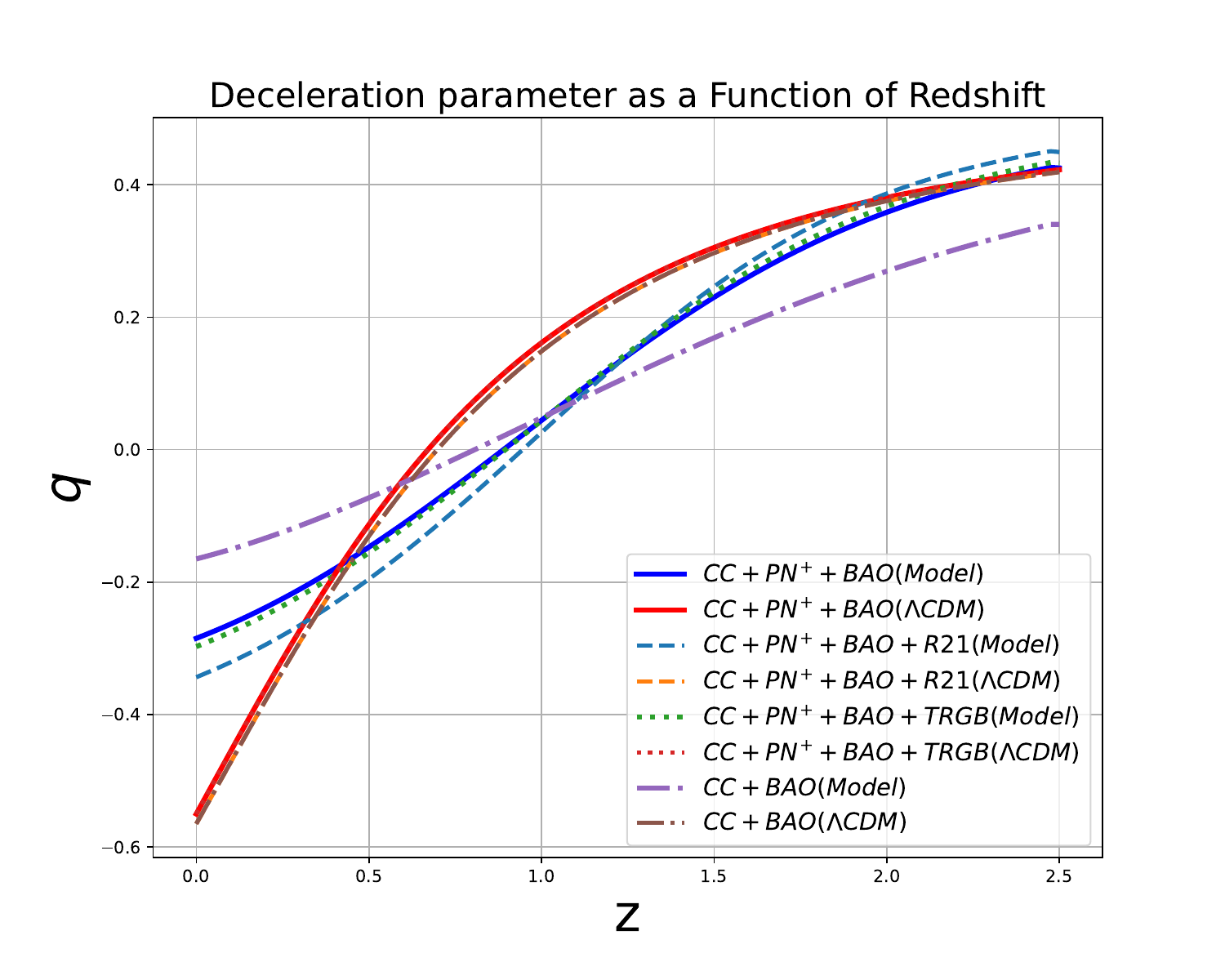}
        \includegraphics[width=40mm]{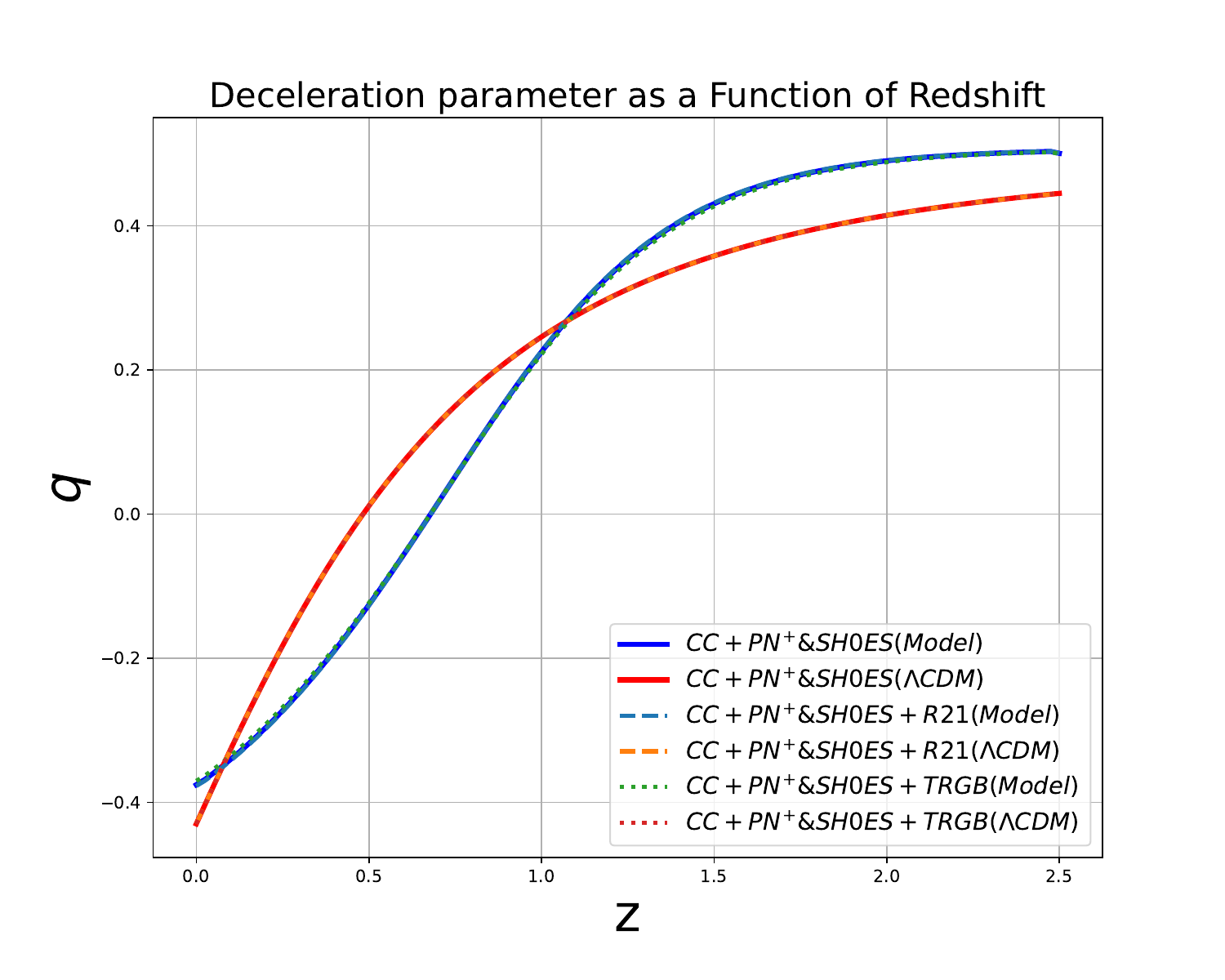}
        \includegraphics[width=40mm]{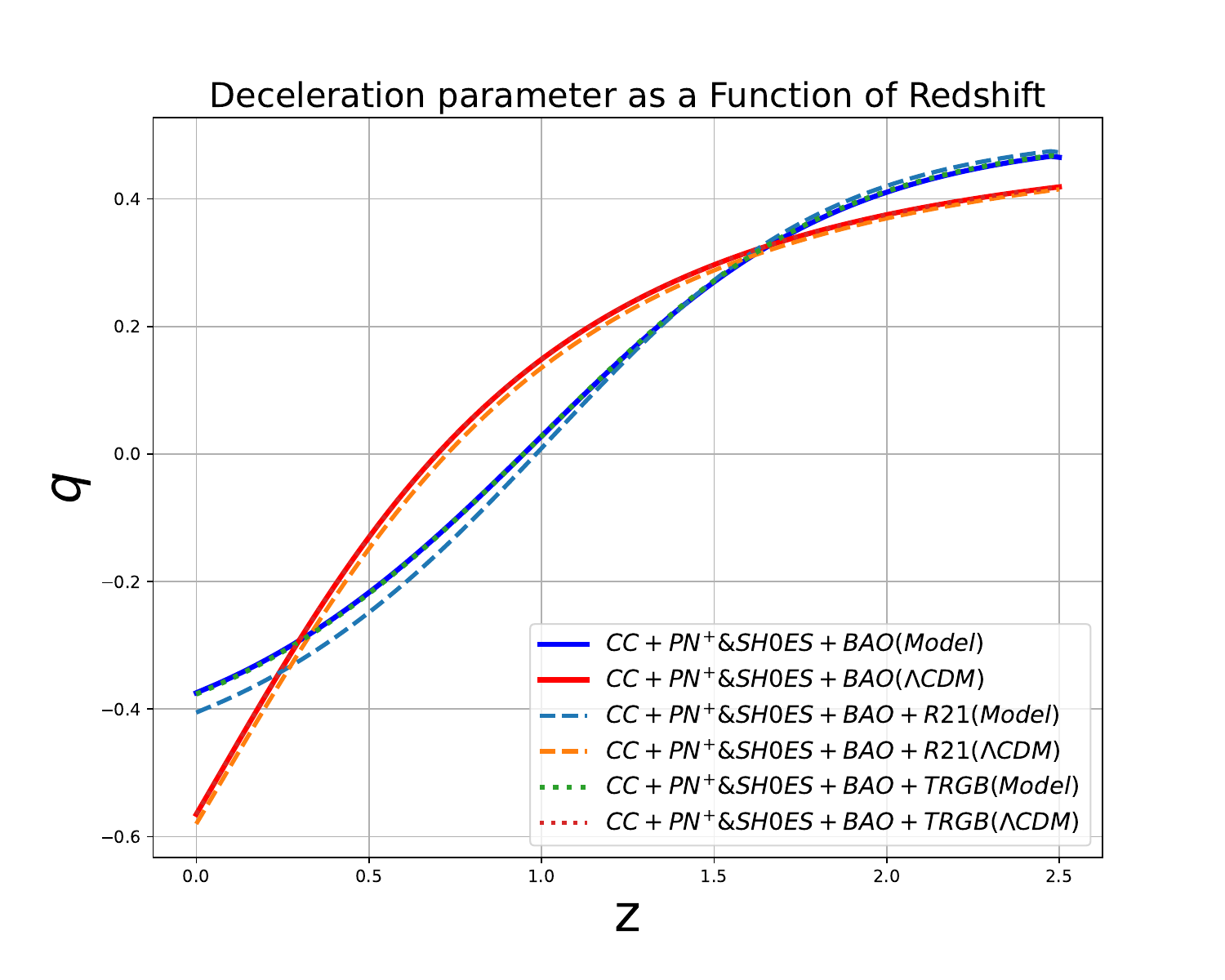}
\caption{Evolutionary behavior of the deceleration parameter and  $\Lambda$CDM model in redshift for the data sets combination: CC, PN$^{+}$ (without SH0ES), PN$^{+}$\&SH0ES (with SH0ES) and BAO. The $H_0$ priors are: R21 and TRGB. } 
\label{plusCCBAOdeceleration}
\end{figure}

In Figs.-\ref{plusCCBAOdeceleration}, we illustrate the evolution of the deceleration parameter for both the selected model and the $\Lambda$CDM model. The selected model indicates a transition from a decelerating period to an accelerating phase of the Universe, implying its potential to represent the accelerated expansion of the Universe. Additionally, we determine the current value of the deceleration parameter. Details regarding the current value of the deceleration parameter and the transition point for various data set combinations can be found in Table~\ref{results}. The outcomes from the selected model concerning the current deceleration parameter value and the transition point are consistent with cosmological observations \cite{PhysRevD.90.044016a, PhysRevResearch.2.013028}.
\begin{figure}[ht]
     \centering
         \includegraphics[width=40mm]{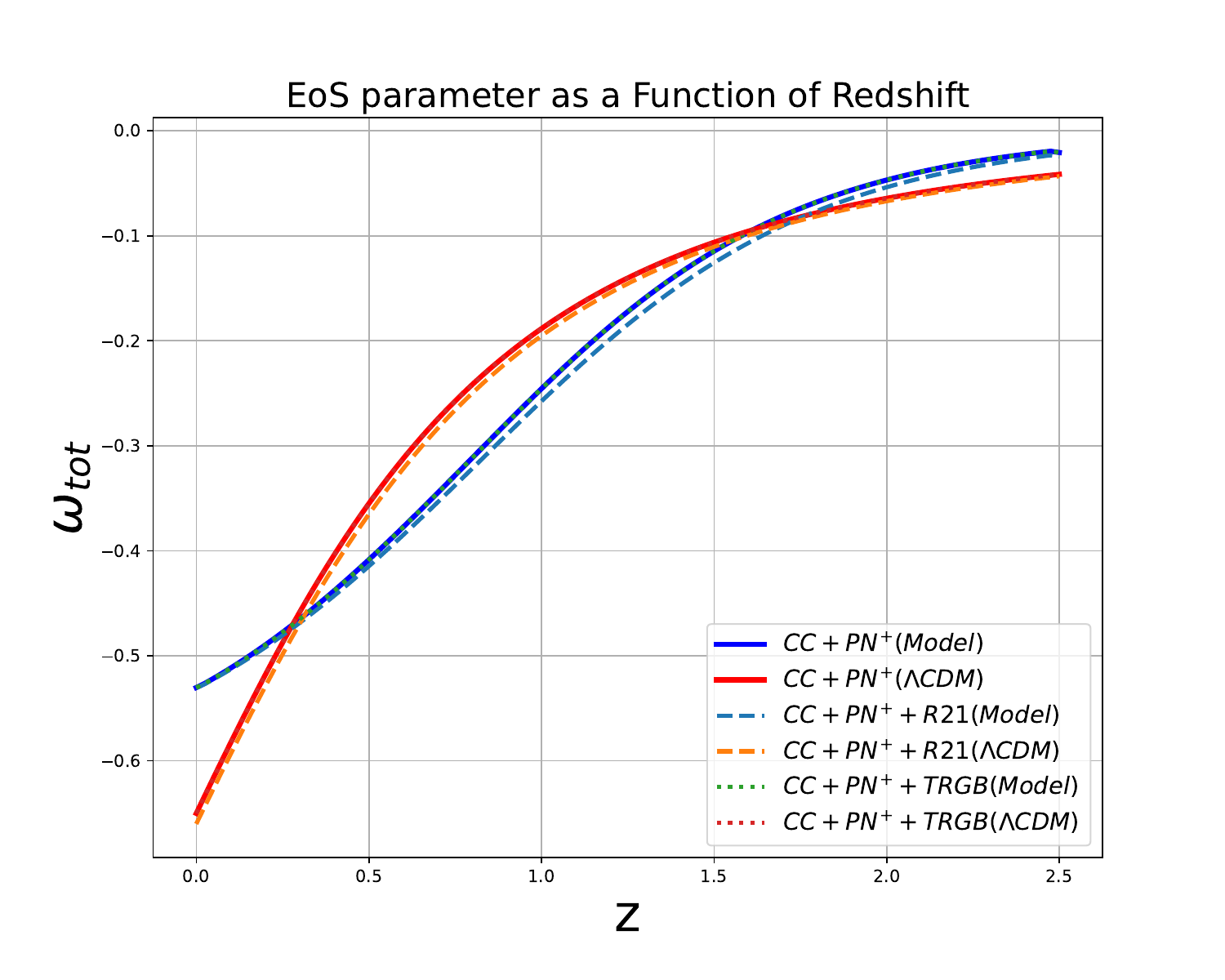}
         \includegraphics[width=40mm]{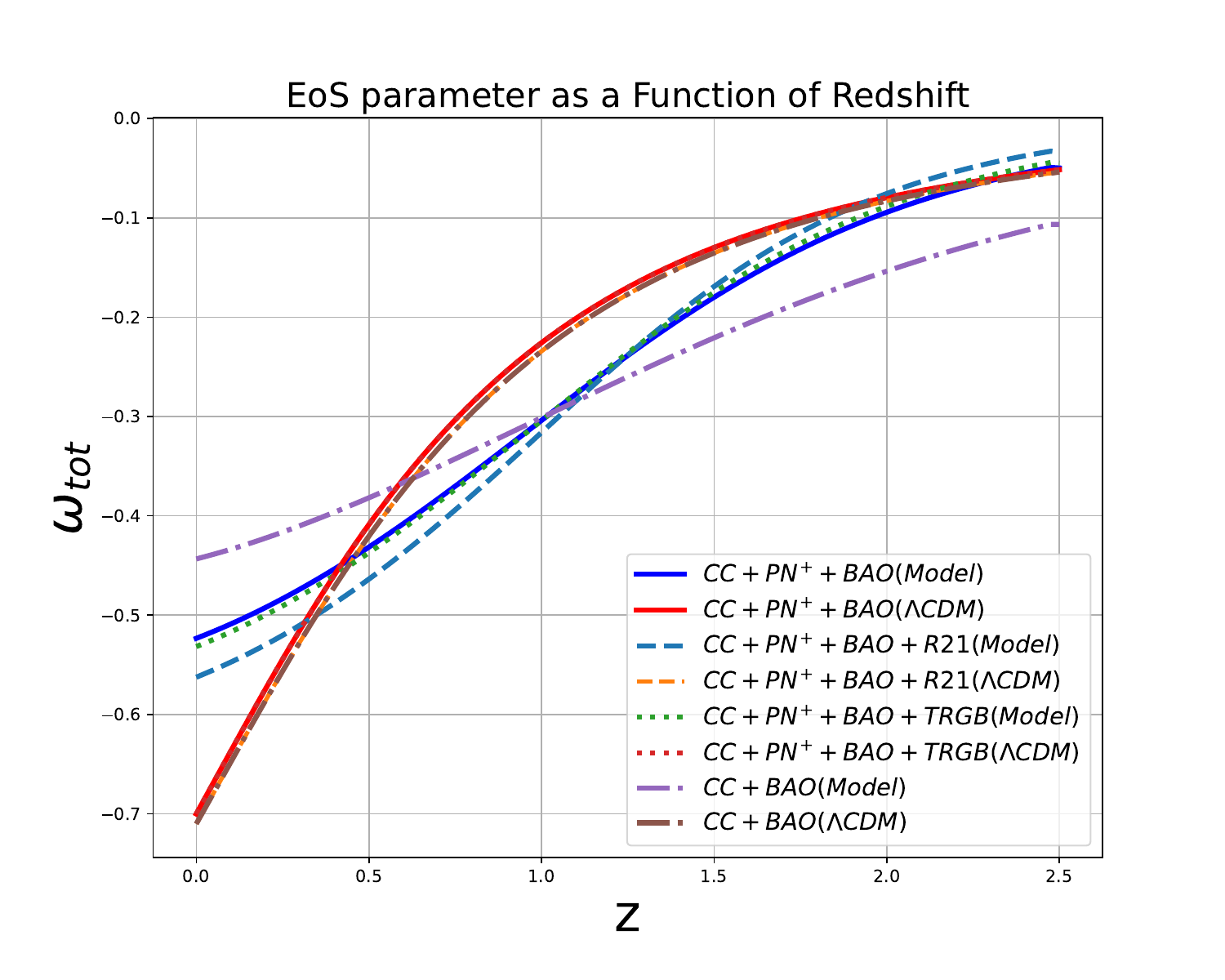}
      \includegraphics[width=40mm]{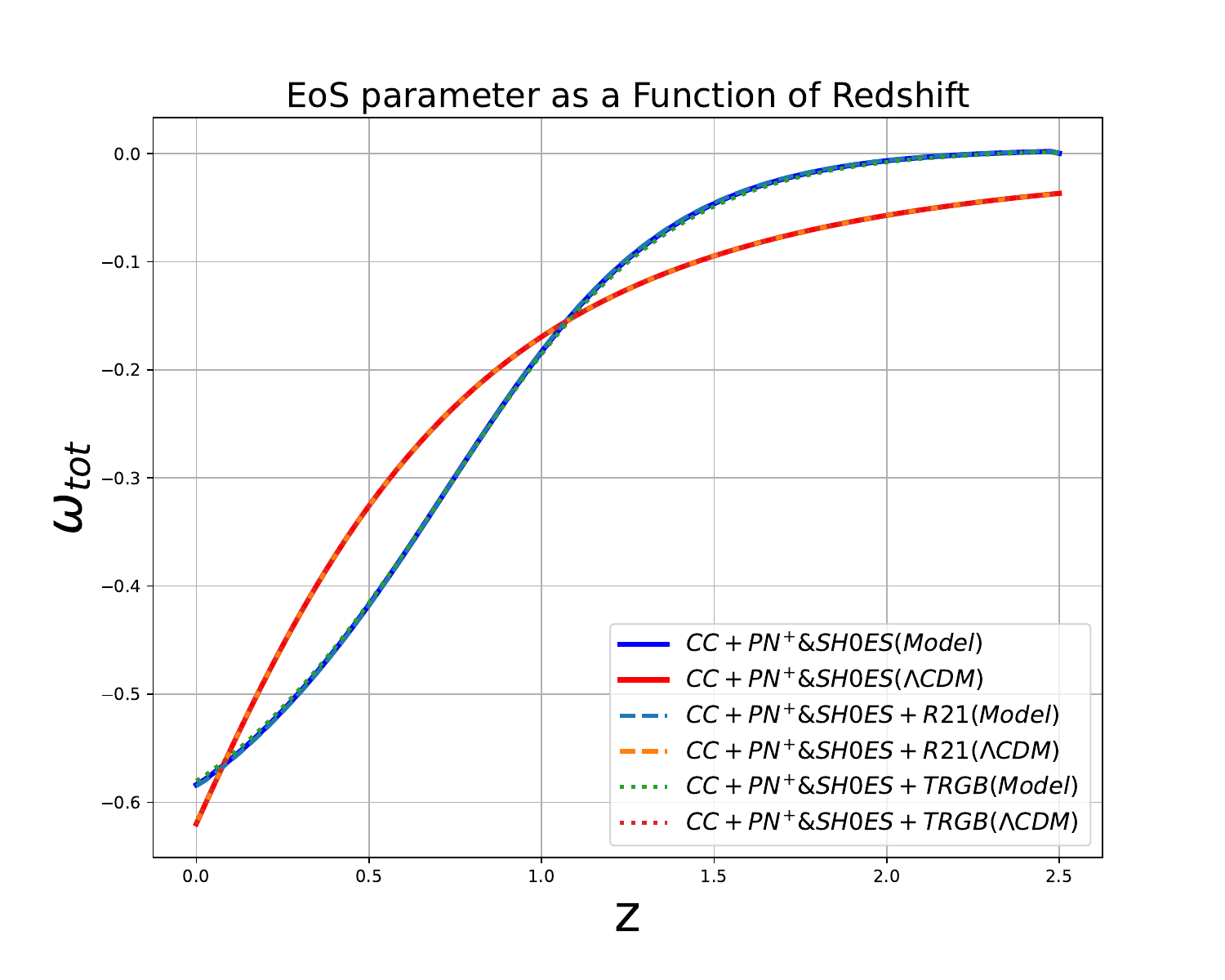}
       \includegraphics[width=40mm]{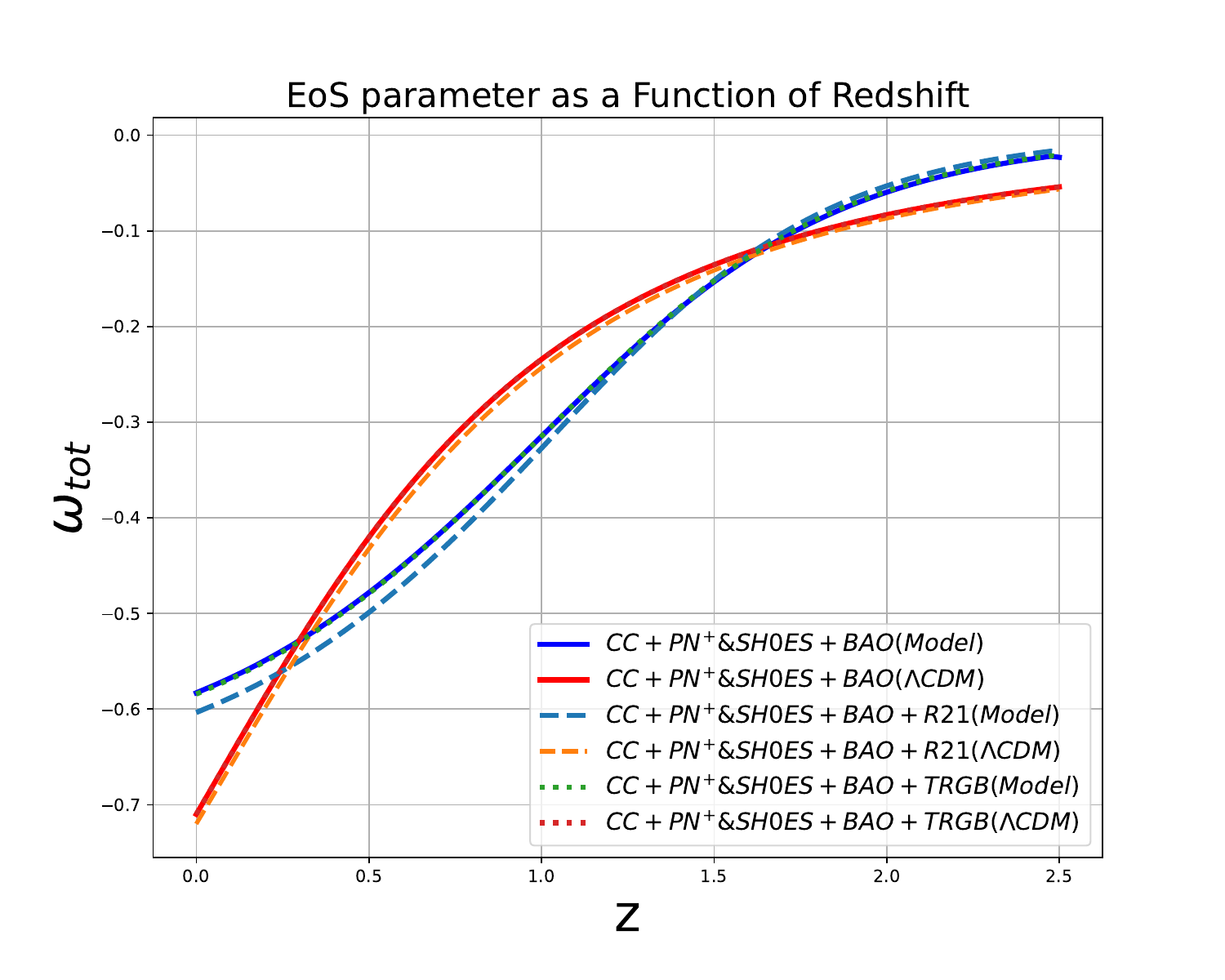}
\caption{Evolutionary behavior of the EoS parameter and  $\Lambda$CDM model in redshift for the data sets combination: CC, PN$^{+}$ (without SH0ES), PN$^{+}$\&SH0ES (with SH0ES) and BAO. The $H_0$ priors are: R21 and TRGB.} 
\label{plusCCBAOEos}
\end{figure}
In Figs.~\ref{plusCCBAOEos}, we depict the evolution of the total EoS parameter for the selected model in comparison with the $\Lambda$CDM model. The EoS parameter indicates that our chosen model shows the quintessence phase of the Universe, as it satisfies the quintessence criterion where \( -1 < \omega_{tot} < -\frac{1}{3} \). A comprehensive summary of the current values of the total EoS parameter is provided in Table \ref{results}. 
\begin{figure}[ht]
     \centering
         \includegraphics[width=40mm]{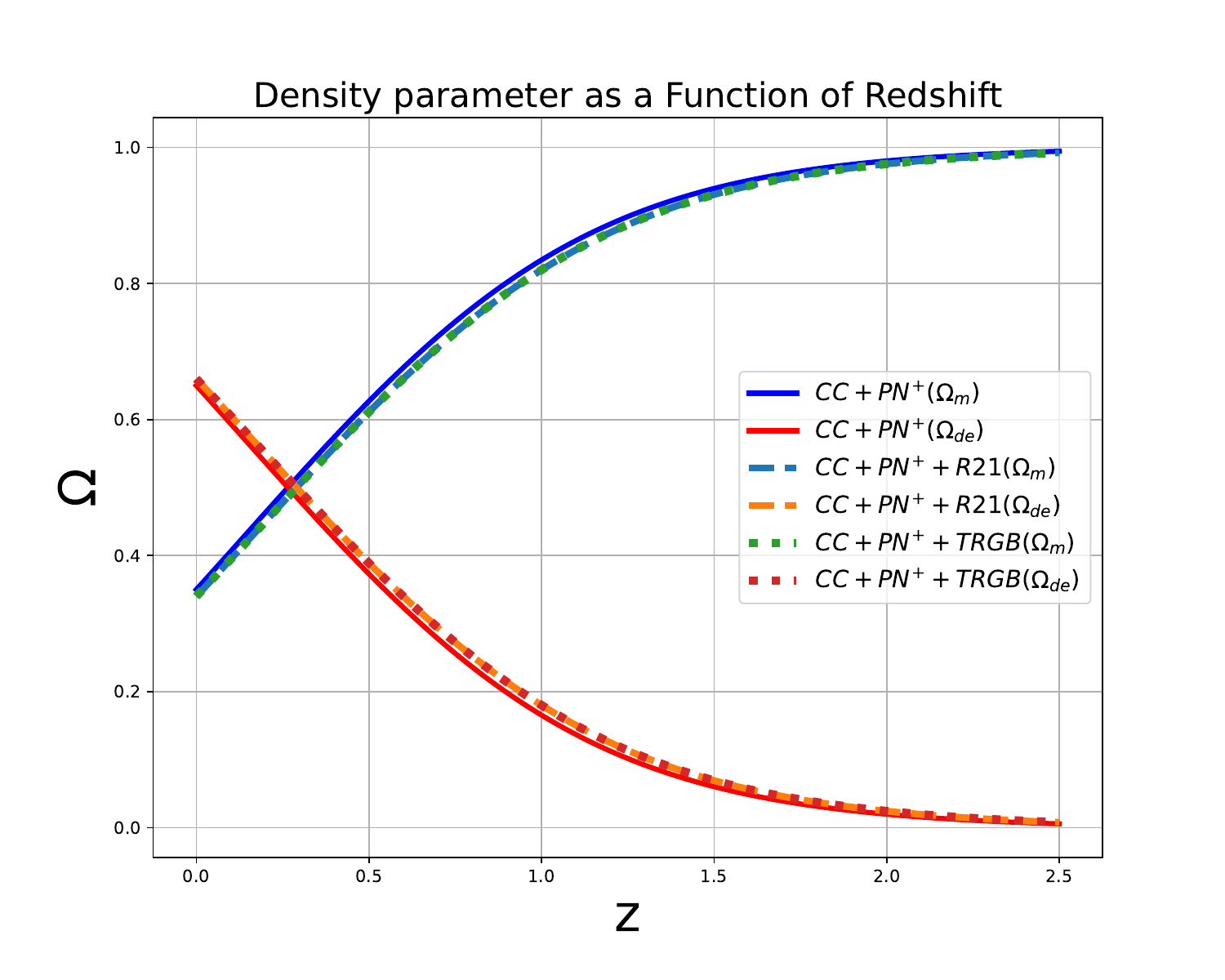}
         \includegraphics[width=40mm]{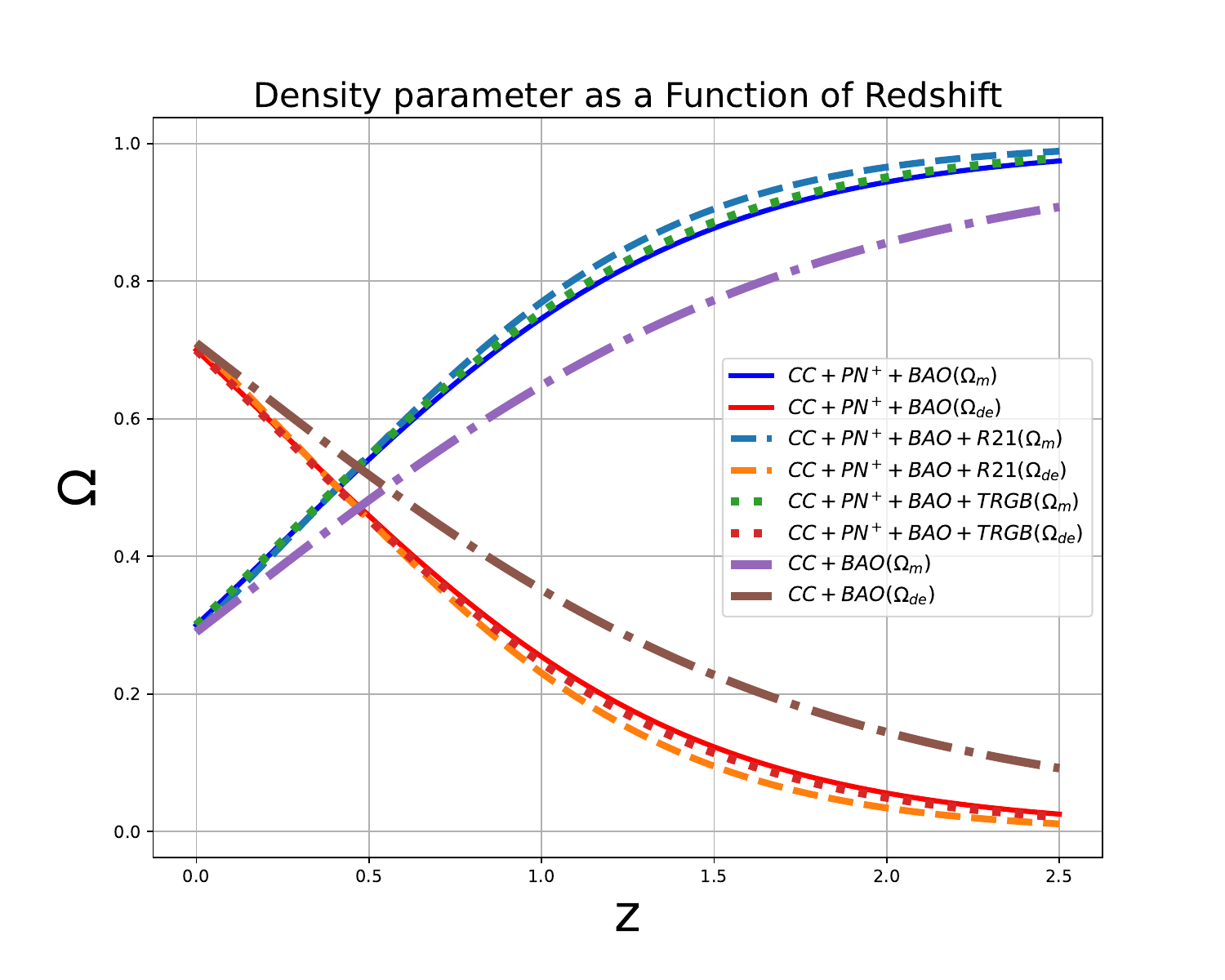}
      \includegraphics[width=40mm]{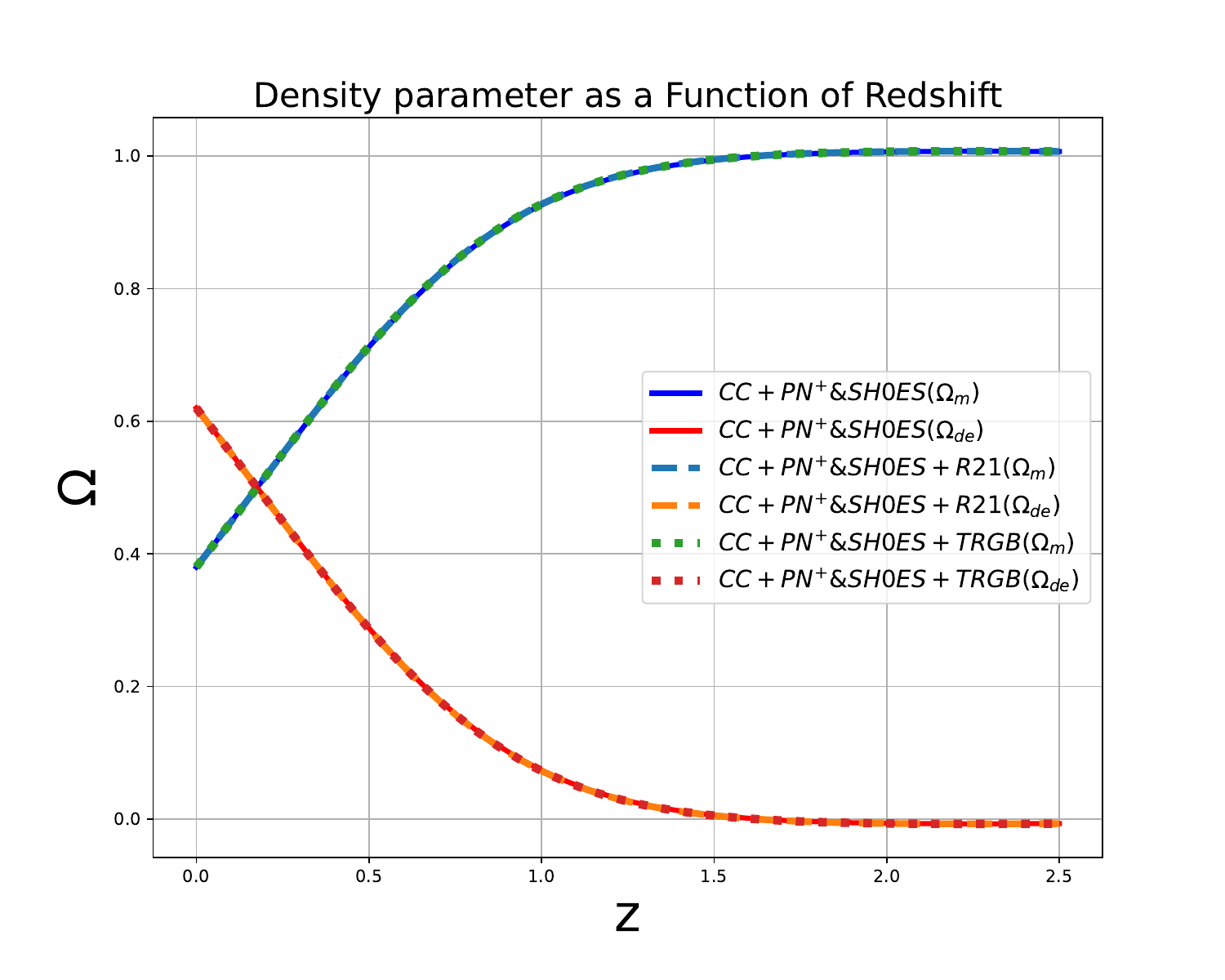}
        \includegraphics[width=40mm]{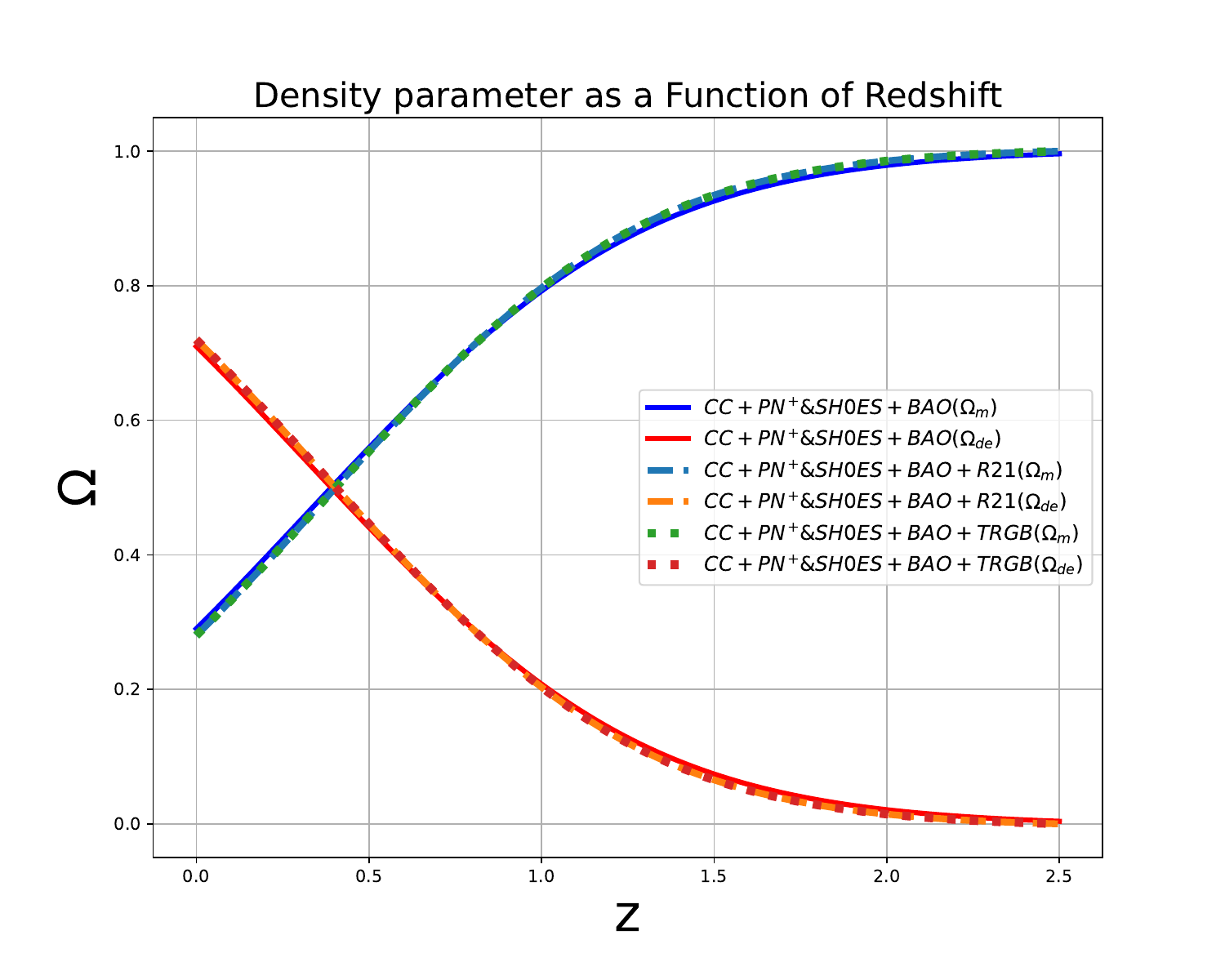}
\caption{Evolutionary behavior of the density parameters and  $\Lambda$CDM model in redshift for the data sets combination: CC, PN$^{+}$ (without SH0ES), PN$^{+}$\&SH0ES (with SH0ES) and BAO. The $H_0$ priors are: R21 and TRGB. }  
\label{plusCCBAOdensity}
\end{figure} 
In Figs.~\ref{plusCCBAOdensity}, we present the evolution of the density parameters for both matter and DE in the Universe as redshift varies. These plots demonstrate the shifting interaction between matter and DE, showing that the density of DE increasingly dominates in later epochs, showing the accelerated expansion. The current values of these density parameters are listed in Table~\ref{results}. During the early stages of the Universe, DM constitutes the main part of the energy density at higher redshifts, far outpacing DE. As redshift decreases, the proportion of DM drops while the influence of DE steadily rises. In the later stages of the Universe, DE becomes the leading component and at lower redshifts, it overtakes DM, driving the accelerated expansion.  

The \( Om(z) \) diagnostic serves as an alternative method to distinguish between various DE cosmological models. This can be represented as \cite{sahniPRD_om, Sahni_2003}.
\begin{equation}\label{Omzequation}
Om(z)= \frac{E^{2}(z)-1}{(1+z)^3-1} \,.   
\end{equation}
Assessing the \( Om(z) \) values across different redshifts can yield insights into the characteristics and dynamics of DE. The procedure for two-point difference diagnostics can be described as follows:
\begin{equation}\label{Omztworedshiftdifference}
Om(z_1-z_2) = Om(z_1)- Om(z_2)\,.   
\end{equation}

\begin{figure}[ht]
     \centering
         \includegraphics[width=40mm]{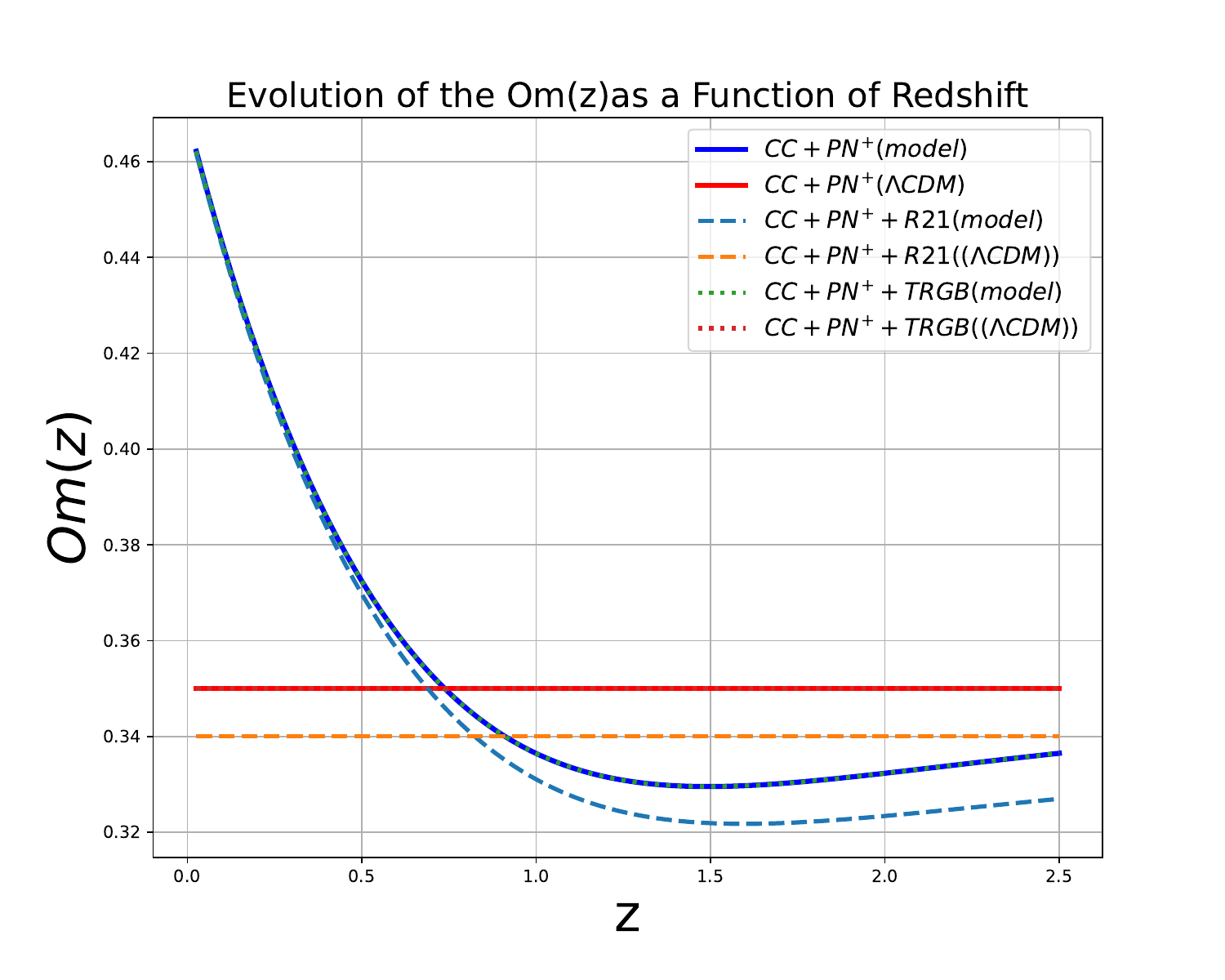}
          \includegraphics[width=40mm]{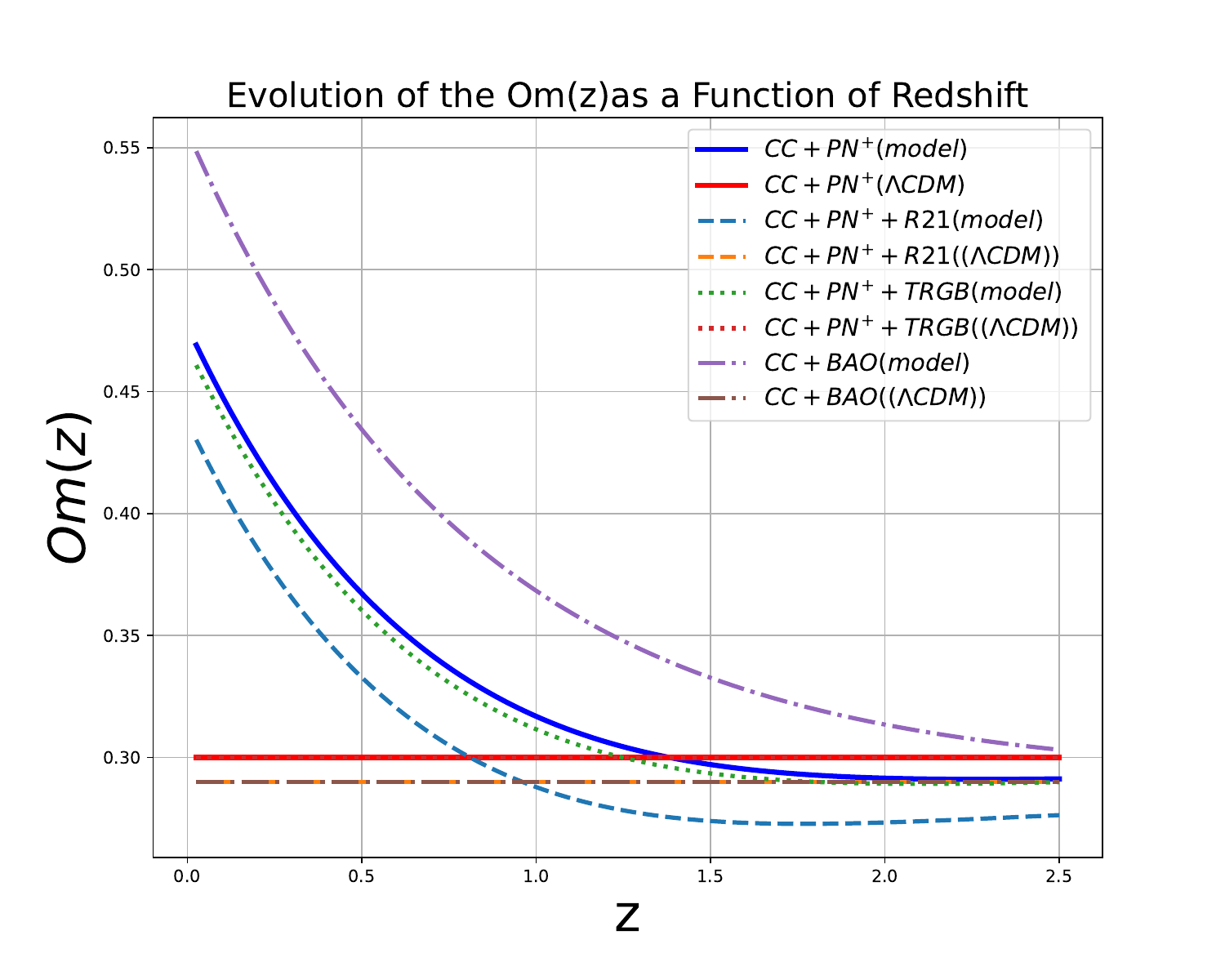}
       \includegraphics[width=40mm]{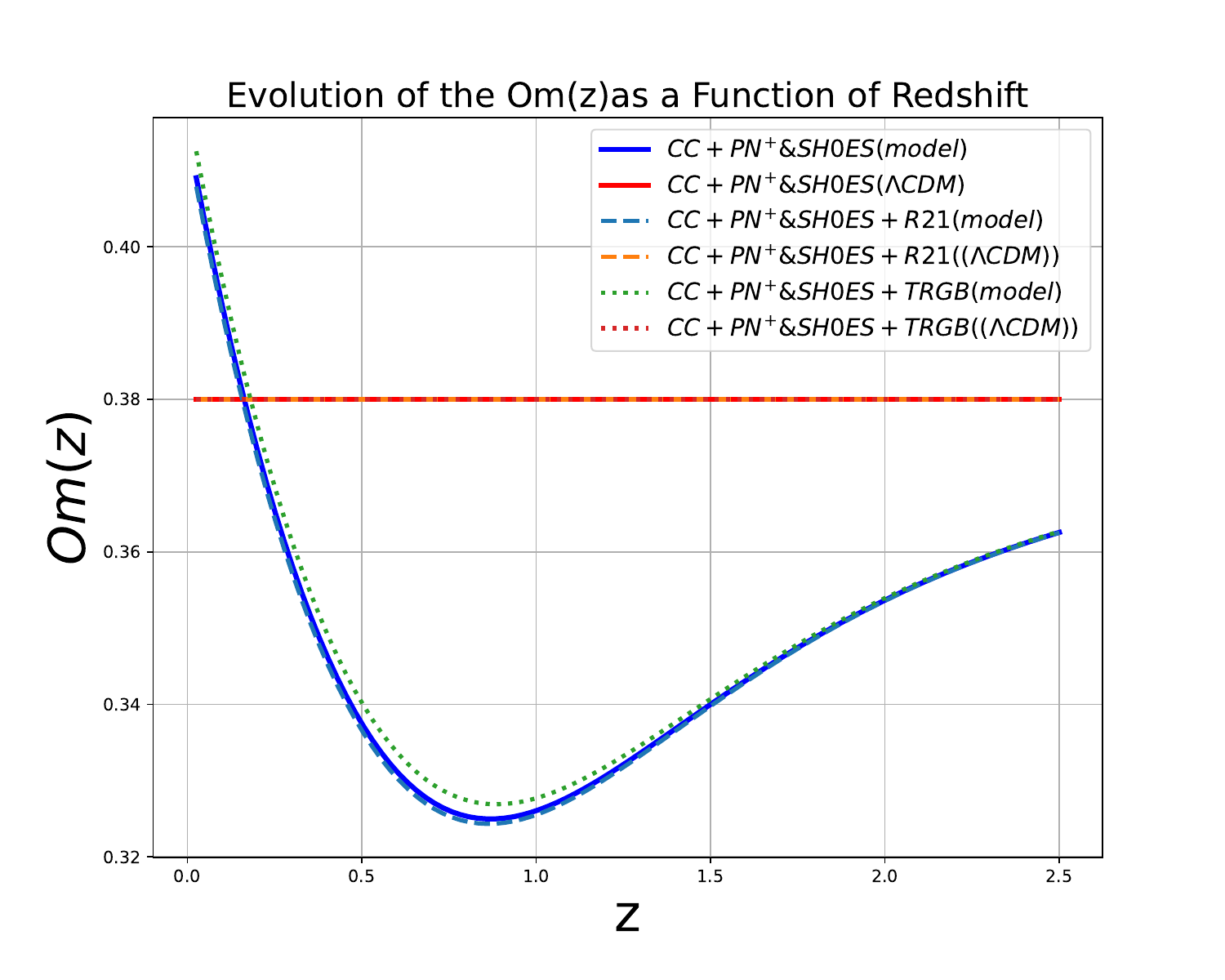}
         \includegraphics[width=40mm]{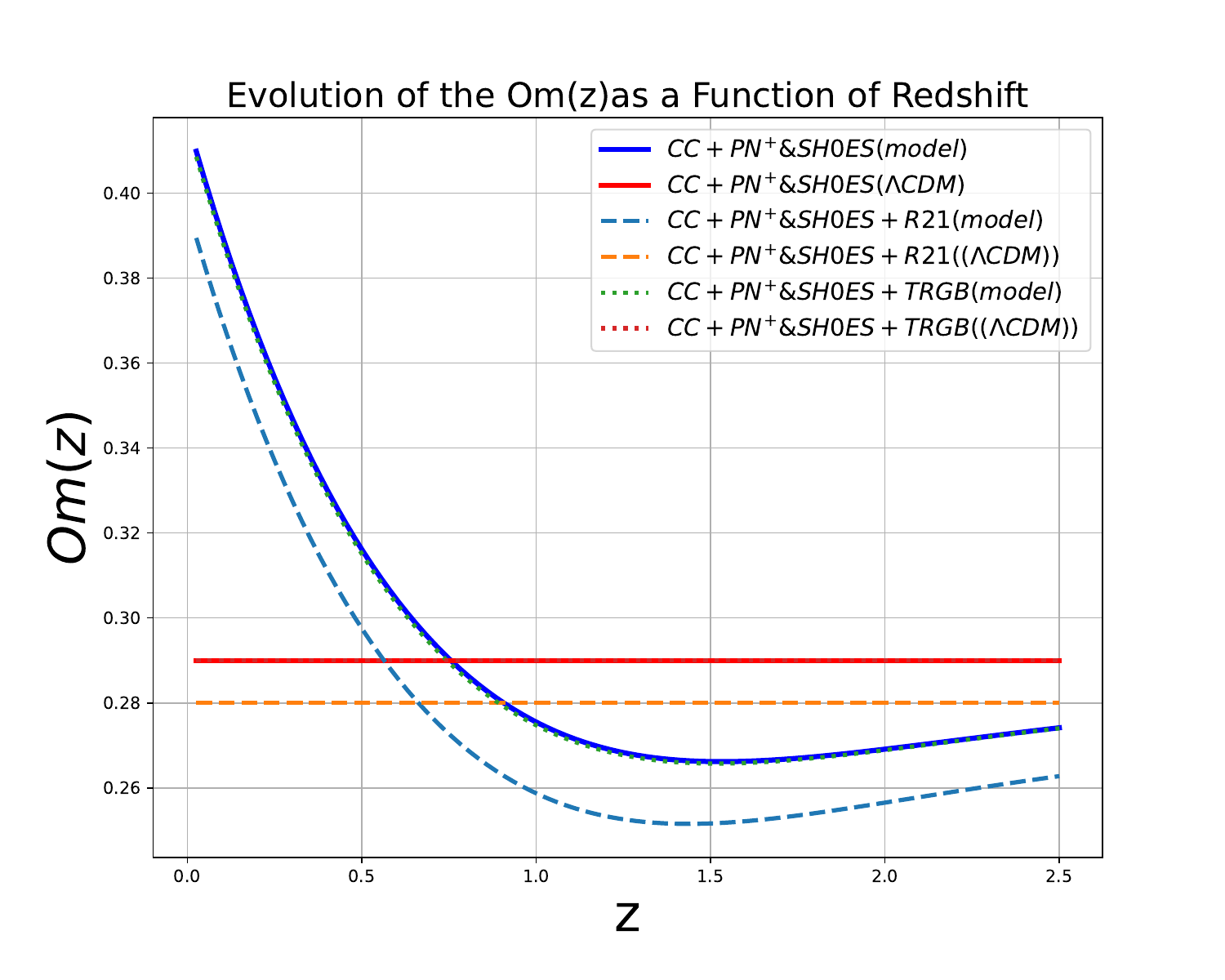}
\caption{Evolutionary behavior of the $Om(z)$ parameter and  $\Lambda$CDM model in redshift for the data sets combination: CC, PN$^{+}$ (without SH0ES), PN$^{+}$\&SH0ES (with SH0ES) and BAO. The $H_0$ priors are: R21 and TRGB. } 
\label{plusFigCCBAOOm}
\end{figure}

If \( Om(z_1, z_2) > 0 \), this suggests the model represents a quintessence scenario, while \( Om(z_1, z_2) < 0 \) implies it exhibits phantom behavior, given that \( z_1 < z_2 \). Additionally, if \( Om(z) \) remains consistent across various redshifts, it indicates a potential connection between DE and the cosmological constant \cite{sahniPRD_om}. We have compared the model discussed here with the $\Lambda$CDM model, as shown in Fig.~\ref{plusFigCCBAOOm}. The \( Om(z) \) parameter displays stability within the redshift range \( 0 < z < 2.5 \). This stability is crucial for understanding the dynamics of the Universe and its accelerated expansion. The slope of \( Om(z) \) is an important indicator for DE models. A positive slope suggests the presence of phantom behavior, which is characterized by an EoS \( \omega < -1 \). Conversely, a negative slope is linked to the quintessence region, where \( \omega > -1 \). In Fig.~\ref {plusFigCCBAOOm}, the slope of the \( Om(z) \) parameter shows a downward trend as redshift increases, suggesting that the impact of DE becomes more significant over time. This reduction indicates a transformation in the dynamics of the Universe, illustrating the evolving roles of matter and DE as the Universe progresses. This decline is consistent with the quintessence phase of the Universe, where the EoS \( \omega \) is greater than \(-1\).

\begin{figure}[H]
 \centering
 \includegraphics[width=140mm]{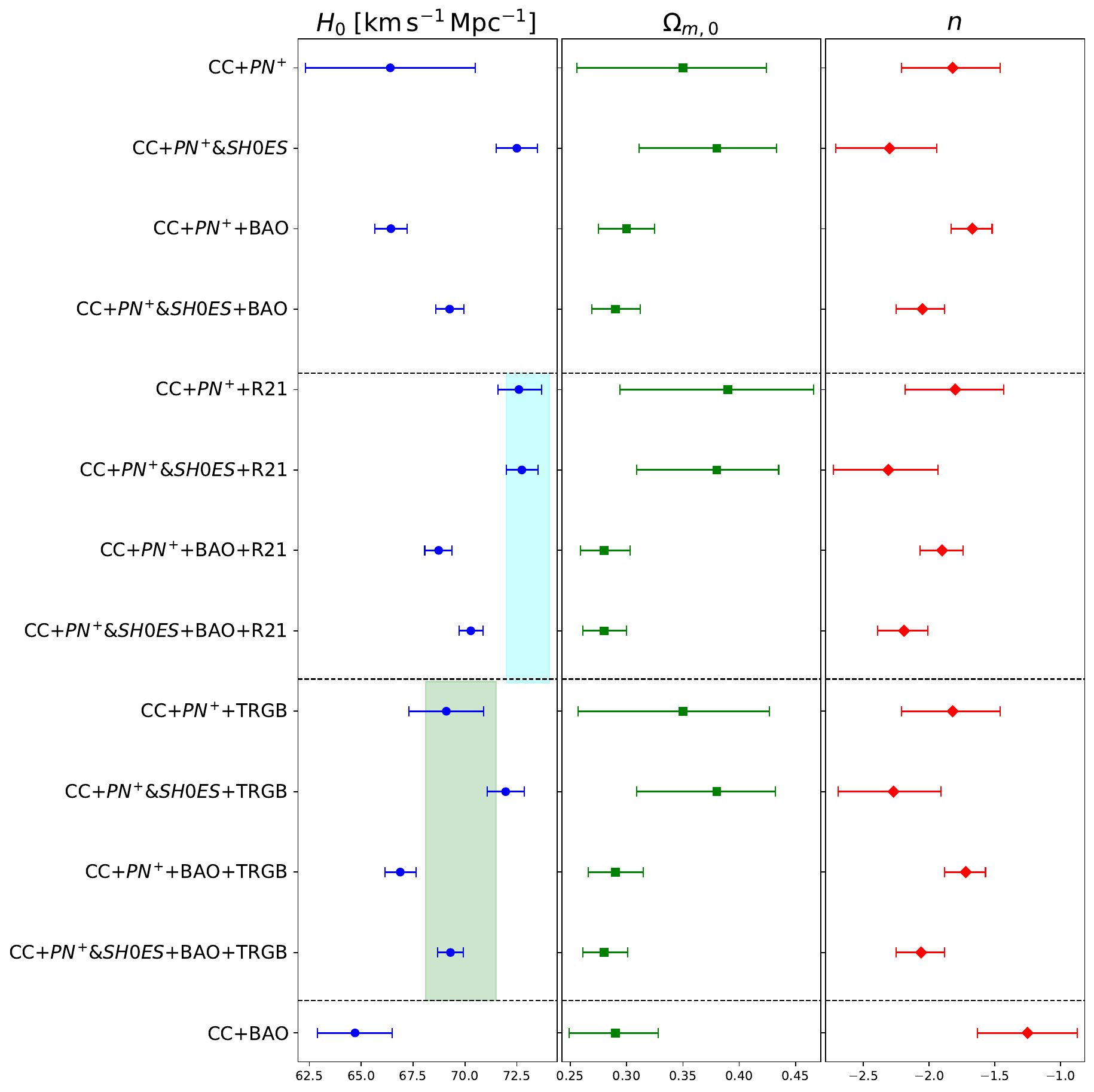} 
 \caption{Whisker plot for the chosen \( f(T, \mathcal{T}) \) model. This plot provides a visual representation of the distributions of several key parameters: the Hubble constant \( H_{0} \), the matter-energy density \( \Omega_{m0} \) and the model parameters \( n \). In the first column, the cyan-shaded region represents the R21 prior, while the green-shaded region indicates the TRGB prior.
} \label{whiskerplot}
 \end{figure}

The whisker plot represents the ranges and central tendencies of the model parameters. In Fig.~\ref{whiskerplot}, the whisker plot effectively distinguishes parameter values derived from different combinations of data sets. Through Fig.~\ref{whiskerplot}, we observe a notable contrast in the values of the parameters ($H_0$, $\Omega_{m0}$, $n$) between the PN$^+$ (without SH0ES) and PN$^+$\& SH0ES data set combinations. Across all combinations of data sets, we find a lower value of $H_0$ in the CC+BAO data set combination where both PN$^+$ and PN$^+$\& SH0ES are absent. Additionally, we can observe how the inclusion of $H_0$ priors influences the values of $H_0$ within the PN$^+$ data set combination, as incorporating the $H_0$ priors increases the $H_0$ values.
\begin{table}[H]
    \centering
    \renewcommand{\arraystretch}{1.2} 
    \begin{tabular}{|p{4.7cm}|p{1cm}|p{1cm}|p{1cm}|p{1cm}|p{1cm}|p{1cm}|p{1cm}|p{1cm}|}
        \hline 
        \textbf{Data set} & \multicolumn{5}{c|}{\textbf{For $f(T, \mathcal{T})$ model}} & \multicolumn{3}{c|}{\textbf{For the $\Lambda$CDM model}} \\
        \hline
        & $q_0$ & $\omega_{0}$ & $z_{tr} $ & $\Omega_{m}^0$ & $\Omega_{de}^0$ & $q_0$ & $\omega_{0}$ & $z_{tr}$\\ [0.5ex]
        \hline
        CC+PN$^{+}$ & $-0.29$ & $-0.53$ &$0.73$  &$0.35$  & $0.65$ & $-0.48$ & $-0.65$ &$0.53$ \\
        \hline
        CC+PN$^{+}$+R21 & $-0.30$ & $-0.53$ & $0.76$ &$0.34$  & $0.66$ & $-0.49$ & $-0.65$ & $0.56$\\
        \hline
        CC+PN$^{+}$+TRGB & $-0.30$ & $-0.53$ &$0.73$  &$0.35$  & $0.65$ & $-0.48$ & $-0.65$ &$0.53$ \\
        \hline
        CC+PN$^{+}$+BAO & $-0.28$ & $-0.52$ &$0.88$  &$0.3$  & $0.7$ & $-0.55$ & $-0.70$ &$0.66$ \\
        \hline
        \begin{tabular}{@{}c@{}}CC+PN$^{+}$+ \\BAO+R21\end{tabular} & $-0.34$ & $-0.56$ &$0.93$  &$0.29$  & $0.71$ & $-0.57$ & $-0.71$ &$0.68$\\
        \hline
        \begin{tabular}{@{}c@{}}CC+PN$^{+}$+ \\BAO+TRGB\end{tabular} & $-0.30$ & $-0.53$ & $0.88$ &$0.3$  & $0.7$ & $-0.55$ & $-0.7$ &$0.66$ \\
        \hline
        CC+BAO & $-0.17$ & $-0.44$ &$0.78$  &$0.29$  & $0.71$ & $-0.565$ & $-0.71$ &$0.68$ \\
        \hline
        CC+$PN^{+}\& SH0ES$ & $-0.37$ & $-0.58$ &$0.66$  &$0.38$  & $0.62$ & $-0.43$ & $-0.62$ &$0.48$ \\
        \hline
        CC+$PN^{+}\& SH0ES$+R21 & $-0.38$ & $-0.58$ & $0.66$ &$0.38$  & $0.62$ & $-0.43$ & $-0.62$ & $0.48$\\
        \hline
        CC+$PN^{+}\& SH0ES$+TRGB & $-0.37$ & $-0.58$ &$0.66$  &$0.38$  & $0.62$ & $-0.43$ & $-0.62$ &$0.48$ \\
        \hline
        CC+$PN^{+}\& SH0ES$+BAO & $-0.37$ & $-0.59$ &$0.93$  &$0.29$  & $0.71$ & $-0.565$ & $-0.71$ &$0.68$ \\
        \hline
        \begin{tabular}{@{}c@{}}CC+$PN^{+}\& SH0ES$+ \\BAO+R21\end{tabular} & $-0.40$ & $-0.60$ &$0.96$  &$0.28$  & $0.72$ & $-0.58$ & $-0.72$ &$0.71$\\
        \hline
        \begin{tabular}{@{}c@{}}CC+$PN^{+}\& SH0ES$+ \\BAO+TRGB\end{tabular} & $-0.37$ & $-0.58$ & $0.93$ &$0.29$  & $0.71$ & $-0.57$ & $-0.71$ &$0.68$ \\
        \hline
    \end{tabular}
    \caption{Present value of the parameters and the transition point. The upper or lower indices $0$ represent the current time at $z=0$.}
    \label{results}
\end{table}

\section{conclusion} \label{conclusion}
In this study, we have presented cosmological model that supports late-time cosmic acceleration of the Universe in the framework of $f(T, \mathcal{T})$ gravity using various cosmological dataset combinations. Our analysis incorporates datasets including the CC, PN$^+$ (without SH0ES), PN$^+$\& SH0ES (with SH0ES) and BAO observations alongside the $H_0$ priors from R21 and TRGB. We specifically focus on a comparative analysis between the PN$^+$ (without SH0ES) and PN$^+$\& SH0ES (with SH0ES) datasets, alongside the BAO datasets and specific combinations of the $H_0$ priors. We present the whisker plot in Fig.-\ref{whiskerplot} to emphasize the differences between these datasets. Additionally, the findings derived from the chosen model have been contrasted with those of the standard \(\Lambda\)CDM model. Our analysis aimed at obtaining the best-fit values of the model parameters using the CC, PN$^{+}$,  PN$^{+}$\& SH0ES and BAO data sets alongside the \(H_{0}\) priors R21 and TRGB.

In examining the PN$^+$ (without SH0ES) and PN$^+$\& SH0ES (with SH0ES) datasets, we have observed that the $H_0$ value is lower for the PN$^+$ (without SH0ES) dataset when compared to the PN$^+$\& SH0ES (with SH0ES) dataset. The increase in the $H_0$ value for the PN$^+$\& SH0ES (with SH0ES) dataset can be attributed to including the SH0ES data points. The SH0ES points elevate the $H_0$ value. In Tables- \ref{tab:model_outputsmodelplus} and \ref{tab:model_outputsmodel}, the results for the PN$^+$ and PN$^+$\& SH0ES datasets are presented, respectively. Table-\ref{tab:model_outputsmodelplus} shows a higher $H_0$ value for the dataset combination CC+PN$^+$+R21 and a lower value for the CC+BAO dataset combination. Consequently, the model yields a lower $H_0$ value without SNIa, SH0ES points and $H_0$ priors. Thus, we can conclude that incorporating the BAO dataset results in a decrease in $H_0$ values. In Table- \ref{tab:model_outputsmodel}, the combination CC+PN$^+$\& SH0ES+R21 yields a higher $H_0$ value. The results from Tables- \ref{tab:model_outputsmodelplus} and \ref{tab:model_outputsmodel} suggest that the CC+PN$^+$\& SH0ES+R21 dataset combination leads to a higher $H_0$ value, whereas the BAO dataset combination results in a lower value. Our result indicates that the highest $H_0$ value for the CC+$PN^{+}$+R21 aligns with the elevated $H_0$ value reported by the SH0ES team (R22), which is stated as $H_{0} = 73.30\pm{1.04} \text{km s}^{-1} \text{ Mpc}^{-1}$ \cite{Riess_2022panplus}. Meanwhile, the $H_0$ value from the BAO dataset inclusion is consistent with the Planck Collaboration, which reports a Hubble constant of $67.4 \pm 0.5  \text{km s}^{-1}  \text{Mpc}^{-1}$ \cite{Aghanim:2018eyx}. In contrast, Aboot et al. \cite{Abbott_2018mnras} present a value of $67.2^{+1.2}_{-1.0}  \text{km s}^{-1}  \text{Mpc}^{-1}$. The $H_0$ findings for PN$^{+}$ and PN$^{+}$\& SH0ES in our study resemble those obtained by Brout et al \cite{Brout_2022pan}. 

Our analysis of linear matter perturbations in \( f(T, \mathcal{T}) \) gravity demonstrates that the growth rate of matter overdensity is influenced by the effective Newton's constant \( G_{\text{eff}} \), modifying the standard evolution of density fluctuations. The results indicate that the modified \( f(T, \mathcal{T}) \) model can help address the \( \sigma_8 \) tension by predicting a lower growth rate \( f\sigma_8(0) \) compared to \( \Lambda \)CDM. Consequently, in Fig.~\ref{sigma8:BAOom}, the results indicated by the red-dot dashed line are approximately 9\% below the \( \Lambda \)CDM model prediction. In contrast, the purple-thick line shows a deviation of roughly 11\% below \( \Lambda \)CDM. These results indicate that the  \( f(T, \mathcal{T}) \) model could provide a better fit to large-scale structure observations, potentially offering an alternative explanation for the discrepancies in cosmic growth measurements as compared to \( \Lambda \)CDM.

Alternatively, we delve deeper into our chosen \( f(T, \mathcal{T}) \) model, which exhibits fascinating characteristics due to its absence of a corresponding \( \Lambda \)CDM limit. In particular, this model has no combination of parameter values that can replicate the precise behavior of the \( \Lambda \)CDM model. From a statistical viewpoint, the AIC and BIC values for the data set combination \text{CC+\( \text{PAN}^{+}\)} and  \text{CC+\( \text{PAN}^{+}\&\text{SH0ES} \)} are shown to be quite close to those of the conventional \( \Lambda \)CDM model, implying that this combination of data supports the \( \Lambda \)CDM model effectively. However, when the BAO data set is incorporated with \text{CC+\( \text{PAN}^{+}\)} and \text{CC+\( \text{PAN}^{+}\&\text{SH0ES} \)}, both the $\Delta$AIC and $\Delta$BIC values rise. It suggests that the \text{CC+\( \text{PAN}^{+}\)+BAO} and \text{CC+\( \text{PAN}^{+}\&\text{SH0ES} \)+BAO} data set combination does not provide compelling evidence in favor of the \( \Lambda \)CDM model.

To explore late-time cosmology, we demonstrate the evolution of background cosmological parameters, including the deceleration parameter, the total equation of state (EoS) parameter, the energy density parameters for both matter and dark energy, and the \( Om(z) \) diagnostic parameter for our chosen \( f(T, \mathcal{T}) \) model in comparison to the standard \( \Lambda \)CDM model. The current values for the deceleration and EoS parameters, energy density, and the transition redshift from deceleration to acceleration are summarized in Table-\ref{results}. The observed behavior of the deceleration parameter suggests that the selected model successfully captures the shift from deceleration in the early Universe to acceleration in later periods. The dynamics of the EoS parameter indicate a quintessence-like characteristic during the late-time phase of the evolution. The energy density parameters reveal a transition from an early Universe dominated by matter to a late-time phase predominantly governed by dark energy. Thus, the model can explain the late-time cosmic phenomena of the Universe.

In the future, this research could be combined with Cosmic Microwave Background (CMB) data, the level of scalar perturbations, and large-scale structure formation and evolution. We expect to gain a more profound insight into the dynamics of inflation and other significant events in the early Universe, which may offer a fresh approach to examining the essential characteristics of cosmic evolution.
\section{Appendix} \label{LCDMapp}
\begin{figure}[H]
     \centering
     \begin{subfigure}[b]{0.49\textwidth}
         \centering
         \includegraphics[width=\linewidth]{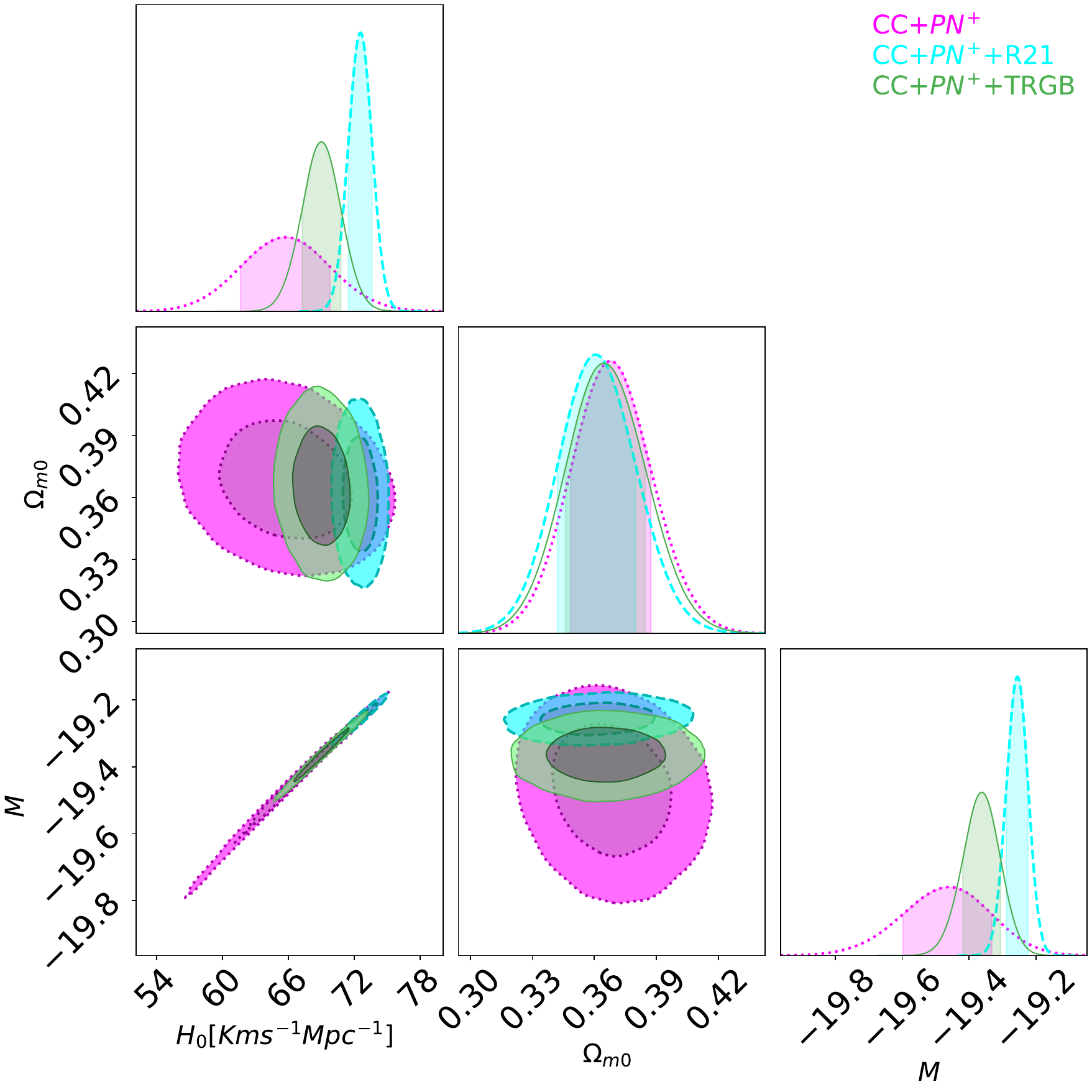}
          \caption{}
         \label{fig:CCMCMCLCDMplus}
     \end{subfigure}
     \hfill
     \begin{subfigure}[b]{0.49\textwidth}
         \centering
         \includegraphics[width=\linewidth]{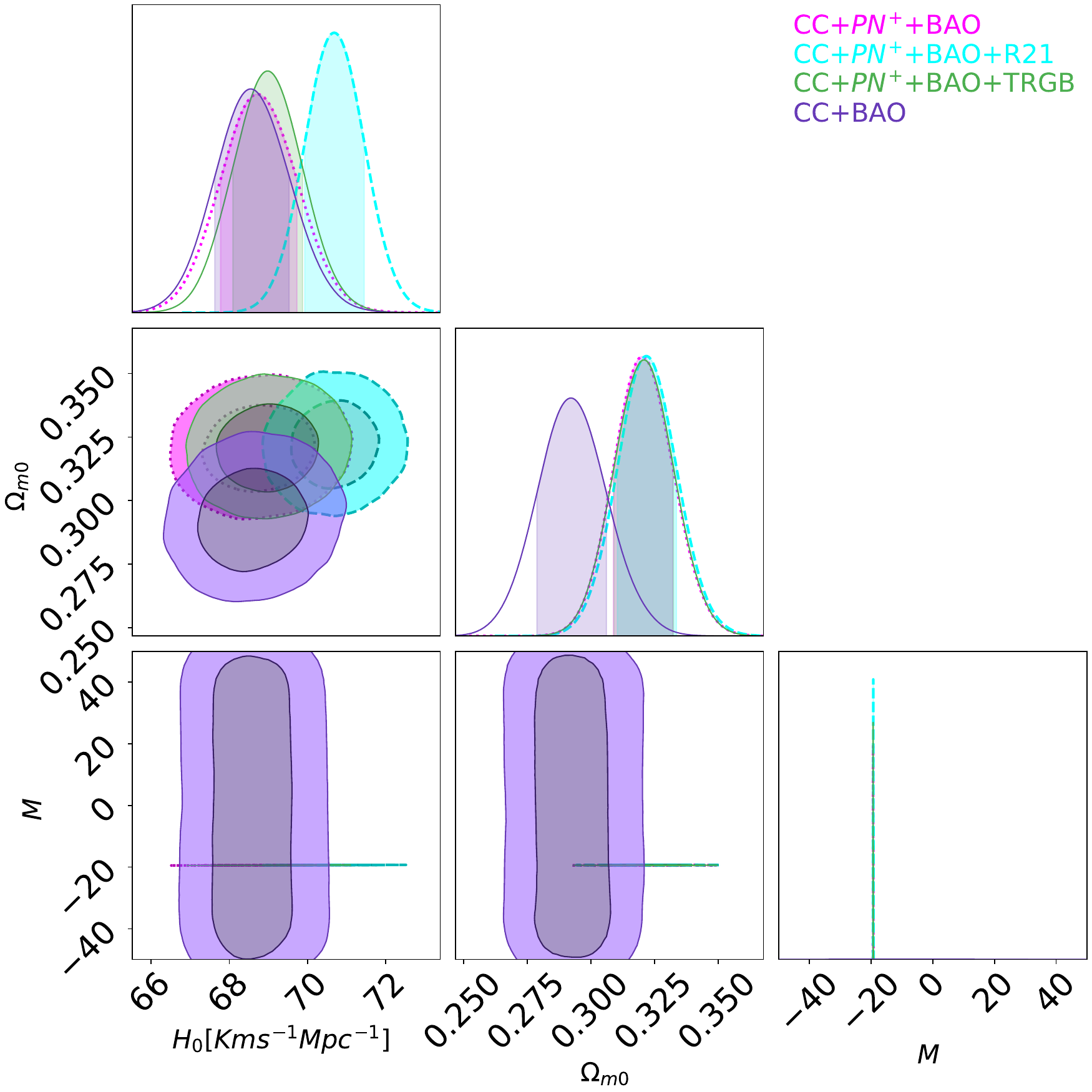} 
         \caption{}
         \label{fig:BAOMCMCLCDMplus}
     \end{subfigure}
\caption{The contour plot of $1\sigma$ and $2\sigma$ uncertainty regions and posterior distribution for the model parameters with the combination of data sets (a) CC, $PN^{+}$ (b) CC, $PN^{+}$ and BAO. The $H_0$ priors are: TRGB (Green) and R21 (Cyan).} 
\label{LCDMMCMCplus}
\end{figure}

\begin{figure}[H]
     \centering
     \begin{subfigure}[b]{0.49\textwidth}
         \centering
         \includegraphics[width=\linewidth]{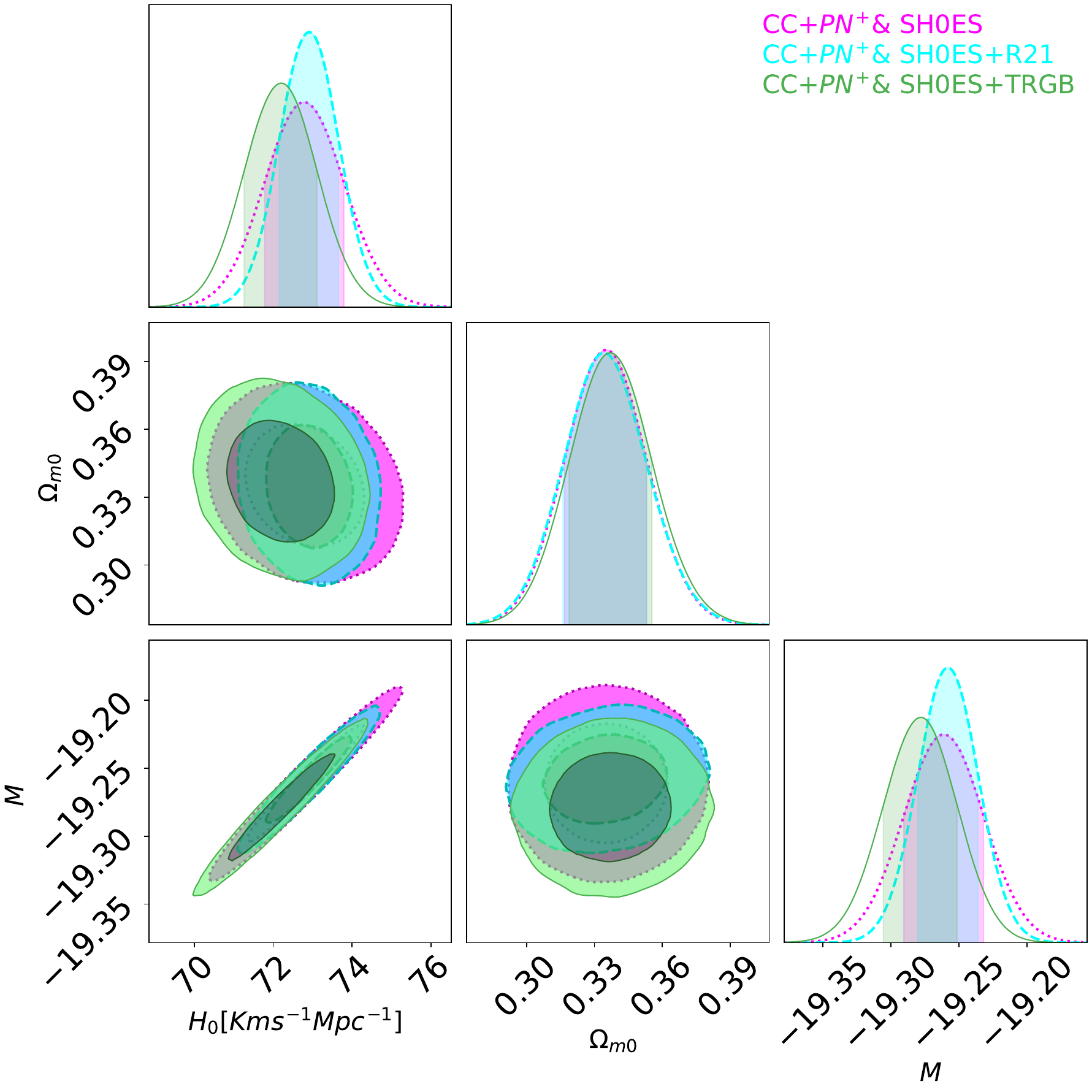}
          \caption{}
         \label{fig:CCMCMCLCDM}
     \end{subfigure}
     \hfill
     \begin{subfigure}[b]{0.49\textwidth}
         \centering
         \includegraphics[width=\linewidth]{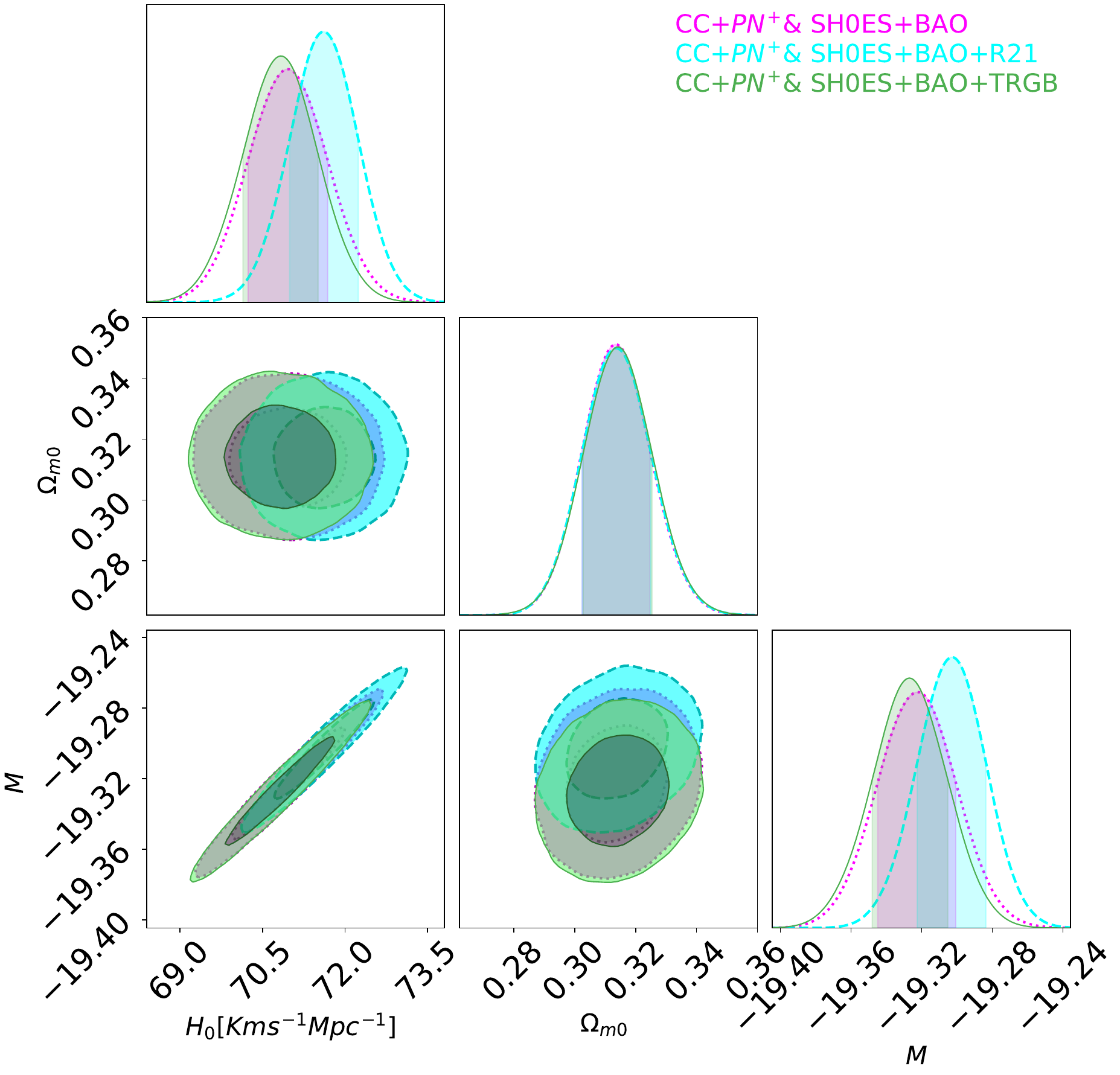} 
         \caption{}
         \label{fig:BAOMCMCLCDM}
     \end{subfigure}
\caption{The contour plot of $1\sigma$ and $2\sigma$ uncertainty regions and posterior distribution for the model parameters with the combination of data sets (a) CC, $PN^{+}\& SH0ES$ (b) CC, $PN^{+}\& SH0ES$ and BAO. The $H_0$ priors are: TRGB (Green) and R21 (Cyan).} 
\label{LCDMMCMC}
\end{figure}

We present the results of the $\Lambda$CDM model highlighting the posterior distributions in Figs. \ref{LCDMMCMCplus} and \ref{LCDMMCMC}. These figures include the $1\sigma$ and $2\sigma$ confidence intervals for various combinations of datasets, providing a comprehensive examination of parameter constraints. Detailed results for each dataset combination are summarized in Tables \ref{tab:model_outputsLCDMplus} and \ref{tab:model_outputsLCDM}. This methodology emphasizes the consistency of our findings with the standard predictions of the $\Lambda$CDM paradigm, facilitating a rigorous assessment on the performance of the model across the selected datasets.

\begin{table}[H]
\renewcommand{\arraystretch}{1.5}
    \centering
    \caption{The results for the $\Lambda$CDM model. The first column identifies the data set combinations and the applied $H_{0}$ priors. The second and third columns present the derived values for $H_{0}$ and $\Omega_{m,0}$ respectively, while the fourth column displays the value of the nuisance parameter. The fifth column provides the minimized $\chi^2_{min}$ values, with the sixth and seventh columns showing the AIC and BIC values respectively. \textbf{(Without SH0ES data points)}}
    \label{tab:model_outputsLCDMplus}
    \begin{tabular}{ccccccc}
        \hline
		$\Lambda$CDM & $H_0$ & $\Omega_{m,0}$ & $M$& $\chi^{2}_{min}$& AIC &BIC \\ 
		\hline
		CC+$PN^{+}$ & $65.7\pm 4.1$ & $0.368^{+0.019}_{-0.020}$ & $-19.46^{+0.13}_{-0.14}$ &1792.30 &  1798.30 &  1802.02  \\ 
		CC+$PN^{+}$ +R21 & $72.6^{+1.0}_{-1.1}$ & $0.361\pm 0.019$ & $-19.256^{+0.032}_{-0.033}$ &1795.64 & 1801.64 &  1805.35\\ 
		CC+$PN^{+}$+TRGB& $69.0^{+1.8}_{-1.7}$ & $0.365^{+0.020}_{-0.019}$ & $-19.362^{+0.055}_{-0.057}$ &1793.20 & 1799.20 &  1802.92 \\ 
		\cline{1-7}
        CC+$PN^{+}$+BAO & $68.73^{+1.00}_{-0.95}$ & $0.320^{+0.012}_{-0.011}$ & $-19.387^{+0.033}_{-0.030}$&1817.75 & 1823.75 &  1827.36 \\ 
		CC+$PN^{+}$+BAO+R21 & $70.68^{+0.76}_{-0.75}$ & $0.322^{+0.012}_{-0.011}$ & $-19.326\pm 0.024$ &1827.16 & 1833.16 &  1836.78\\ 
		 CC+$PN^{+}$+BAO+TRGB & $68.99^{+0.88}_{-0.90}$ & $0.321^{+0.011}_{-0.012}$ & $-19.379\pm 0.029$ &1818.00 & 1824.00 &  1827.61\\
          CC+BAO & $68.56^{+0.97}_{-0.93}$ & $0.292^{+0.014}_{-0.013}$ & - &20.14 & 26.14 &    29.76\\
		\hline
    \end{tabular}
\end{table}
\begin{table}[H]
\renewcommand{\arraystretch}{1.5}
    \centering
    \caption{The results for the $\Lambda$CDM model. The first column identifies the data set combinations and the applied $H_{0}$ priors. The second and third columns present the derived values for $H_{0}$ and $\Omega_{m,0}$ respectively, while the fourth column displays the value of the nuisance parameter. The fifth column provides the minimized $\chi^2_{min}$ values, with the sixth and seventh columns showing the AIC and BIC values respectively. \textbf{(With SH0ES data points)}}
    \label{tab:model_outputsLCDM}
    \begin{tabular}{ccccccc}
        \hline
		$\Lambda$CDM  & $H_0$ & $\Omega_{m,0}$ & $M$& $\chi^{2}_{min}$& AIC &BIC \\ 
		\hline
		CC+$PN^{+}\&$ SH0ES & $72.8\pm 1.0$ & $0.335\pm 0.018$ & $-19.261^{+0.029}_{-0.030}$  &1539.22 &  1545.22 &  1548.93  \\ 
		CC+$PN^{+}\&$ SH0ES+R21 & $72.91\pm 0.75$ & $0.334^{+0.019}_{-0.018}$ & $-19.258\pm 0.022$ &1539.25 & 1545.25 &  1548.96\\ 
		CC+$PN^{+}\&$ SH0ES+TRGB & $72.22^{+0.89}_{-0.95}$ & $0.337^{+0.019}_{-0.018}$ & $-19.278\pm 0.027$ &1541.18 & 1547.18 &  1550.89 \\ 
		\cline{1-7}
        CC+$PN^{+}\&$ SH0ES+BAO & $70.95^{+0.74}_{-0.71}$ & $0.313\pm 0.011$ & $-19.322^{+0.021}_{-0.023}$&1567.17 & 1573.17 &  1576.90 \\ 
		CC+$PN^{+}\&$ SH0ES+BAO+R21 & $71.60^{+0.64}_{-0.61}$ & $0.314\pm 0.011$ & $-19.303^{+0.019}_{-0.020}$ &1569.96 & 1575.96 &  1579.67\\ 
		CC+$PN^{+}\&$ SH0ES+BAO+TRGB & $70.82^{+0.69}_{-0.68}$ & $0.315^{+0.011}_{-0.012}$ & $-19.327^{+0.022}_{-0.021}$ &1567.50 & 1573.50 &  1577.23 \\ 
		\hline
    \end{tabular}
\end{table}


\section*{Acknowledgements}
LKD acknowledges the financial support provided by University Grants Commission (UGC) through Senior Research Fellowship UGC Ref. No.: 191620180688 to carry out the research work. BM acknowledges the support of Anusandhan National Research Foundation(ANRF), Science \& Engineering Research Board(SERB), DST for the grant (File No: CRG/2023/000475). The authors gratefully acknowledge the computing time provided on the high-performance computing facility, Sharanga, at the Birla Institute of Technology and Science - Pilani, Hyderabad Campus. 

\section{References} \label{SEC-VI} 
\bibliographystyle{utphys}
\bibliography{references}

\providecommand{\href}[2]{#2}\begingroup\raggedright\begin{thebibliography}{10}

\bibitem{Riess:1998cb}
{\bf Supernova Search Team} Collaboration, A.~G. Riess {\em et al.},
  ``{Observational evidence from supernovae for an accelerating universe and a
  cosmological constant},'' \href{http://dx.doi.org/10.1086/300499}{{\em
  Astron. J.} {\bf 116} (1998)  1009--1038},
  \href{http://arxiv.org/abs/astro-ph/9805201}{{\tt arXiv:astro-ph/9805201}}.

\bibitem{Perlmutter:1998np}
{\bf Supernova Cosmology Project} Collaboration, S.~Perlmutter {\em et al.},
  ``{Measurements of $\Omega$ and $\Lambda$ from 42 high redshift
  supernovae},'' \href{http://dx.doi.org/10.1086/307221}{{\em Astrophys. J.}
  {\bf 517} (1999)  565--586},
  \href{http://arxiv.org/abs/astro-ph/9812133}{{\tt arXiv:astro-ph/9812133}}.

\bibitem{Bennett_2003a}
C.~L. Bennett {\em et al.}, ``{First-Year Wilkinson Microwave Anisotropy Probe
  (WMAP)* Observations: Preliminary Maps and Basic Results},''
  \href{http://dx.doi.org/10.1086/377253}{{\em The Astrophysical Journal
  Supplement Series} {\bf 148} (2003) no.~1, 1},
  \href{http://arxiv.org/abs/astro-ph/0302207}{{\tt arXiv:astro-ph/0302207}}.

\bibitem{Tegmark_2004a}
M.~Tegmark {\em et al.}, ``Cosmological parameters from {SDSS} and {WMAP},''
  \href{http://dx.doi.org/10.1103/physrevd.69.103501}{{\em Physical Review D}
  {\bf 69} (2004) no.~10, }, \href{http://arxiv.org/abs/astro-ph/0310723}{{\tt
  arXiv:astro-ph/0310723}}.

\bibitem{Alam_2017ab}
S.~Alam {\em et al.}, ``{The clustering of galaxies in the completed SDSS-III
  Baryon Oscillation Spectroscopic Survey: cosmological analysis of the DR12
  galaxy sample},'' \href{http://dx.doi.org/10.1093/mnras/stx721}{{\em Monthly
  Notices of the Royal Astronomical Society} {\bf 470} (2017) no.~3,
  2617--2652}, \href{http://arxiv.org/abs/1607.03155}{{\tt arXiv:1607.03155
  [astro-ph.CO]}}.

\bibitem{Scolnic_2018}
D.~M. Scolnic {\em et al.}, ``{The Complete Light-curve Sample of
  Spectroscopically Confirmed SNe Ia from Pan-STARRS1 and Cosmological
  Constraints from the Combined Pantheon Sample},''
  \href{http://dx.doi.org/10.3847/1538-4357/aab9bb}{{\em The Astrophysical
  Journal} {\bf 859} (2018) no.~2, 101},
  \href{http://arxiv.org/abs/1710.00845}{{\tt arXiv:1710.00845 [astro-ph.CO]}}.

\bibitem{Riess:2019cxk}
A.~G. Riess {\em et al.}, ``{Large Magellanic Cloud Cepheid Standards Provide a
  1\% Foundation for the Determination of the Hubble Constant and Stronger
  Evidence for Physics beyond $\Lambda$CDM},''
  \href{http://dx.doi.org/10.3847/1538-4357/ab1422}{{\em Astrophys. J.} {\bf
  876} (2019) no.~1, 85}, \href{http://arxiv.org/abs/1903.07603}{{\tt
  arXiv:1903.07603 [astro-ph.CO]}}.

\bibitem{Benisty_2021}
D.~Benisty and D.~Staicova, ``Testing late-time cosmic acceleration with
  uncorrelated baryon acoustic oscillation dataset,''
  \href{http://dx.doi.org/10.1051/0004-6361/202039502}{{\em {Astronomy \&
  Astrophysics}} {\bf 647} (2021)  A38},
  \href{http://arxiv.org/abs/2009.10701}{{\tt arXiv:2009.10701 [astro-ph.CO]}}.

\bibitem{Cawthon_2022DES}
{\bf DES Collaboration} Collaboration, ``{Dark Energy Survey Year 3 results:
  calibration of lens sample redshift distributions using clustering redshifts
  with BOSS/eBOSS},'' \href{http://dx.doi.org/10.1093/mnras/stac1160}{{\em
  Monthly Notices of the Royal Astronomical Society} {\bf 513} (2022) no.~4,
  5517--5539}, \href{http://arxiv.org/abs/2012.12826}{{\tt arXiv:2012.12826
  [astro-ph.CO]}}. \url{http://dx.doi.org/10.1093/mnras/stac1160}.

\bibitem{Gupta_2023}
R.~P. Gupta, ``{JWST early Universe observations and $\Lambda$CDM cosmology},''
  \href{http://dx.doi.org/10.1093/mnras/stad2032}{{\em Monthly Notices of the
  Royal Astronomical Society} {\bf 524} (2023) no.~3, 3385--3395},
  \href{http://arxiv.org/abs/2309.13100}{{\tt arXiv:2309.13100 [astro-ph.CO]}}.

\bibitem{plank2013}
{\bf Planck Collaboration} Collaboration, ``{Planck2013 results. XVI.
  Cosmological parameters},''
  \href{http://dx.doi.org/10.1051/0004-6361/201321591}{{\em {Astronomy \&
  Astrophysics}} {\bf 571} (2014)  A16},
  \href{http://arxiv.org/abs/1303.5076}{{\tt arXiv:1303.5076 [astro-ph.CO]}}.

\bibitem{Riess_2022panplus}
A.~G. Riess {\em et al.}, ``{A Comprehensive Measurement of the Local Value of
  the Hubble Constant with 1 km s$^{-1}$ Mpc$^{-1}$ Uncertainty from the Hubble
  Space Telescope and the SH0ES Team},''
  \href{http://dx.doi.org/10.3847/2041-8213/ac5c5b}{{\em The Astrophysical
  Journal Letters} {\bf 934} (2022) no.~1, L7},
  \href{http://arxiv.org/abs/2112.04510}{{\tt arXiv:2112.04510 [astro-ph.CO]}}.

\bibitem{wang_H0LiCOWmnras}
Wang {\em et al.}, ``{H0LiCOW XIII. A 2.4 $\%$ measurement of $H_0$ from lensed
  quasars: 5.3$\sigma$ tension between early and late-Universe probes},''
  \href{http://dx.doi.org/10.1093/mnras/stz3094}{{\em Monthly Notices of the
  Royal Astronomical Society} {\bf 498} (2019) no.~1, 1420--1439},
  \href{http://arxiv.org/abs/1907.04869}{{\tt arXiv:1907.04869 [astro-ph.CO]}}.

\bibitem{Freedman_2019TRGB}
W.~L. Freedman {\em et al.}, ``{The Carnegie-Chicago Hubble Program. VIII. An
  Independent Determination of the Hubble Constant Based on the Tip of the Red
  Giant Branch*},'' \href{http://dx.doi.org/10.3847/1538-4357/ab2f73}{{\em The
  Astrophysical Journal} {\bf 882} (2019) no.~1, 34},
  \href{http://arxiv.org/abs/1907.05922}{{\tt arXiv:1907.05922 [astro-ph.CO]}}.

\bibitem{Aghanim:2018eyx}
{\bf Planck} Collaboration, N.~Aghanim {\em et al.}, ``{Planck 2018 results.
  VI. Cosmological parameters},''
  \href{http://dx.doi.org/10.1051/0004-6361/201833910}{{\em Astron. Astrophys.}
  {\bf 641} (2020)  A6}, \href{http://arxiv.org/abs/1807.06209}{{\tt
  arXiv:1807.06209 [astro-ph.CO]}}. [Erratum: Astron.Astrophys. 652, C4
  (2021)].

\bibitem{Abbott_2018mnras}
{\bf DES Collaboration} Collaboration, ``{Dark Energy Survey Year 1 Results: A
  Precise H0 Estimate from DES Y1, BAO, and D/H Data},''
  \href{http://dx.doi.org/10.1093/mnras/sty1939}{{\em Monthly Notices of the
  Royal Astronomical Society} {\bf 480} (2018) no.~3, 3879--3888},
  \href{http://arxiv.org/abs/1711.00403}{{\tt arXiv:1711.00403 [astro-ph.CO]}}.

\bibitem{Di_Valentino_2017}
E.~Di~Valentino, A.~Melchiorri, and O.~Mena, ``{Can interacting dark energy
  solve the $H0$ tension?},''
  \href{http://dx.doi.org/10.1103/physrevd.96.043503}{{\em Physical Review D}
  {\bf 96} (2017) no.~4, }, \href{http://arxiv.org/abs/1704.08342}{{\tt
  arXiv:1704.08342 [astro-ph.CO]}}.

\bibitem{Di_Valentino_2021}
E.~Di~Valentino {\em et al.}, ``{In the realm of the Hubble tension - a review
  of solutions},'' \href{http://dx.doi.org/10.1088/1361-6382/ac086d}{{\em
  Classical and Quantum Gravity} {\bf 38} (2021) no.~15, 153001},
  \href{http://arxiv.org/abs/2103.01183}{{\tt arXiv:2103.01183 [astro-ph.CO]}}.

\bibitem{Ata_2017BAO}
M.~Ata {\em et al.}, ``{The clustering of the SDSS-IV extended Baryon
  Oscillation Spectroscopic Survey DR14 quasar sample: first measurement of
  baryon acoustic oscillations between redshift 0.8 and 2.2},''
  \href{http://dx.doi.org/10.1093/mnras/stx2630}{{\em Mon. Not. Roy. Astron.
  Soc.} {\bf 473} (2017) no.~4, 4773--4794},
  \href{http://arxiv.org/abs/1705.06373}{{\tt arXiv:1705.06373 [astro-ph.CO]}}.

\bibitem{Asgari_2021}
M.~Asgari {\em et al.}, ``{KiDS-1000 cosmology: Cosmic shear constraints and
  comparison between two point statistics},''
  \href{http://dx.doi.org/10.1051/0004-6361/202039070}{{\em {Astronomy \&
  Astrophysics}} {\bf 645} (2021)  A104},
  \href{http://arxiv.org/abs/2007.15633}{{\tt arXiv:2007.15633 [astro-ph.CO]}}.

\bibitem{Ruiz_Zapatero_2021}
J.~Ruiz-Zapatero {\em et al.}, ``{Geometry versus growth: Internal consistency
  of the flat $\Lambda$CDM model with KiDS-1000},''
  \href{http://dx.doi.org/10.1051/0004-6361/202141350}{{\em {Astronomy \&
  Astrophysics}} {\bf 655} (2021)  A11},
  \href{http://arxiv.org/abs/2105.09545}{{\tt arXiv:2105.09545 [astro-ph.CO]}}.

\bibitem{Ghirardini_2024}
V.~Ghirardini {\em et al.}, ``{The SRG/eROSITA all-sky survey: Cosmology
  constraints from cluster abundances in the western Galactic hemisphere},''
  \href{http://dx.doi.org/10.1051/0004-6361/202348852}{{\em {Astronomy \&
  Astrophysics}} {\bf 689} (2024)  A298},
  \href{http://arxiv.org/abs/2402.08458}{{\tt arXiv:2402.08458 [astro-ph.CO]}}.

\bibitem{Nojiri2006}
S.~Nojiri and S.~D. Odintsov, ``{Modified $f(R)$ gravity consistent with
  realistic cosmology: From a matter dominated epoch to a dark energy
  universe},'' \href{http://dx.doi.org/10.1103/PhysRevD.74.086005}{{\em Phys.
  Rev. D} {\bf 74} (2006)  086005},
  \href{http://arxiv.org/abs/hep-th/0608008}{{\tt arXiv:hep-th/0608008}}.

\bibitem{NOJIRI2007238a}
S.~Nojiri and S.~D. Odintsov, ``{Unifying inflation with {$\Lambda$CDM} epoch
  in modified f(R) gravity consistent with Solar System Test},''
  \href{http://dx.doi.org/10.1016/j.physletb.2007.10.027}{{\em Physics Letters
  B} {\bf 657} (2007) no.~4, 238--245},
  \href{http://arxiv.org/abs/0707.1941}{{\tt arXiv:0707.1941 [hep-th]}}.

\bibitem{Sotiriou:2008rp}
T.~P. Sotiriou and V.~Faraoni, ``{$f(R)$ Theories Of Gravity},''
  \href{http://dx.doi.org/10.1103/RevModPhys.82.451}{{\em Rev. Mod. Phys.} {\bf
  82} (2010)  451--497}, \href{http://arxiv.org/abs/0805.1726}{{\tt
  arXiv:0805.1726 [gr-qc]}}.

\bibitem{Capozziello:2011et}
S.~Capozziello and M.~De~Laurentis, ``{Extended Theories of Gravity},''
  \href{http://dx.doi.org/10.1016/j.physrep.2011.09.003}{{\em Phys. Rept.} {\bf
  509} (2011)  167--321}, \href{http://arxiv.org/abs/1108.6266}{{\tt
  arXiv:1108.6266 [gr-qc]}}.

\bibitem{Maluf:1994j}
J.~W. Maluf, ``{Hamiltonian formulation of the teleparallel description of
  general relativity},'' \href{http://dx.doi.org/10.1063/1.530774}{{\em J.
  Math. Phys.} {\bf 35} (1994)  335--343},
  \href{http://arxiv.org/abs/gr-qc/0002059}{{\tt arXiv:gr-qc/0002059}}.

\bibitem{Aldrovandi:2013wha}
R.~Aldrovandi and J.~G. Pereira,
  \href{http://dx.doi.org/10.1007/978-94-007-5143-9}{{\em {Teleparallel
  Gravity}: {An Introduction}}}.
\newblock Springer, 2013.

\bibitem{Weitzenbock1923}
R.~Weitzenb\"{o}ock, {\em `Invariantentheorie'}.
\newblock Noordhoff, Gronningen, 1923.

\bibitem{Bengochea:2008gz}
G.~R. Bengochea and R.~Ferraro, ``{Dark torsion as the cosmic speed-up},''
  \href{http://dx.doi.org/10.1103/PhysRevD.79.124019}{{\em Phys. Rev. D} {\bf
  79} (2009)  124019}, \href{http://arxiv.org/abs/0812.1205}{{\tt
  arXiv:0812.1205 [astro-ph]}}.

\bibitem{Ferraro:2008ey}
R.~Ferraro and F.~Fiorini, ``{On Born-Infeld Gravity in Weitzenbock
  spacetime},'' \href{http://dx.doi.org/10.1103/PhysRevD.78.124019}{{\em Phys.
  Rev. D} {\bf 78} (2008)  124019}, \href{http://arxiv.org/abs/0812.1981}{{\tt
  arXiv:0812.1981 [gr-qc]}}.

\bibitem{Linder:2010py}
E.~V. Linder, ``{Einstein's Other Gravity and the Acceleration of the
  Universe},'' \href{http://dx.doi.org/10.1103/PhysRevD.81.127301}{{\em Phys.
  Rev. D} {\bf 81} (2010)  127301}, \href{http://arxiv.org/abs/1005.3039}{{\tt
  arXiv:1005.3039 [astro-ph.CO]}}. [Erratum: Phys.Rev.D 82, 109902 (2010)].

\bibitem{Chen:2010va}
S.-H. Chen {\em et al.}, ``{Cosmological perturbations in $f(T)$ gravity},''
  \href{http://dx.doi.org/10.1103/PhysRevD.83.023508}{{\em Phys. Rev. D} {\bf
  83} (2011)  023508}, \href{http://arxiv.org/abs/1008.1250}{{\tt
  arXiv:1008.1250 [astro-ph.CO]}}.

\bibitem{Cai_2016}
Y.-F. Cai {\em et al.}, ``${f(T)}$ teleparallel gravity and cosmology,''
  \href{http://dx.doi.org/10.1088/0034-4885/79/10/106901}{{\em Reports on
  Progress in Physics} {\bf 79} (2016) no.~10, 106901}.

\bibitem{Duchaniya:2022rqu}
L.~K. Duchaniya {\em et al.}, ``{Dynamical stability analysis of accelerating
  f(T) gravity models},''
  \href{http://dx.doi.org/10.1140/epjc/s10052-022-10406-w}{{\em Eur. Phys. J.
  C} {\bf 82} (2022) no.~5, 448}, \href{http://arxiv.org/abs/2202.08150}{{\tt
  arXiv:2202.08150 [gr-qc]}}.

\bibitem{Briffa_2024}
R.~Briffa {\em et al.}, ``{Growth of structures using redshift space distortion
  in $f(T)$ cosmology},'' \href{http://dx.doi.org/10.1093/mnras/stae103}{{\em
  Monthly Notices of the Royal Astronomical Society} {\bf 528} (2024) no.~2,
  2711--2727}, \href{http://arxiv.org/abs/2310.09159}{{\tt arXiv:2310.09159
  [gr-qc]}}.

\bibitem{Duchaniya_2024ab}
L.~K. Duchaniya, K.~Gandhi, and B.~Mishra, ``{Attractor behavior of $f(T)$
  modified gravity and the cosmic acceleration},''
  \href{http://dx.doi.org/10.1016/j.dark.2024.101461}{{\em Physics of the Dark
  Universe} {\bf 44} (2024)  101461},
  \href{http://arxiv.org/abs/2303.09076}{{\tt arXiv:2303.09076 [gr-qc]}}.

\bibitem{Kofinas:2014owa}
G.~Kofinas and E.~N. Saridakis, ``{Teleparallel equivalent of Gauss-Bonnet
  gravity and its modifications},''
  \href{http://dx.doi.org/10.1103/PhysRevD.90.084044}{{\em Phys. Rev. D} {\bf
  90} (2014)  084044}, \href{http://arxiv.org/abs/1404.2249}{{\tt
  arXiv:1404.2249 [gr-qc]}}.

\bibitem{Kofinas:2014daa}
G.~Kofinas and E.~N. Saridakis, ``{Cosmological applications of $F(T,T_G)$
  gravity},'' \href{http://dx.doi.org/10.1103/PhysRevD.90.084045}{{\em Phys.
  Rev. D} {\bf 90} (2014)  084045}, \href{http://arxiv.org/abs/1408.0107}{{\tt
  arXiv:1408.0107 [gr-qc]}}.

\bibitem{Escamilla-Rivera:2019ulu}
C.~Escamilla-Rivera and J.~Levi~Said, ``{Cosmological viable models in $f(T,B)$
  theory as solutions to the $H_0$ tension},''
  \href{http://dx.doi.org/10.1088/1361-6382/ab939c}{{\em Class. Quant. Grav.}
  {\bf 37} (2020) no.~16, 165002}, \href{http://arxiv.org/abs/1909.10328}{{\tt
  arXiv:1909.10328 [gr-qc]}}.

\bibitem{Kadam_2023ab}
S.~A. Kadam, N.~P. Thakkar, and B.~Mishra, ``Dynamical system analysis in
  teleparallel gravity with boundary term,''
  \href{http://dx.doi.org/10.1140/epjc/s10052-023-11937-6}{{\em The European
  Physical Journal C} {\bf 83} (2023) no.~9, },
  \href{http://arxiv.org/abs/2306.06677}{{\tt arXiv:2306.06677 [gr-qc]}}.

\bibitem{Gonzalez-Espinoza:2020jss}
M.~Gonzalez-Espinoza and G.~Otalora, ``{Cosmological dynamics of dark energy in
  scalar-torsion $f(T,\phi )$ gravity},''
  \href{http://dx.doi.org/10.1140/epjc/s10052-021-09270-x}{{\em Eur. Phys. J.
  C} {\bf 81} (2021) no.~5, 480}, \href{http://arxiv.org/abs/2011.08377}{{\tt
  arXiv:2011.08377 [gr-qc]}}.

\bibitem{Gonzalez-Espinoza:2021mwr}
M.~Gonzalez-Espinoza, G.~Otalora, and J.~Saavedra, ``{Stability of scalar
  perturbations in scalar-torsion f(T,\ensuremath{\phi}) gravity theories in
  the presence of a matter fluid},''
  \href{http://dx.doi.org/10.1088/1475-7516/2021/10/007}{{\em JCAP} {\bf 10}
  (2021)  007}, \href{http://arxiv.org/abs/2101.09123}{{\tt arXiv:2101.09123
  [gr-qc]}}.

\bibitem{Duchaniya_2023noet}
L.~K. Duchaniya, B.~Mishra, and J.~L. Said, ``{Noether symmetry approach in
  scalar-torsion $f(T,\phi )$ gravity},''
  \href{http://dx.doi.org/10.1140/epjc/s10052-023-11792-5}{{\em The European
  Physical Journal C} {\bf 83} (2023) no.~7, },
  \href{http://arxiv.org/abs/2210.11944}{{\tt arXiv:2210.11944 [gr-qc]}}.

\bibitem{Duchaniya_2023tphi}
L.~K. Duchaniya {\em et al.}, ``{Dynamical systems analysis in $f(T, \phi)$
  gravity},'' \href{http://dx.doi.org/10.1140/epjc/s10052-022-11155-6}{{\em The
  European Physical Journal C} {\bf 83} (2023) no.~1, },
  \href{http://arxiv.org/abs/2209.03414}{{\tt arXiv:2209.03414 [gr-qc]}}.

\bibitem{Harko_2014a}
T.~Harko {\em et al.}, ``{$f(T,\mathcal{T})$ gravity and cosmology},''
  \href{http://dx.doi.org/10.1088/1475-7516/2014/12/021}{{\em Journal of
  Cosmology and Astroparticle Physics} {\bf 2014} (2014) no.~12, 021},
  \href{http://arxiv.org/abs/1405.0519}{{\tt arXiv:1405.0519 [gr-qc]}}.

\bibitem{Momeni_2014}
D.~Momeni and R.~Myrzakulov, ``{Cosmological reconstruction of $f(T,
  \mathcal{T})$ gravity},''
  \href{http://dx.doi.org/10.1142/s0219887814500777}{{\em International Journal
  of Geometric Methods in Modern Physics} {\bf 11} (2014) no.~08, 1450077},
  \href{http://arxiv.org/abs/1405.5863}{{\tt arXiv:1405.5863 [gr-qc]}}.

\bibitem{Jackson2016a}
G.~Farrugia and J.~L. Said, ``{Growth factor in $f(T,\mathcal{T})$ gravity},''
  \href{http://dx.doi.org/10.1103/PhysRevD.94.124004}{{\em Phys. Rev. D} {\bf
  94} (2016)  124004}, \href{http://arxiv.org/abs/1612.00974}{{\tt
  arXiv:1612.00974 [gr-qc]}}.

\bibitem{Junior_2016}
E.~L.~B. Junior {\em et al.}, ``{Reconstruction, thermodynamics and stability
  of the {$\Lambda$CDM} model in $f(T,{ \mathcal T })$ gravity},''
  \href{http://dx.doi.org/10.1088/0264-9381/33/12/125006}{{\em Classical and
  Quantum Gravity} {\bf 33} (2016) no.~12, 125006},
  \href{http://arxiv.org/abs/1501.00621}{{\tt arXiv:1501.00621 [gr-qc]}}.

\bibitem{Pace_2017}
M.~Pace and J.~L. Said, ``{Quark Stars in $f(T, \mathcal{T})$-Gravity},''
  \href{http://dx.doi.org/10.1140/epjc/s10052-017-4637-8}{{\em The European
  Physical Journal C} {\bf 77} (2017) no.~2, },
  \href{http://arxiv.org/abs/1701.04761}{{\tt arXiv:1701.04761 [gr-qc]}}.

\bibitem{Duchaniya_2024tt}
L.~K. Duchaniya, S.~V. Lohakare, and B.~Mishra, ``{Cosmological models in $f(T,
  \mathcal{T})$ gravity and the dynamical system analysis},''
  \href{http://dx.doi.org/10.1016/j.dark.2023.101402}{{\em Physics of the Dark
  Universe} {\bf 43} (2024)  101402},
  \href{http://arxiv.org/abs/2302.07132}{{\tt arXiv:2302.07132 [gr-qc]}}.

\bibitem{Vagnozzi_2020newph}
S.~Vagnozzi, ``{New physics in light of the $H_0$ tension: An alternative
  view},'' \href{http://dx.doi.org/10.1103/physrevd.102.023518}{{\em Physical
  Review D} {\bf 102} (2020) no.~2, 023518},
  \href{http://arxiv.org/abs/1907.07569}{{\tt arXiv:1907.07569 [astro-ph.CO]}}.

\bibitem{Freedman_2021apjh0tension}
W.~L. Freedman, ``Measurements of the hubble constant: Tensions in
  perspective,'' \href{http://dx.doi.org/10.3847/1538-4357/ac0e95}{{\em The
  Astrophysical Journal} {\bf 919} (2021) no.~1, 16},
  \href{http://arxiv.org/abs/2106.15656}{{\tt arXiv:2106.15656 [astro-ph]}}.
  \url{http://dx.doi.org/10.3847/1538-4357/ac0e95}.

\bibitem{Abdalla:2022yfr}
E.~Abdalla {\em et al.}, ``{Cosmology intertwined: A review of the particle
  physics, astrophysics, and cosmology associated with the cosmological
  tensions and anomalies},''
  \href{http://dx.doi.org/10.1016/j.jheap.2022.04.002}{{\em Journal of High
  Energy Astrophysics} {\bf 34} (2022)  49--211},
  \href{http://arxiv.org/abs/2203.06142}{{\tt arXiv:2203.06142 [astro-ph.CO]}}.

\bibitem{Moresco_2022_H0}
M.~Moresco {\em et al.}, ``Unveiling the universe with emerging cosmological
  probes,'' \href{http://dx.doi.org/10.1007/s41114-022-00040-z}{{\em Living
  Reviews in Relativity} {\bf 25} (2022) no.~1, },
  \href{http://arxiv.org/abs/2201.07241}{{\tt arXiv:2201.07241 [astro-ph.CO]}}.

\bibitem{Brout_2022pan}
D.~Brout {\em et al.}, ``{The Pantheon+ Analysis: Cosmological Constraints},''
  \href{http://dx.doi.org/10.3847/1538-4357/ac8e04}{{\em The Astrophysical
  Journal} {\bf 938} (2022) no.~2, 110},
  \href{http://arxiv.org/abs/2202.04077}{{\tt arXiv:2202.04077 [astro-ph.CO]}}.

\bibitem{Briffa_2022}
R.~Briffa {\em et al.}, ``{Impact of $H_0$ priors on $f(T)$ late time
  cosmology},'' \href{http://dx.doi.org/10.1140/epjp/s13360-022-02725-4}{{\em
  The European Physical Journal Plus} {\bf 137} (2022) no.~5, },
  \href{http://arxiv.org/abs/2108.03853}{{\tt arXiv:2108.03853 [astro-ph.CO]}}.

\bibitem{Vagnozzi_2023}
S.~Vagnozzi, ``{Seven Hints That Early-Time New Physics Alone Is Not Sufficient
  to Solve the Hubble Tension},''
  \href{http://dx.doi.org/10.3390/universe9090393}{{\em Universe} {\bf 9}
  (2023) no.~9, 393}, \href{http://arxiv.org/abs/2308.16628}{{\tt
  arXiv:2308.16628 [astro-ph.CO]}}.

\bibitem{Briffa_2023mnras}
R.~Briffa {\em et al.}, ``{Constraints on $f(T)$ cosmology with Pantheon+},''
  \href{http://dx.doi.org/10.1093/mnras/stad1384}{{\em Monthly Notices of the
  Royal Astronomical Society} {\bf 522} (2023) no.~4, 6024--6034},
  \href{http://arxiv.org/abs/2303.13840}{{\tt arXiv:2303.13840 [gr-qc]}}.

\bibitem{Capozziello_2024_ten}
S.~Capozziello, G.~Sarracino, and G.~De~Somma, ``{A Critical Discussion on the
  $H0$ Tension},'' \href{http://dx.doi.org/10.3390/universe10030140}{{\em
  Universe} {\bf 10} (2024) no.~3, 140},
  \href{http://arxiv.org/abs/2403.12796}{{\tt arXiv:2403.12796 [gr-qc]}}.

\bibitem{divalentino2025cosmoversewhitepaperaddressing}
E.~D. Valentino {\em et al.}, ``{The CosmoVerse White Paper: Addressing
  observational tensions in cosmology with systematics and fundamental
  physics},'' \href{http://arxiv.org/abs/2504.01669}{{\tt arXiv:2504.01669
  [astro-ph.CO]}}.

\bibitem{misner1973gravitation}
C.~Misner, K.~Thorne, and J.~Wheeler, {\em Gravitation}.
\newblock No.~pt. 3 in Gravitation. W. H. Freeman, 1973.
\newblock \url{https://books.google.com.mt/books?id=w4Gigq3tY1kC}.

\bibitem{Foreman_Mackey_2013}
D.~Foreman-Mackey {\em et al.}, ``emcee: The mcmc hammer,''
  \href{http://dx.doi.org/10.1086/670067}{{\em Publications of the Astronomical
  Society of the Pacific} {\bf 125} (2013) no.~925, 306--312},
  \href{http://arxiv.org/abs/1202.3665}{{\tt arXiv:1202.3665 [astro-ph.CO]}}.

\bibitem{Jimenez_2002}
R.~Jimenez and A.~Loeb, ``Constraining cosmological parameters based on
  relative galaxy ages,'' \href{http://dx.doi.org/10.1086/340549}{{\em The
  Astrophysical Journal} {\bf 573} (2002) no.~1, 37--42},
  \href{http://arxiv.org/abs/astro-ph/0106145}{{\tt arXiv:astro-ph/0106145
  [astro-ph.CO]}}.

\bibitem{Zhang_2014hz}
Z.~Cong {\em et al.}, ``{Four new observational H(z) data from luminous red
  galaxies in the Sloan Digital Sky Survey data release seven},''
  \href{http://dx.doi.org/10.1088/1674-4527/14/10/002}{{\em Research in
  Astronomy and Astrophysics} {\bf 14} (2014) no.~10, 1221},
  \href{http://arxiv.org/abs/1207.4541}{{\tt arXiv:1207.4541 [astro-ph.CO]}}.
  \url{https://dx.doi.org/10.1088/1674-4527/14/10/002}.

\bibitem{Jimenez_2003cmb}
R.~Jimenez {\em et al.}, ``{Constraints on the Equation of State of Dark Energy
  and the Hubble Constant from Stellar Ages and the Cosmic Microwave
  Background},'' \href{http://dx.doi.org/10.1086/376595}{{\em The Astrophysical
  Journal} {\bf 593} (2003) no.~2, 622--629},
  \href{http://arxiv.org/abs/astro-ph/0302560}{{\tt arXiv:astro-ph/0302560
  [astro-ph.CO]}}.

\bibitem{Moresco_2016hubb}
M.~Moresco {\em et al.}, ``{A 6$\%$ measurement of the Hubble parameter at
  $z\sim 0.45$: direct evidence of the epoch of cosmic re-acceleration},''
  \href{http://dx.doi.org/10.1088/1475-7516/2016/05/014}{{\em Journal of
  Cosmology and Astroparticle Physics} {\bf 2016} (2016) no.~05, 014--014},
  \href{http://arxiv.org/abs/1601.01701}{{\tt arXiv:1601.01701 [astro-ph.CO]}}.

\bibitem{Simon_2005prd}
J.~Simon, L.~Verde, and R.~Jimenez, ``Constraints on the redshift dependence of
  the dark energy potential,''
  \href{http://dx.doi.org/10.1103/physrevd.71.123001}{{\em Physical Review D}
  {\bf 71} (2005) no.~12, }, \href{http://arxiv.org/abs/astro-ph/0412269}{{\tt
  arXiv:astro-ph/0412269 [astro-ph.CO]}}.

\bibitem{M_Moresco_2012JCAP}
M.~Moresco {\em et al.}, ``{Improved constraints on the expansion rate of the
  Universe up to z$ \sim$1.1 from the spectroscopic evolution of cosmic
  chronometers},'' \href{http://dx.doi.org/10.1088/1475-7516/2012/08/006}{{\em
  Journal of Cosmology and Astroparticle Physics} {\bf 2012} (2012) no.~08,
  006--006}, \href{http://arxiv.org/abs/1201.3609}{{\tt arXiv:1201.3609
  [astro-ph.CO]}}.

\bibitem{Daniel_Stern_2010jcap}
D.~Stern {\em et al.}, ``{Cosmic chronometers: constraining the equation of
  state of dark energy. I: H(z) measurements},''
  \href{http://dx.doi.org/10.1088/1475-7516/2010/02/008}{{\em Journal of
  Cosmology and Astroparticle Physics} {\bf 2010} (2010) no.~02, 008--008},
  \href{http://arxiv.org/abs/0907.3149}{{\tt arXiv:0907.3149 [astro-ph.CO]}}.

\bibitem{Moresco_2015mnras}
M.~Moresco, ``{Raising the bar: new constraints on the Hubble parameter with
  cosmic chronometers at z$\sim$2},''
  \href{http://dx.doi.org/10.1093/mnrasl/slv037}{{\em Monthly Notices of the
  Royal Astronomical Society} {\bf 450} (2015) no.~1, 16--20},
  \href{http://arxiv.org/abs/1503.01116}{{\tt arXiv:1503.01116 [astro-ph.CO]}}.

\bibitem{Moresco_2020covariance}
M.~Moresco {\em et al.}, ``Setting the stage for cosmic chronometers. ii.
  impact of stellar population synthesis models systematics and full covariance
  matrix,'' \href{http://dx.doi.org/10.3847/1538-4357/ab9eb0}{{\em The
  Astrophysical Journal} {\bf 898} (2020) no.~1, 82},
  \href{http://arxiv.org/abs/2003.07362}{{\tt arXiv:2003.07362 [astro-ph.CO]}}.

\bibitem{Brout_2022panplus}
D.~Brout {\em et al.}, ``{The Pantheon+ Analysis: SuperCal-fragilistic Cross
  Calibration, Retrained SALT2 Light-curve Model, and Calibration Systematic
  Uncertainty},'' \href{http://dx.doi.org/10.3847/1538-4357/ac8bcc}{{\em The
  Astrophysical Journal} {\bf 938} (2022) no.~2, 111},
  \href{http://arxiv.org/abs/2112.03864}{{\tt arXiv:2112.03864 [astro-ph.CO]}}.

\bibitem{Scolnic_2022panplus}
D.~Scolnic {\em et al.}, ``{The Pantheon+ Analysis: The Full Data Set and
  Light-curve Release},''
  \href{http://dx.doi.org/10.3847/1538-4357/ac8b7a}{{\em The Astrophysical
  Journal} {\bf 938} (2022) no.~2, 113},
  \href{http://arxiv.org/abs/2112.03863}{{\tt arXiv:2112.03863 [astro-ph.CO]}}.

\bibitem{Conley_2010Apjs}
A.~Conley {\em et al.}, ``Supernova constraints and systematic uncertainties
  from the first three years of the supernova legacy survey,''
  \href{http://dx.doi.org/10.1088/0067-0049/192/1/1}{{\em The Astrophysical
  Journal Supplement Series} {\bf 192} (2010) no.~1, 1},
  \href{http://arxiv.org/abs/1104.1443}{{\tt arXiv:1104.1443 [astro-ph.CO]}}.

\bibitem{Beutler_2011baosixdegree}
F.~Beutler {\em et al.}, ``{The 6dF Galaxy Survey: baryon acoustic oscillations
  and the local Hubble constant: 6dFGS: BAOs and the local Hubble constant},''
  \href{http://dx.doi.org/10.1111/j.1365-2966.2011.19250.x}{{\em Monthly
  Notices of the Royal Astronomical Society} {\bf 416} (2011) no.~4,
  3017--3032}, \href{http://arxiv.org/abs/1106.3366}{{\tt arXiv:1106.3366
  [astro-ph.CO]}}.

\bibitem{du_Mas_des_Bourboux_2017}
H.~du~Mas~des Bourboux {\em et al.}, ``{Baryon acoustic oscillations from the
  complete SDSS-III Ly$\alpha$-quasar cross-correlation function at z = 2.4},''
  \href{http://dx.doi.org/10.1051/0004-6361/201731731}{{\em {Astronomy $\&$;
  Astrophysics}} {\bf 608} (2017)  A130},
  \href{http://arxiv.org/abs/1708.02225}{{\tt arXiv:1708.02225 [astro-ph.CO]}}.

\bibitem{Ross_2015}
A.~J. Ross {\em et al.}, ``{The clustering of the SDSS DR7 main Galaxy sample
  I. A 4 per cent distance measure at z=0.15},''
  \href{http://dx.doi.org/10.1093/mnras/stv154}{{\em Monthly Notices of the
  Royal Astronomical Society} {\bf 449} (2015) no.~1, 835--847},
  \href{http://arxiv.org/abs/1409.3242}{{\tt arXiv:1409.3242 [astro-ph.CO]}}.

\bibitem{Zhao_2018sdss_IV}
G.-B. Zhao {\em et al.}, ``The clustering of the sdss-iv extended baryon
  oscillation spectroscopic survey dr14 quasar sample: a tomographic
  measurement of cosmic structure growth and expansion rate based on optimal
  redshift weights,'' \href{http://dx.doi.org/10.1093/mnras/sty2845}{{\em
  Monthly Notices of the Royal Astronomical Society} {\bf 482} (2018) no.~3,
  3497--3513}, \href{http://arxiv.org/abs/1801.03043}{{\tt arXiv:1801.03043
  [astro-ph.CO]}}.

\bibitem{Fixsen_2009temcmb}
D.~J. Fixsen, ``The temperature of the cosmic microwave background,''
  \href{http://dx.doi.org/10.1088/0004-637x/707/2/916}{{\em The Astrophysical
  Journal} {\bf 707} (2009) no.~2, 916--920},
  \href{http://arxiv.org/abs/0911.1955}{{\tt arXiv:0911.1955 [astro-ph.CO]}}.

\bibitem{Lue_2004}
A.~Lue, R.~Scoccimarro, and G.~D. Starkman, ``{Probing Newton's constant on
  vast scales: Dvali-Gabadadze-Porrati gravity, cosmic acceleration, and large
  scale structure},'' \href{http://dx.doi.org/10.1103/physrevd.69.124015}{{\em
  Physical Review D} {\bf 69} (2004) no.~12, },
  \href{http://arxiv.org/abs/astro-ph/0401515}{{\tt arXiv:astro-ph/0401515
  [astro-ph.CO]}}.

\bibitem{Linder_2005growth}
E.~V. Linder, ``Cosmic growth history and expansion history,''
  \href{http://dx.doi.org/10.1103/physrevd.72.043529}{{\em Physical Review D}
  {\bf 72} (2005) no.~4, }, \href{http://arxiv.org/abs/astro-ph/0507263}{{\tt
  arXiv:astro-ph/0507263 [astro-ph.CO]}}.

\bibitem{Uzan_2007}
J.-P. Uzan, ``The acceleration of the universe and the physics behind it,''
  \href{http://dx.doi.org/10.1007/s10714-006-0385-z}{{\em General Relativity
  and Gravitation} {\bf 39} (2007) no.~3, 307--342},
  \href{http://arxiv.org/abs/astro-ph/0605313}{{\tt arXiv:astro-ph/0605313
  [astro-ph.CO]}}.

\bibitem{Gannouji_2009}
R.~Gannouji, B.~Moraes, and D.~Polarski, ``{The growth of matter perturbations
  in $f(R)$ models},''
  \href{http://dx.doi.org/10.1088/1475-7516/2009/02/034}{{\em Journal of
  Cosmology and Astroparticle Physics} {\bf 2009} (2009) no.~02, 034--034},
  \href{http://arxiv.org/abs/0809.3374}{{\tt arXiv:0809.3374 [astro-ph.CO]}}.

\bibitem{Tsujikawa_2008}
S.~Tsujikawa, K.~Uddin, and R.~Tavakol, ``{Density perturbations in $f(R)$
  gravity theories in metric and Palatini formalisms},''
  \href{http://dx.doi.org/10.1103/physrevd.77.043007}{{\em Physical Review D}
  {\bf 77} (2008) no.~4, }, \href{http://arxiv.org/abs/0712.0082}{{\tt
  arXiv:0712.0082 [astro-ph.CO]}}.

\bibitem{Basilakos_2013}
S.~Basilakos, S.~Nesseris, and L.~Perivolaropoulos, ``Observational constraints
  on viable $f(r)$ parametrizations with geometrical and dynamical probes,''
  \href{http://dx.doi.org/10.1103/physrevd.87.123529}{{\em Physical Review D}
  {\bf 87} (2013) no.~12, }, \href{http://arxiv.org/abs/1302.6051}{{\tt
  arXiv:1302.6051 [astro-ph.CO]}}.

\bibitem{Anagnostopoulos_2019}
F.~K. Anagnostopoulos, S.~Basilakos, and E.~N. Saridakis, ``{Bayesian analysis
  of $f(T)$ gravity using $f \sigma_8$ data },''
  \href{http://dx.doi.org/10.1103/physrevd.100.083517}{{\em Physical Review D}
  {\bf 100} (2019) no.~8, }, \href{http://arxiv.org/abs/1907.07533}{{\tt
  arXiv:1907.07533 [astro-ph.CO]}}.

\bibitem{Peebles1993}
P.~J.~E. Peebles, {\em Principles of Physical Cosmology, Princeton University
  Press, Princeton New Jersey}.
\newblock 1993.

\bibitem{Sola:2017znb}
J.~Sol\`a, A.~G\'omez-Valent, and J.~de~Cruz~P\'erez, ``{The $H_0$ tension in
  light of vacuum dynamics in the Universe},''
  \href{http://dx.doi.org/10.1016/j.physletb.2017.09.073}{{\em Phys. Lett. B}
  {\bf 774} (2017)  317--324}, \href{http://arxiv.org/abs/1705.06723}{{\tt
  arXiv:1705.06723 [astro-ph.CO]}}.

\bibitem{Gonzalez-Espinoza:2018gyl}
M.~Gonzalez-Espinoza, G.~Otalora, J.~Saavedra, and N.~Videla, ``{Growth of
  matter overdensities in non-minimal torsion-matter coupling theories},'' {\em
  Eur. Phys. J. C} {\bf 78} (2018) no.~10, 799,
  \href{http://arxiv.org/abs/1808.01941}{{\tt arXiv:1808.01941 [gr-qc]}}.

\bibitem{Gomez-Valent:2018nib}
A.~G\'omez-Valent and J.~Sol\`a~Peracaula, ``{Density perturbations for running
  vacuum: a successful approach to structure formation and to the
  $\sigma_8$-tension},'' \href{http://dx.doi.org/10.1093/mnras/sty1028}{{\em
  Mon. Not. Roy. Astron. Soc.} {\bf 478} (2018) no.~1, 126--145},
  \href{http://arxiv.org/abs/1801.08501}{{\tt arXiv:1801.08501 [astro-ph.CO]}}.

\bibitem{DeFelice:2010aj}
A.~De~Felice and S.~Tsujikawa, ``{f(R) theories},''
  \href{http://dx.doi.org/10.12942/lrr-2010-3}{{\em Living Rev. Rel.} {\bf 13}
  (2010)  3}, \href{http://arxiv.org/abs/1002.4928}{{\tt arXiv:1002.4928
  [gr-qc]}}.

\bibitem{Kazantzidis:2018rnb}
L.~Kazantzidis and L.~Perivolaropoulos, ``{Evolution of the $f\sigma_8$ tension
  with the Planck15/$\Lambda$CDM determination and implications for modified
  gravity theories},'' \href{http://dx.doi.org/10.1103/PhysRevD.97.103503}{{\em
  Phys. Rev. D} {\bf 97} (2018) no.~10, 103503},
  \href{http://arxiv.org/abs/1803.01337}{{\tt arXiv:1803.01337 [astro-ph.CO]}}.

\bibitem{DiValentino_2018a}
{Di Valentino, Eleonora and others}, ``{Reducing the $H_0$ and $\sigma_8$
  tensions with dark matter-neutrino interactions},''
  \href{http://dx.doi.org/10.1103/physrevd.97.043513}{{\em Physical Review D}
  {\bf 97} (2018) no.~4, }, \href{http://arxiv.org/abs/1710.02559}{{\tt
  arXiv:1710.02559 [astro-ph.CO]}}.

\bibitem{Poulin_2023}
V.~Poulin, J.~L. Bernal, E.~D. Kovetz, and M.~Kamionkowski, ``Sigma-8 tension
  is a drag,'' \href{http://dx.doi.org/10.1103/physrevd.107.123538}{{\em
  Physical Review D} {\bf 107} (2023) no.~12, },
  \href{http://arxiv.org/abs/2209.06217}{{\tt arXiv:2209.06217 [astro-ph.CO]}}.

\bibitem{PhysRevD.90.044016a}
S.~Capozziello {\em et al.}, ``{Cosmographic bounds on the cosmological
  deceleration-acceleration transition redshift in $f(\mathcal{R})$ gravity},''
  \href{http://dx.doi.org/10.1103/PhysRevD.90.044016}{{\em Phys. Rev. D} {\bf
  90} (2014)  044016}, \href{http://arxiv.org/abs/1403.1421}{{\tt
  arXiv:1403.1421 [gr-qc]}}.

\bibitem{PhysRevResearch.2.013028}
D.~Camarena and V.~Marra, ``Local determination of the hubble constant and the
  deceleration parameter,''
  \href{http://dx.doi.org/10.1103/PhysRevResearch.2.013028}{{\em Phys. Rev.
  Res.} {\bf 2} (2020)  013028}, \href{http://arxiv.org/abs/1906.11814}{{\tt
  arXiv:1906.11814 [astro-ph.CO]}}.

\bibitem{sahniPRD_om}
V.~Sahni, A.~Shafieloo, and A.~A. Starobinsky, ``Two new diagnostics of dark
  energy,'' \href{http://dx.doi.org/10.1103/PhysRevD.78.103502}{{\em Phys. Rev.
  D} {\bf 78} (2008)  103502}, \href{http://arxiv.org/abs/0807.3548}{{\tt
  arXiv:0807.3548 [astro-ph.CO]}}.

\bibitem{Sahni_2003}
V.~Sahni {\em et al.}, ``Statefinder-a new geometrical diagnostic of dark
  energy,'' \href{http://dx.doi.org/10.1134/1.1574831}{{\em Journal of
  Experimental and Theoretical Physics Letters} {\bf 77} (2003) no.~5,
  201--206}, \href{http://arxiv.org/abs/astro-ph/0201498}{{\tt
  arXiv:astro-ph/0201498 [astro-ph.CO]}}.

\end{thebibliography}\endgroup
\end{document}